\documentclass[preprint2]{emulateapj-rtx4} 
\pdfoutput=1
\usepackage{amsmath,natbib,graphicx}
\usepackage{epsf}
\input{epsf}
\usepackage{epstopdf}
\usepackage{multirow}
\usepackage{epsfig}
\usepackage{color}
\usepackage{lscape}
\usepackage{booktabs}

\usepackage{ifthen}

\DeclareGraphicsExtensions{.jpg,.pdf,.png,.eps,.ps}
\graphicspath{{figures/}}

\DeclareGraphicsExtensions{.jpg,.pdf,.png,.eps,.ps}

\newcommand{\clustera}{SPT-CL J0516-5430}
\newcommand{\clusterb}{SPT-CL J2022-6323}
\newcommand{\clusterc}{SPT-CL J2030-5638}
\newcommand{\clusterd}{SPT-CL J2032-5627}
\newcommand{\clustere}{SPT-CL J2135-5726}

\newcommand{\rhocrit}{\mbox{$\rho_{\mathrm{crit}}$}}

\newcommand{\ltsima}{$\; \buildrel < \over \sim \;$}
\newcommand{\ltsim}{\lower.5ex\hbox{\ltsima}}

\newcommand{\be}{\begin{equation}}
\newcommand{\ee}{\end{equation}}
\newcommand{\bea}{\begin{eqnarray}}
\newcommand{\eea}{\end{eqnarray}}

\newcommand{\degs}{deg$^2$}

\newcommand{\Msun}{\mbox{$M_\odot$}}
\newcommand{\degree}{\ensuremath{^\circ}}

\newcommand{\cfhti}{\ensuremath{i_{\mathrm{CFHT}}}}

\newcommand{\sdssg}{\ensuremath{g_{\mathrm{SDSS}}}}
\newcommand{\sdssr}{\ensuremath{r_{\mathrm{SDSS}}}}
\newcommand{\sdssi}{\ensuremath{i_{\mathrm{SDSS}}}}

\newcommand{\cmcg}{\ensuremath{g_{\mathrm{CMC}}}}
\newcommand{\cmcr}{\ensuremath{r_{\mathrm{CMC}}}}
\newcommand{\cmci}{\ensuremath{i_{\mathrm{CMC}}}}

\newcommand{\zphot}{\ensuremath{z_{\mathrm{phot}}}}

\newcommand{\zcluster}{\ensuremath{z_{\mathrm{l}}}}

\def\KICPChicago{1}
\def\AAUChicago{2}
\def\Leiden{3}
\def\PhysicsUChicago{4}
\def\McGill{5}
\def\UChicago{6}
\def\NCSA{7}
\def\CfA{8}
\def\MIT{9}
\def\Harvard{10}
\def\Munich{11}
\def\ExcellenceCluster{12}
\def\EFIChicago{13}
\def\Miss{14}
\def\ANL{15}
\def\NIST{16}
\def\PUC{17}
\def\Berkeley{18}
\def\UFlorida{19}
\def\Colorado{20}
\def\NASA{21}
\def\Davis{22}
\def\LBNL{23}
\def\Caltech{24}
\def\Arizona{25}
\def\Michigan{26}
\def\MPE{27}
\def\CaseWestern{28}
\def\Minnesota{29}
\def\STScI{30}
\def\SAIC{31}
\def\Yale{32}
\def\Bonn{33}
\def\BCCP{34}

\begin{document}

\title{Weak-Lensing Mass Measurements of Five Galaxy Clusters\\in the
  South Pole Telescope Survey Using Magellan/Megacam}
% \title{First Direct Weak Lensing Test of the Calibration\\of Galaxy
%   Cluster Masses in the South Pole Telescope Survey}

\slugcomment{Submitted to \apj}

\author{
F.~W.~High\altaffilmark{\KICPChicago,\AAUChicago}, 
H.~Hoekstra\altaffilmark{\Leiden},
N.~Leethochawalit\altaffilmark{\PhysicsUChicago}, 
T.~de~Haan\altaffilmark{\McGill},
L.~Abramson\altaffilmark{\AAUChicago},
K.~A.~Aird\altaffilmark{\UChicago},
R.~Armstrong\altaffilmark{\NCSA},
M.~L.~N.~Ashby\altaffilmark{\CfA},
M.~Bautz\altaffilmark{\MIT},
M.~Bayliss\altaffilmark{\Harvard}, 
G.~Bazin\altaffilmark{\Munich,\ExcellenceCluster},
B.~A.~Benson\altaffilmark{\KICPChicago,\EFIChicago},
L.~E.~Bleem\altaffilmark{\KICPChicago,\PhysicsUChicago},
M.~Brodwin\altaffilmark{\Miss},
J.~E.~Carlstrom\altaffilmark{\KICPChicago,\AAUChicago,\PhysicsUChicago,\EFIChicago,\ANL}, 
C.~L.~Chang\altaffilmark{\KICPChicago,\EFIChicago,\ANL}, 
H.~M. Cho\altaffilmark{\NIST}, 
A.~Clocchiatti\altaffilmark{\PUC},
M.~Conroy\altaffilmark{\CfA},
T.~M.~Crawford\altaffilmark{\KICPChicago,\AAUChicago},
A.~T.~Crites\altaffilmark{\KICPChicago,\AAUChicago},
S.~Desai\altaffilmark{\Munich,\ExcellenceCluster},
M.~A.~Dobbs\altaffilmark{\McGill},
J.~P.~Dudley\altaffilmark{\McGill},
R.~J.~Foley\altaffilmark{\CfA}, 
W.~R.~Forman\altaffilmark{\CfA},
E.~M.~George\altaffilmark{\Berkeley},
M.~D.~Gladders\altaffilmark{\KICPChicago,\AAUChicago},
A.~H.~Gonzalez\altaffilmark{\UFlorida},
N.~W.~Halverson\altaffilmark{\Colorado},
N.~L.~Harrington\altaffilmark{\Berkeley},
G.~P.~Holder\altaffilmark{\McGill},
W.~L.~Holzapfel\altaffilmark{\Berkeley},
S.~Hoover\altaffilmark{\KICPChicago,\EFIChicago},
J.~D.~Hrubes\altaffilmark{\UChicago},
C.~Jones\altaffilmark{\CfA},
M.~Joy\altaffilmark{\NASA},
R.~Keisler\altaffilmark{\KICPChicago,\PhysicsUChicago},
L.~Knox\altaffilmark{\Davis},
A.~T.~Lee\altaffilmark{\Berkeley,\LBNL},
E.~M.~Leitch\altaffilmark{\KICPChicago,\AAUChicago},
J.~Liu\altaffilmark{\Munich,\ExcellenceCluster},
M.~Lueker\altaffilmark{\Berkeley,\Caltech},
D.~Luong-Van\altaffilmark{\UChicago},
A.~Mantz\altaffilmark{\KICPChicago},
D.~P.~Marrone\altaffilmark{\Arizona},
M.~McDonald\altaffilmark{\MIT},
J.~J.~McMahon\altaffilmark{\KICPChicago,\EFIChicago,\Michigan},
J.~Mehl\altaffilmark{\KICPChicago,\AAUChicago},
S.~S.~Meyer\altaffilmark{\KICPChicago,\AAUChicago,\PhysicsUChicago,\EFIChicago},
L.~Mocanu\altaffilmark{\KICPChicago,\AAUChicago},
J.~J.~Mohr\altaffilmark{\Munich,\ExcellenceCluster,\MPE},
T.~E.~Montroy\altaffilmark{\CaseWestern},
S.~S.~Murray\altaffilmark{\CfA},
T.~Natoli\altaffilmark{\KICPChicago,\PhysicsUChicago},
D.~Nurgaliev\altaffilmark{\Harvard},
S.~Padin\altaffilmark{\KICPChicago,\AAUChicago,\Caltech},
T.~Plagge\altaffilmark{\KICPChicago,\AAUChicago},
C.~Pryke\altaffilmark{\Minnesota}, 
C.~L.~Reichardt\altaffilmark{\Berkeley},
A.~Rest\altaffilmark{\STScI},
J.~Ruel\altaffilmark{\Harvard},
J.~E.~Ruhl\altaffilmark{\CaseWestern}, 
B.~R.~Saliwanchik\altaffilmark{\CaseWestern}, 
A.~Saro\altaffilmark{\Munich},
J.~T.~Sayre\altaffilmark{\CaseWestern}, 
K.~K.~Schaffer\altaffilmark{\KICPChicago,\EFIChicago,\SAIC}, 
L.~Shaw\altaffilmark{\McGill,\Yale},
T.~Schrabback\altaffilmark{\Bonn},
E.~Shirokoff\altaffilmark{\Berkeley,\Caltech}, 
J.~Song\altaffilmark{\Michigan},
H.~G.~Spieler\altaffilmark{\LBNL},
B.~Stalder\altaffilmark{\CfA},
Z.~Staniszewski\altaffilmark{\CaseWestern},
A.~A.~Stark\altaffilmark{\CfA}, 
K.~Story\altaffilmark{\KICPChicago,\PhysicsUChicago},
C.~W.~Stubbs\altaffilmark{\CfA,\Harvard}, 
R.~\v{S}uhada\altaffilmark{\Munich},
S.~Tokarz\altaffilmark{\CfA},
A.~van~Engelen\altaffilmark{\McGill},
K.~Vanderlinde\altaffilmark{\McGill},
J.~D.~Vieira\altaffilmark{\KICPChicago,\PhysicsUChicago,\Caltech},
A. Vikhlinin\altaffilmark{\CfA},
R.~Williamson\altaffilmark{\KICPChicago,\AAUChicago}, 
O.~Zahn\altaffilmark{\Berkeley,\BCCP},
and
A.~Zenteno\altaffilmark{\Munich,\ExcellenceCluster}
}

\altaffiltext{\KICPChicago}{Kavli Institute for Cosmological Physics, University of Chicago, 5640 South Ellis Avenue, Chicago, IL 60637}
\altaffiltext{\AAUChicago}{Department of Astronomy and Astrophysics, University of Chicago, 5640 South Ellis Avenue, Chicago, IL 60637}
\altaffiltext{\Leiden}{Leiden Observatory, Leiden University, Leiden, The Netherlands}
\altaffiltext{\PhysicsUChicago}{Department of Physics,
University of Chicago,
5640 South Ellis Avenue, Chicago, IL 60637}
\altaffiltext{\McGill}{Department of Physics,
McGill University,
3600 Rue University, Montreal, Quebec H3A 2T8, Canada}
\altaffiltext{\UChicago}{University of Chicago,
5640 South Ellis Avenue, Chicago, IL 60637}
\altaffiltext{\NCSA}{National Center for Supercomputing Applications,
University of Illinois, 1205 West Clark Street, Urbana, IL 61801}
\altaffiltext{\CfA}{Harvard-Smithsonian Center for Astrophysics,
60 Garden Street, Cambridge, MA 02138}
\altaffiltext{\MIT}{MIT Kavli Institute for Astrophysics and Space
Research, Massachusetts Institute of Technology, 77 Massachusetts Avenue,
Cambridge, MA 02139}
\altaffiltext{\Harvard}{Department of Physics, Harvard University, 17 Oxford Street, Cambridge, MA 02138}
\altaffiltext{\Munich}{Department of Physics,
Ludwig-Maximilians-Universit\"{a}t,
Scheinerstr.\ 1, 81679 M\"{u}nchen, Germany}
\altaffiltext{\ExcellenceCluster}{Excellence Cluster Universe,
Boltzmannstr.\ 2, 85748 Garching, Germany}
\altaffiltext{\EFIChicago}{Enrico Fermi Institute,
University of Chicago,
5640 South Ellis Avenue, Chicago, IL 60637}
\altaffiltext{\Miss}{Department of Physics and Astronomy, University of Missouri, 5110 Rockhill Road, Kansas City, MO 64110}
\altaffiltext{\ANL}{Argonne National Laboratory, 9700 S. Cass Avenue, Argonne, IL, USA 60439}
\altaffiltext{\NIST}{NIST Quantum Devices Group, 325 Broadway Mailcode 817.03, Boulder, CO, USA 80305}
\altaffiltext{\PUC}{Departamento de Astronom\'{\i}a y Astrof\'{\i}sica, Pontificia Universidad Cat\'olica, Casilla 306, Santiago 22, Chile}
\altaffiltext{\Berkeley}{Department of Physics,
University of California, Berkeley, CA 94720}
\altaffiltext{\UFlorida}{Department of Astronomy, University of Florida, Gainesville, FL 32611}
\altaffiltext{\Colorado}{Department of Astrophysical and Planetary Sciences and Department of Physics,
University of Colorado,
Boulder, CO 80309}
\altaffiltext{\NASA}{Department of Space Science, VP62,
NASA Marshall Space Flight Center,
Huntsville, AL 35812}
\altaffiltext{\Davis}{Department of Physics, 
University of California, One Shields Avenue, Davis, CA 95616}
\altaffiltext{\LBNL}{Physics Division,
Lawrence Berkeley National Laboratory,
Berkeley, CA 94720}
\altaffiltext{\Caltech}{California Institute of Technology, 1200 E. California Blvd., Pasadena, CA 91125}
\altaffiltext{\Arizona}{Steward Observatory, University of Arizona, 933 North Cherry Avenue, Tucson, AZ 85721}
\altaffiltext{\Michigan}{Department of Physics, University of Michigan, 450 Church Street, Ann  
Arbor, MI, 48109}
\altaffiltext{\MPE}{Max-Planck-Institut f\"{u}r extraterrestrische Physik,
Giessenbachstr.\ 85748 Garching, Germany}
\altaffiltext{\CaseWestern}{Physics Department, Center for Education and Research in Cosmology 
and Astrophysics, 
Case Western Reserve University,
Cleveland, OH 44106}
\altaffiltext{\Minnesota}{Physics Department, University of Minnesota, 116 Church Street S.E., Minneapolis, MN 55455}
\altaffiltext{\STScI}{Space Telescope Science Institute, 3700 San Martin
Dr., Baltimore, MD 21218}
\altaffiltext{\SAIC}{Liberal Arts Department, 
School of the Art Institute of Chicago, 
112 S Michigan Ave, Chicago, IL 60603}
\altaffiltext{\Yale}{Department of Physics, Yale University, P.O. Box 208210, New Haven,
CT 06520-8120}
\altaffiltext{\Bonn}{Argelander-Institut f\"{u}r Astronomie, Universit\"{a}t Bonn, Auf dem H\"{u}gel 71, 53121, Bonn, Germany}
\altaffiltext{\BCCP}{Berkeley Center for Cosmological Physics,
Department of Physics, University of California, and Lawrence Berkeley
National Labs, Berkeley, CA 94720}

\shorttitle{Weak Lensing Measurements of Five SPT Clusters}

\shortauthors{F.~W.~High et al.}

\email{fwhigh@kicp.uchicago.edu}

\begin{abstract}

  We use weak gravitational lensing to measure the masses of five
  galaxy clusters selected from the South Pole Telescope (SPT) survey,
  with the primary goal of comparing these with the SPT
  Sunyaev--Zel'dovich (SZ) and X-ray based mass estimates. The clusters
  span redshifts $0.28 < z < 0.43$ and have masses $M_{500} >
  2\times10^{14} h^{-1} M_\odot$, and three of the five clusters were
  discovered by the SPT survey. We observed the clusters in the $g'r'i'$
  passbands with the Megacam imager on the Magellan Clay 6.5m
  telescope.  We measure a mean ratio of weak lensing (WL) aperture masses
  to inferred aperture masses from the SZ data, both within an
  aperture of $R_{500,\mathrm{SZ}}$ derived from the SZ mass, of $1.04
  \pm 0.18$. We measure a mean ratio of spherical WL masses
  evaluated at $R_{500,\mathrm{SZ}}$ to spherical SZ masses of $1.07
  \pm 0.18$, and a mean ratio of spherical WL masses
  evaluated at $R_{500,\mathrm{WL}}$ to spherical SZ masses of $1.10
  \pm 0.24$. 
  % We test each weak-lensing mass statistic in mock catalogs based on
  % N-body simulations under simple assumptions, and measure mean biases
  % between about $-5\%$ and $-10\%$, depending on the statistic, which
  % is consistent with a number of previous works.
  We explore potential sources of systematic
  error in the mass comparisons and conclude that all are subdominant
  to the statistical uncertainty, with dominant terms being cluster
  concentration uncertainty and $N$-body simulation calibration bias. 
  Expanding the sample of SPT clusters
  with WL observations has the potential to significantly
  improve the SPT cluster mass calibration and the resulting
  cosmological constraints from the SPT cluster survey. These are the
  first WL detections using Megacam on the Magellan Clay
  telescope.

\end{abstract}


\defcitealias{andersson10}{Andersson10}
\defcitealias{zhang08}{Zhang08}
\defcitealias{vanderlinde10}{V10}
\defcitealias{high10}{H10}
\defcitealias{mortonson11}{M11}
\defcitealias{navarro97}{NFW}
\defcitealias{duffy08}{D08}
\defcitealias{reichardt12}{R12}
\defcitealias{benson11}{B11}
\defcitealias{kaiser95b}{KSB}
\defcitealias{hurleywalker12}{AMI Consortium: Hurley-Walker et al.\ (2012)}	
\defcitealias{planck12}{Planck Collaboration: Aghanim et al.\ (2012)}	

%\tableofcontents

\section{Introduction}
\label{sec:intro}

The abundance of galaxy clusters as a function of mass and redshift is
sensitive to dark energy and other cosmological parameters through the
growth function of large-scale structure (LSS) and the cosmological volume
element \citep[e.g.,][]{wang98,haiman01,holder01b,weller03}.  As
emphasized by the Dark Energy Task Force \citep{albrecht06}, the
abundance of clusters provides constraints on dark energy that are
complementary to those of distance-redshift relations, such as
standard candles and rulers including Type Ia supernovae (SNe) and
baryon acoustic oscillations (BAO).  Recent results using this method
have shown that cluster surveys can significantly improve the best
current constraints on cosmological parameters, particularly the dark
energy equation of state
\citep[][]{vikhlinin09,mantz10b,rozo10,benson11,reichardt12}.

The Sunyaev--Zel'dovich (SZ) effect offers a novel way to search for
high-redshift, massive clusters, which are particularly useful for
constraining cosmology \citep[e.g.,][]{carlstrom02}.  The SZ effect is
the inverse Compton scattering of cosmic microwave background (CMB)
photons by the hot electrons in the intracluster medium.  SZ
observables are nearly redshift independent, and moreover, are
expected from simulations and observations to trace total cluster mass
with low intrinsic scatter \citep[e.g.,][]{kravtsov06a,benson11}.  To
extract constraints on cosmological parameters, the cluster redshifts
must be measured with optical and infrared follow-up observations and
the cluster masses must be estimated using accurately calibrated
proxies.

The South Pole Telescope \citep[SPT;][]{carlstrom11} is a
millimeter-wavelength telescope that recently completed a 2500 \degs \
SZ cluster survey.  A catalog from the first $720\deg^2$ of the survey
has been released that includes 224 cluster candidates, 158 of which
have confirmed optical or infrared galaxy cluster counterparts with
redshifts as high as $z = 1.37$, a median redshift of $0.55$, and a
median mass of $M_{500} \approx 2.3 \times 10^{14}h^{-1}\Msun$
\citep[][hereafter \citetalias{reichardt12}]{reichardt12}. Using the
100 cluster candidates at $z > 0.3$ above the SPT $95 \%$-purity
threshold, the SPT cluster data have been combined with
CMB+BAO+$H_0$+SNe data to provide constraints on the dark energy
equation of state of $w=-1.010\pm0.058$, a factor of 1.3 improvement
over the constraints without the cluster abundance data.  However,
this improvement was limited by the $\sim10\%$ uncertainty in the SPT
cluster mass calibration.

The method to determine the masses of clusters in the SPT survey is
described in detail by \citet[][hereafter
\citetalias{benson11}]{benson11}.  In brief, the mass estimates are
computed from the SPT SZ data, and are calibrated to scale with total
mass using
a combination of X-ray observations and cluster
simulations.  For a subset of the SPT survey cluster sample, X-ray
observations have been obtained to measure $Y_X$, the product of the
X-ray derived gas mass and core-excised temperature.  This is used in
combination with a $Y_X$-mass relation that is calibrated at
low-redshift ($z < 0.3$) using X-ray derived hydrostatic mass
estimates of relaxed clusters \citep{vikhlinin09b}.  The calibration
of the $Y_X$--mass relation is expected to be accurate to $10$\% based
on simulations \citep{nagai07}, and has been empirically verified to
have this level of accuracy using weak lensing (WL) observations of the
same clusters \citep{hoekstra07}.

WL here refers to the subtle tangential shearing of
extended sources behind cluster halos, at projected distances well
outside the Einstein radius \citep[for a review, see][]{bartelmann01}.
Gravitational lensing is sensitive to total projected mass and has the
benefit of being insensitive to the dynamical state of the lens: the
observables are independent of whether the cluster gas or galaxies are
in hydrostatic equilibrium.  WL mass estimates have been
used to test the accuracy of X-ray mass estimates
\citep[e.g.,][]{hoekstra07, mahdavi08, zhang08}; however, these works
have not explicitly cross-checked the $Y_X$-mass calibration of
\citet{vikhlinin09b}, and have used clusters predominantly at low
redshift ($z < 0.3$) that are outside the SPT survey region.

As the first steps toward a WL-based calibration of the SPT cluster
sample, we have observed five clusters from the SPT survey with the
Magellan Clay-Megacam CCD imager.  The five clusters span redshifts
$0.28<z<0.43$ and were selected from the SPT catalog of
\citetalias{reichardt12} from a subset of clusters with existing or
scheduled observations by either the {\it Chandra} or {\it XMM-Newton}
X-ray space-telescopes.  
From this SZ--plus--¯X-ray sample, we randomly selected five clusters
that were observable during awarded telescope time and which were at
$0.3 \lesssim z \lesssim 0.5$.
The X-ray analyses for these clusters are in progress.
We have also obtained multi-object spectroscopy of cluster
members for precise redshift measurements of four of the clusters.

The primary goal of this paper is to provide the first direct
comparison of the SZ and X-ray derived mass estimates of
\citetalias{reichardt12} to WL mass measurements.  We employ
aperture masses, which are largely insensitive to the properties of
cluster cores, and have been shown in previous works to scale well
with other, low-scatter mass observables
\citep[e.g.,][]{hoekstra07,mahdavi08}.  We also use spherical masses
by fitting the WL shear data to analytic profiles.

% Section \ref{sec:data} describes the WL imaging data, including
% observations with Megacam, image reductions, and photometry.  Section
% \ref{sec:shear} summarizes the WL theory used in the shear
% measurement, including our adopted method of estimating it as well as
% the PSF performance of Clay-Megacam.  Section \ref{sec:wlmasses}
% describes how mass is estimated from shear catalogs, including
% source-redshift estimation techniques.  Shear profiles are shown in
% Section \ref{app:figures}.  Section \ref{sec:sz} summarizes the SZ
% mass estimates from SPT survey data.  Section \ref{sec:comp} compares
% total cluster mass estimates from the WL to the SZ masses.
% %% Section \ref{sec:error} is an error analysis.  
% We conclude with Section \ref{sec:conclusion}.

Magnitudes are in the Sloan Digital Sky Survey (SDSS) AB system unless
otherwise noted. We adopt a flat $\Lambda$CDM 
cosmology with $\Omega_{\mathrm{M}} =
0.27$ and $H_0 = 70\,\mathrm{km}\,\mathrm{s}^{-1}\,\mathrm{Mpc}^{-1}$
\citep{komatsu11}. Masses $M_{\Delta}$ are defined at radius
$R_{\Delta}$, where the mean interior density is $\Delta$ times the
critical density of the universe at the cluster's epoch, $\rhocrit(z)
= 3H^2(z)/8\pi G$, and $H(z)$ is the Hubble parameter.

\section{The Cluster Sample}
\label{sec:clusters}

In this section, we provide an overview of the larger sample of
clusters from which we selected targets for WL follow-up.
We briefly discuss SZ detection with SPT, the SZ cluster center
determination, and SZ mass estimation. We then discuss the cluster
spectroscopic redshift measurements of four of the five systems
selected for WL observations. Table \ref{tab:clusters}
summarizes these basic cluster data.

\begin{deluxetable*}{clcccc}
%\tabletypesize{\scriptsize}
%\rotate
\tablecaption{The cluster sample\label{tab:clusters}}
\tablewidth{0pt}
\tablehead{
\colhead{Cluster Name} & 
\colhead{$z_l$} &
\colhead{$N$} &
\colhead{$\xi$} &
\colhead{SZ R.A.} &
\colhead{SZ Decl.} \\
\colhead{~¯} &
\colhead{~} &
\colhead{~} &
\colhead{~} &
\colhead{(deg J2000)} &
\colhead{(deg J2000)} 
}
\startdata
\clustera\tablenotemark{a} & 0.294(1) & 48 & 9.42 & 79.1480 & $-54.5062$ \\
\clusterb\tablenotemark{b} & 0.383(1) & 37 & 6.58 & 305.5235 & $-63.3973$ \\
\clusterc\tablenotemark{b} & 0.40(4)  & \nodata & 5.47 & 307.7067 & $-56.6352$ \\
\clusterd\tablenotemark{c} & 0.284(1) & 31 & 8.14 & 308.0800 & $-56.4557$ \\
\clustere\tablenotemark{b} & 0.427(1) & 33 & 10.43 & 323.9158 & $-57.4415$ 
\enddata
\tablecomments{Basic data for the five clusters we have targeted for
  weak lensing analysis.  \\{Column 1:} \citetalias{reichardt12}
  cluster designation.  \\{Column 2:} mean cluster redshift as
  measured from the ensemble of cluster members. The number in
  parentheses is the uncertainty in the last digit.  All redshifts are
  spectroscopic except for \clusterc, which was photometrically
  derived (\citetalias[][]{reichardt12}; \citealt{song12}). \\{
    Column 3:} number of cluster galaxies for which we successfully
  measured spectroscopic redshifts. \\{Column 4:} peak
  signal-to-noise ratio of the SZ detection. \\{Column 5:} right
  ascension of the SZ signal-to-noise ratio centroid.  \\{Column 6:}
  declination of the SZ signal-to-noise ratio centroid.}
\tablenotetext{a}{Discovered by \citet{abell89} where it was
  designated as AS0520. The spectroscopic redshift was measured
  to be $z_l=0.2950$ \citep{leccardi08}.}  \tablenotetext{b}{Not known
  prior to \citetalias{reichardt12}.}  \tablenotetext{c}{A number of
  structures have been identified near this cluster by other
  authors. See Section \ref{sec:unrelaxed}, \citetalias{reichardt12},
  and \citet{song12} for discussions.}
\end{deluxetable*}

\subsection{SZ Cluster Detection}
\label{sec:szdetect}

The clusters were selected from the \citetalias{reichardt12} SPT
cluster survey catalog.  That work contains details of the sample, the
SPT data from which it was extracted, and the cluster extraction
process.  In summary, $\sim$720 \degs \ of sky were surveyed by the
SPT in the 2008 and 2009 observing seasons to a depth such that the
median mass of a cluster detected is $M_{500} \approx 2.3 \times
10^{14}h^{-1}\Msun$.  Cluster candidates are extracted from the data
using a multifrequency matched filter \citep{melin06}.  Twelve
different matched filters are used spanning a range of angular scales
for the assumed cluster profile, and the cluster candidates are ranked
by the maximum detection significance across all filter scales,
defined as $\xi$.  All candidates with $\xi \ge 4.5$ are included in
the catalog, with $\xi$ also used as the primary observable to
determine the cluster mass.
%{\bf Info on SPT, the 2008-2009 survey, how data
%  were processed, clusters detected, and feel free to send the reader
%  to the literature.}  
SZ significance maps for the five clusters discussed in this work
are shown in the Appendix.

%{\bf Insert words on SZ position determination and accuracy here.} 
We use the SZ detection positions as the cluster centers in the
baseline WL analysis, although we explore the effect of using
%Brightest Cluster Galaxy (BCG) positions and WL-reconstructed
%convergence-field peak
other positions as well.  The statistical uncertainty in SZ-determined
positions is a function of the cluster size, the SPT beam size
($\mathrm{FWHM}=1\farcm6$ and $1\farcm19$ at $95\mathrm{GHz}$ and
$150\mathrm{GHz}$ respectively), and the significance of detection.
In the limit that clusters are point sources in the SPT data, the rms
positional uncertainty is $\sim \theta_\mathrm{FWHM}/\xi$, where
$\theta_\mathrm{FWHM}$ is the beam width \citep{ivison07}.  For
resolved clusters, the uncertainty is $\sim
\sqrt{\theta_\mathrm{FWHM}^2+(k \theta_{c})^2}/\xi$, where $\theta_c$
is the cluster core size, and $k$ is a factor of order unity
\citet{story11,song12}.
For clusters such as those described in this
work (with $\xi \sim 8$ and $z \sim 0.4$), this uncertainty is
estimated to be $10\arcsec$--$15\arcsec$.
% Astrophysical effects such as recent
% mergers can significantly increase the rms offset between the center
% of the gas responsible for the SZ signal and the dark matter chiefly
% responsible for weak lensing.  However, \citet{song12} show that the
% offsets between the SZ-determined centroid and the position of the BCG
% for the \citetalias{reichardt12} clusters is consistent with the
% statistical uncertainty from SPT centroid measurements, implying that
% merger-driven offsets are subdominant over the full sample.

\subsection{SZ Mass Estimates}
\label{sec:sz}

In this work, we use the mass estimates from \citetalias{reichardt12}.
The SPT mass calibration and method to estimate cluster masses from
the SZ and X-ray data is described in detail in \citetalias{benson11}
and \citetalias{reichardt12}.  In summary, a probability density
function of each cluster's mass estimate was calculated at each point
in a Markov Chain Monte Carlo (MCMC) that varied both the cluster
scaling relations and cosmological parameters assuming a $\Lambda$CDM
cosmology and using the CMB+BAO+SNe+SPT$_{\mathrm{CL}}$ data, where
SPT$_{\mathrm{CL}}$ denotes the added cluster data set.  In effect,
this step is calculating the posterior probability given the
measurement uncertainties and the expected distribution of galaxy
cluster masses for that specific cosmology and scaling relation.  The
resulting masses are SZ plus X-ray posterior mass estimates where
applicable; even for clusters without X-ray data, the mass
normalization from clusters with X-ray data affects the SZ scaling
relation parameters explored by the chain.  The probability density
functions for different points in the chain are combined to obtain a
mass estimate that has been fully marginalized over all cosmological
and scaling relation parameters.  The final products are labeled
$M_{500,\mathrm{SZ}}$ here.

The uncertainty on $M_{500,\mathrm{SZ}}$ is dominated by a $\sim 15\%$
intrinsic scatter in the SZ-mass scaling relation, a $\sim 10\%$
uncertainty due to the finite detection significance of the cluster in
the SZ maps, and a $\sim 10\%$ systematic uncertainty associated with
the normalization of the SZ--mass scaling relation.  Together, these
and other sources of uncertainty yield a $\sim 20\%$ uncertainty on
$M_{500,\mathrm{SZ}}$ for each cluster.

\subsection{Spectroscopic Redshifts}
\label{sec:mos}

We have targeted four of the clusters for multi-object spectroscopy
observations, the details of which will appear in J.\ Ruel et al.\ (in
preparation).  In summary, we used GISMO (Gladders Image Slicing Multi-slit
Option) on IMACS \citep[Inamori Magellan Areal
Camera and Spectrograph;][]{dressler03,osip08} at the Magellan Baade
$6.5\,\mathrm{m}$ 
telescope in 2010 September and October. These observations used the
$f/4$ camera, the $z1\,430-675$ filter, and the $300\,\mathrm{lines}\,\mathrm{mm}^{-1}$
grating. Conditions were photometric and the seeing varied between
$0\farcs5$ and $0\farcs9$.  Reduction of the raw spectra was done with
the COSMOS package\footnote{\url{http://code.obs.carnegiescience.edu/cosmos}}
(Carnegie Observatories System for Multi-Object Spectroscopy);
redshifts were measured by cross-correlation with the ``fabtemp97''
template in RVSAO \citep[Radial Velocity Smithsonian Astrophysical
Observatory package,][]{kurtz98} and checked for agreement with
visually-identified features.  Outliers were rejected by iterative
clipping at $3\sigma$. The redshift of each cluster was calculated
using the robust biweight estimator \citep{beers90} and the confidence
interval by bootstrap resampling.  These results are given in Table
\ref{tab:clusters}.

The redshift of \clusterc\ is photometrically derived from the red
sequence of cluster galaxies, the details of which are described by
\citet{song12} and \citetalias{reichardt12}.

\section{Weak Lensing Data}
\label{sec:data}

In this section we describe the acquisition, reduction, and
calibration of the images and photometry from which weak lensing
masses are measured.

\subsection{Observations}

We imaged through $g'$, $r'$, and $i'$ passbands using Megacam on the
Magellan Clay $6.5\,\mathrm{m}$ telescope at Las Campanas Observatory,
Chile \citep{mcleod98}.  Megacam was previously commissioned on the
Multiple Mirror Telescope, where it was used to study weak lensing by
galaxy clusters in the northern hemisphere \citep{israel10,israel11}.
Megacam consists of a $9\times4$ CCD array producing a $25\arcmin
\times 25\arcmin$ field-of-view. We operated read-out in $2\times2$
binning mode for an effective pixel scale of $0\farcs16$.  The 
$r'$-band images are used for shape measurements (Section \ref{sec:shear}),
and the added $g'$ and $i'$ bands are used in concert with deep
photometric-redshift catalogs from external surveys to prune and
characterize the source population using magnitude and color
information (Section \ref{sec:sigmacrit}).

Observations of \clustera\ occurred at the end of the nights of
2010 October 6 and 7. A total of $1760\,\mathrm{s}$ of exposure time
was obtained in the $r'$-band using a $2\times2$ square dither pattern
of $6\arcsec$ on a side, plus the same $2\times2$ pattern executed
about $25\arcsec$ north and $56\arcsec$ west, for a total of eight
individual exposures of $220\,\mathrm{s}$ each. The two star guiders,
situated on opposite sides of the Megacam focal plane, were
simultaneously operational for every $r'$-band exposure, which resulted
in uniform and stable point-spread function (PSF) FWHM patterns across the entire field, as
monitored upon readout during observation.  Twelve hundred seconds of
total exposure time was obtained in the $g'$-band, and
$3600\,\mathrm{s}$ in the $i'$-band, each with dithers that covered the
chip gaps.  Conditions were clear and stable with good seeing.

The rest of the clusters were observed over 3 second-half nights on
2011 May 31--June 2.  For each cluster, a total of $1800\,\mathrm{s}$
of exposure time was obtained in the $r'$-band using a three-point
diagonal linear dither pattern that covered the chip gaps.  In the $g'$
band, we obtained $1200\,\mathrm{s}$ of exposures with the same
three-point dither pattern, and in $i'$ we integrated for
$2400\,\mathrm{s}$ using a five-point diagonal linear dither pattern.
Conditions in this run were intermittently cloudy, resulting in
approximately 50\% unsuitable time.  Seeing was sub-arcsecond in the
$r'$-band for these four clusters.

For all five clusters, special care was taken to observe in the $r'$-band in
stable, good seeing conditions under the clearest available skies.  We
executed the dither patterns in immediate succession and monitored the
seeing.  The result was $\lesssim15\%$ variation in $r'$-band seeing in
all dithers for any given cluster.  Because of this and the uniform
PSF pattern afforded by Clay and Megacam (Section \ref{sec:shear}), we
coadd images without homogenizing to a common PSF in any way.

\subsection{Image Reductions}

The images are reduced at the Smithsonian Astrophysical Observatory
(SAO) Telescope Data Center using the SAO Megacam reduction pipeline.
The pipeline uses publicly available utilities from IRAF, SExtractor
\citep{bertin96}, and Swarp programs\footnote{SExtractor and Swarp are
  hosted at \url{http://www.astromatic.net/}.} \citep{bertin02}, as
well as in-house routines.

Basic CCD processing includes overscan correction, trimming, and bad
pixel removal.  Cosmic rays are removed using L.A.Cosmic
\citep{vandokkum01}.  Flat-field images for each filter were generated
from sets of twilight sky observations, taken during clear dawn or
dusk twilight, and then applied to the data.

To correct for any scattered light remaining after flat-fielding, an
illumination correction is also performed.  The illumination
correction image for a given filter is made by using 12 exposures
from a dithered pattern designed to expose the same stars over the
full span of the focal plane.  Where these data are not photometric,
i.e., there is too much scatter in the residuals of the dithered
exposures, we instead use just one exposure of the SDSS Stripe 82
field and match to the catalog of \citet{ivezic07}.

A fringe correction is performed for the $i'$ filter.  We make
fringe-frames by combining large numbers of $i'$ science-field
exposures.

In the final step, two passes are made of the world coordinate system
(WCS).  The first pass fits star positions to a reference catalog, the
Two-Micron All Sky Survey \citep[2MASS,][]{skrutskie06} catalog.  Once
that is successfully done, the second pass computes the WCS relative
to a source catalog made from a target exposure.  All exposures of the
same target are fit relative to the same reference, regardless of the
filter used.  Generally for this study, the stars used in the final
WCS solutions number in the two- to three-thousands, which provides
astrometry accurate typically to $0\farcs02$ rms.

The final product of the SAO Megacam reduction pipeline is the
multi-extension file for each exposure.  We run Swarp to mosaic and
coadd exposures for each target and filter, weighting by the relative
zeropoints of each frame.

\subsection{Photometry}
\label{sec:phot}

SExtractor is used to find sources in the images and perform
photometry.  We operate SExtractor in dual-image mode, with the 
$r'$-band image serving as the detection image.  In this mode, objects are
detected in the $r'$-band image while photometry is done in the $g'$,
$r'$, or $i'$-band image, and these final catalogs are joined to produce
a catalog of colors and magnitudes in the instrumental system.  We use
\verb|MAG_AUTO| photometry.
The colors of stars and galaxies are then calibrated using Stellar
Locus Regression \citep[SLR,][]{high09}.  SLR calibrates colors by
fitting the instrumental stellar locus to that of $\sim 10^5$ stars in
the SDSS.  Cross matching with the 2MASS allows us to solve for the
zeropoints of individual bands as well to produce calibrated
magnitudes.  The resulting photometry is effectively dereddened, as
this is an inherent feature of the method, so we do not apply any
additional Galactic extinction corrections.

\begin{figure}
\plotone{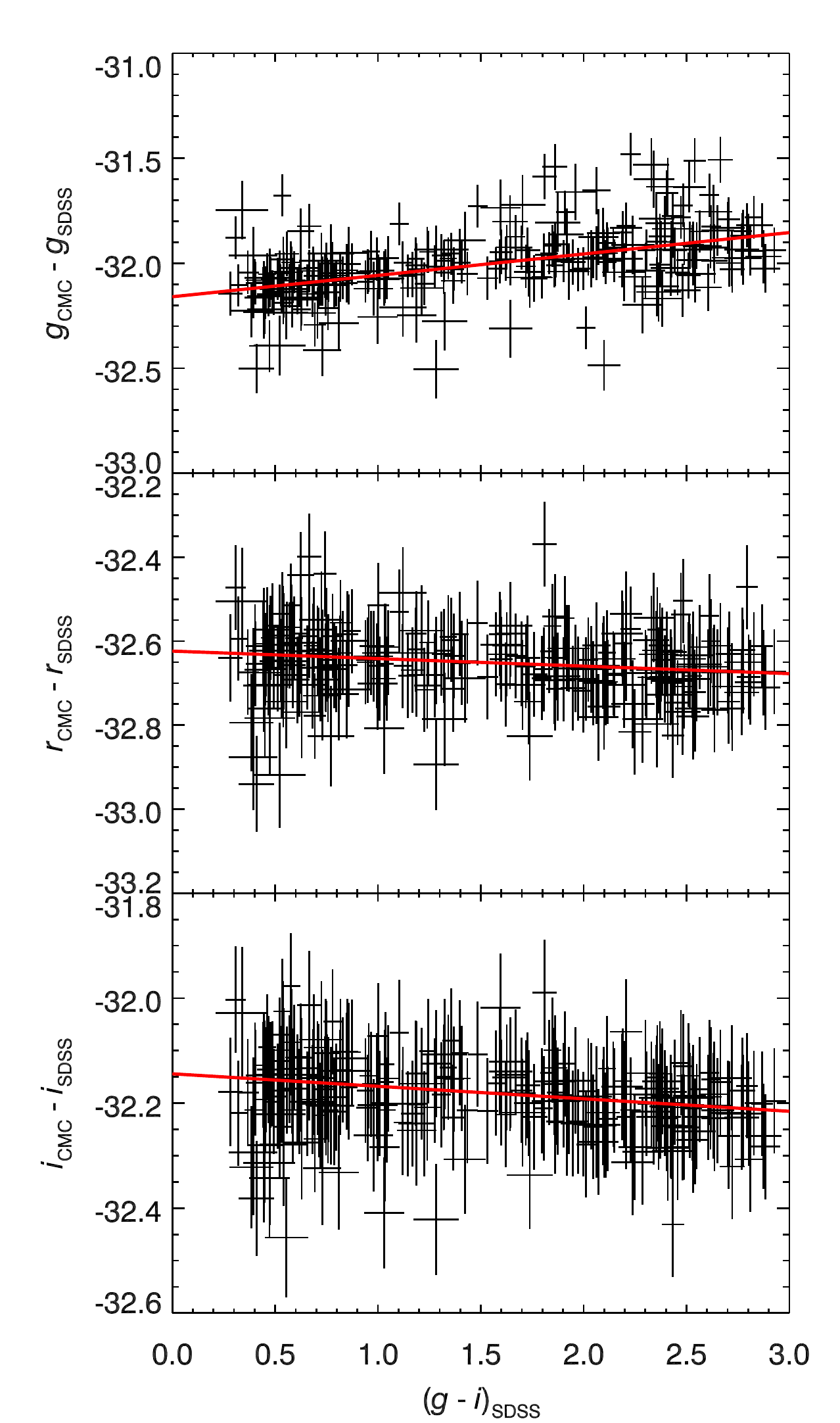}
\caption{Clay-Megacam color-term measurement.  Data points are the
  photometry of point sources that have been matched between the
  Clay-Megacam and IMACS catalogs, the latter of which has already
  been transformed to the SDSS system. The best-fit lines, whose
  slopes are equal to the color-term coefficients, are also
  shown.\label{fig:ct}}
\end{figure}

Transforming Clay-Megacam photometry to the SDSS system with SLR
requires estimating color terms.  We use $gri$-band photometry of
\clustera\ on the SDSS photometric system, acquired with IMACS
on Magellan Baade, operated in $f/2$ imaging mode.  These
IMACS data are used in the analysis of \citet{high10}, and are
described in detail in that work.  We cross-match point sources
with the IMACS catalogs, obtaining the color transformations
\begin{align}
\cmcg-\sdssg & = C_g + 0.100 (g-i)_{\mathrm{SDSS}} \\
\cmcr-\sdssr & = C_r - 0.022 (g-i)_{\mathrm{SDSS}} \\
\cmci-\sdssi & = C_i - 0.025 (g-i)_{\mathrm{SDSS}},
\end{align}
where zeropoints $C$ are nuisance parameters left free in the fit in
addition to the slopes, and CMC denotes Clay-Megacam\footnote{We
  emphasize that $\cmcg \equiv g'$ and $\sdssg \equiv g$ in this work,
  and so forth for the other bands.}.  These measurements are
illustrated in Figure \ref{fig:ct}.  We chose the multiplier $g-i$
because it gives maximal leverage on the color-term measurement over
the full range of stellar temperatures, and thus over the catalog's
color space.

The systematic uncertainty of the color calibration is estimated at
$0.01\,\mathrm{mag}$ to $0.03\,\mathrm{mag}$ in $g-r$ and $r-i$,
likely dominated by the non-uniformity of Galactic extinction,
flat-fielding, and details of the photometry \citep[see][]{high09}.
Magnitude calibrations are uncertain at $\sim 0.05\,\mathrm{mag}$ in
all bands, dominated by the overall accuracy of the 2MASS point-source
catalog.  These levels of uncertainty are sufficient for our purposes.

%%% From LB and MB
% To estimate the depth of the coadded images we use the Monte-Carlo
% based technique described in \citet{ashby09}. In brief, we measure
% the sky in various apertures at 1500 random positions in each coadded
% image. To measure the sky noise we fit a Gaussian function to the
% resulting flux distribution, after excluding the bright tail which is
% biased by real sources in the image.  In Table \ref{tab:depth} we
% report the magnitude of a source that would be detected at a
% signal-to-noise ratio of 5 in a $3\arcsec$ diameter aperture. 

Table \ref{tab:depth} lists the basic imaging information for the five
clusters.  Depths are estimated from the median magnitude of sources
whose signal-to-noise ratio is 5.  Seeing is estimated in the
WL band, $r'$.

\begin{deluxetable}{ccccc}
%\tabletypesize{\scriptsize}
%\rotate
\tablecaption{Imaging data\label{tab:depth}}
\tablewidth{0pt}
\tablehead{
\colhead{Cluster Name} & 
\multicolumn{3}{c}{Magnitude of $5\sigma$ Point Source} &
\colhead{$r$ Seeing} \\
\cmidrule(lr){2-4}
\colhead{~} &
\colhead{$g$} &
\colhead{$r$} &
\colhead{$i$} &
\colhead{~}
}
\startdata
% \clustera &  25.7 & 25.1 & 24.5 & $0\farcs67$ \\
% \clusterb &  25.3 & 24.8 & 24.1 & $0\farcs89$ \\ 
% \clusterc &  25.6 & 24.8 & 24.2 & $0\farcs81$ \\
% \clusterd &  25.3 & 24.7 & 23.5 & $0\farcs82$ \\
% \clustere &  25.5 & 24.8 & 24.3 & $0\farcs91$ 
% %% \clusterf &  24.9 & 24.3 & 23.6 & $0\farcs95$ \\
\clustera & 27.1 & 26.7 & 26.0 & $0\farcs67$ \\
\clusterb & 26.3 & 26.2 & 25.3 & $0\farcs88$ \\ 
\clusterc & 26.4 & 26.2 & 25.4 & $0\farcs80$ \\
\clusterd & 25.9 & 25.8 & 24.4 & $0\farcs82$ \\
\clustere & 26.6 & 26.1 & 25.4 & $0\farcs89$ 
%% \clusterf &  25.33 & 25.35 & 24.78 & $0\farcs95$ \\
\enddata
\tablecomments{Basic properties of the imaging of the five clusters.
  \\{Column 1:} \citetalias{reichardt12} cluster designation.
  \\{Columns 2--4:} the median magnitude of sources whose
  signal-to-noise ratio is 5. \\{Column 5:} the median of the
  stellar FWHM across the entire coadded image.}
\end{deluxetable}

We test the effect of photometric zeropoint errors on the final WL-SZ
mass ratios.  Systematic errors in photometry enter into the mass
analysis through the estimation of the critical surface density for
each cluster (Section \ref{sec:sigmacrit}).  We estimate this quantity
using publicly available photometric redshift catalogs of external
fields, to which we apply the same photometric cuts as are applied to
the Megacam catalogs.  If there is an offset in the Megacam $i'$-band
zeropoint relative to that of the standard catalog, then we
effectively probe a population that is different than that from which
we infer the source redshift distribution.  We test the effect of
photometric error, $\delta_i$, by cutting the standard photo-$z$
catalogs at $\cfhti > 24-\delta_i$ and repeating the full
analysis.  The level of photometric accuracy estimated here (5\%)
causes changes in the WL-SZ mass ratios at the sub-percent level,
which is significantly subdominant to the statistical uncertainty and
the largest systematic uncertainties.

\section{Creating Shear Catalogs}
\label{sec:shear}

In this section, we summarize the standard theoretical framework on
which WL mass measurements rest, including the two primary
quantities that must be estimated from data: the critical surface
density and the reduced shear.

\subsection{Tangential Shear}
\label{sec:sheartheory}

Weak gravitational lensing of extended sources by spherically
symmetric mass overdensities induces a mean shear in a direction
oriented tangentially to the center of mass.  Tangential shear,
$\gamma_+$, is calculated from the Cartesian components of shear,
$(\gamma_1,\gamma_2)$, as
\begin{equation}
\gamma_+ = -\gamma_1\cos(2\phi) - \gamma_2 \sin(2\phi)
\end{equation}
\citep[see for example][Section 2, for an overview of WL
shear]{mellier99}.  Indices $i\in\{1,2\}$ correspond to horizontal and
vertical image coordinates.  Here, $\gamma_1$ is the component along
the horizontal axis (position angle $\phi=0\degree$) and $\gamma_2$ is
the shear at position angle $\phi = 45\degree$.  Cross shear is
calculated as
\begin{equation}
\gamma_\times = -\gamma_1\sin(2\phi) + \gamma_2 \cos(2\phi);
\end{equation}
this is the shear component oriented at $45\degree$ with respect to
$\gamma_+$.
The azimuthally averaged cross shear $\langle\gamma_\times\rangle$ as
a function of radius provides a diagnostic for residual
systematics, because no astrophysical effects,
including lensing, produce such a signal. As a consequence, a non-zero
$\langle\gamma_\times\rangle$ indicates the presence of some types of
residual systematic error, though we note that this is not an
exhaustive test.

The mean tangential shear as a function of radial distance in the
plane of the sky at the cluster redshift, $R$, depends on the
projected surface density, $\Sigma(R)$, as
\begin{equation}
  \langle\gamma_+\rangle (R) =
  \frac{\langle{\Sigma}\rangle({<R})-\Sigma(R)}{\Sigma_{\mathrm{crit}}}
\end{equation}
\citep{miralda-escude95a}. This depends on the critical surface
density,
\begin{equation}
  \label{eqn:sigmacrit}
  \Sigma_{\mathrm{crit}} = \frac{c^2}{4\pi G}\frac{1}{D_{\mathrm{l}} \beta },
\end{equation}
where $c$ is the speed of light, $G$ is the gravitational constant,
and $\beta \equiv D_{ls}/D_{s}$ is the lensing efficiency.  Quantities
$D$ are angular-diameter distances, and $l$ indicates the lens (the
cluster) while $s$ indicates sources.

The observable quantity is not the shear but the reduced shear, $g$,
which relates to the shear as 
\begin{equation}
\gamma = (1-\kappa)g
\end{equation}
via the convergence, $\kappa = \Sigma/\Sigma_{\mathrm{crit}}$.
Estimating true shear therefore requires estimating convergence, which
we perform in radial bins, as discussed in Section \ref{sec:profiles}.

The two key quantities in measuring mass from WL data are thus the
critical surface density and the reduced shear.  We describe these two
steps in the following sections.

\subsection{Cluster-galaxy Decontamination and Critical Surface
  Density Estimation}
\label{sec:sigmacrit}

Calculating WL masses requires estimating the critical surface density
(Equation \eqref{eqn:sigmacrit}), which is a geometric quantity
containing ratios of angular diameter distances between the observer,
the lens, and sources. This requires redshift information for the lens
and sources.  Three bands are not sufficient for estimating source
redshifts in the Clay-Megacam data themselves, so we rely on
photometric redshift catalogs of other, non-overlapping surveys that
have integrated to equal or greater depths, under the assumption that
the mean underlying galaxy population is the same everywhere in the
sky.  The Canada-France-Hawaii Telescope Legacy Survey (CFHTLS) Deep
field catalogs are sufficient for this purpose \citep[][]{coupon09}.
For the $\Sigma_{\mathrm{crit}}$ estimate from the photo-$z$ catalog
to accurately reflect the population in the Clay-Megacam images on
average, cluster galaxies must be removed from the Clay-Megacam
catalogs.

\begin{figure*}
\epsscale{1}
\plotone{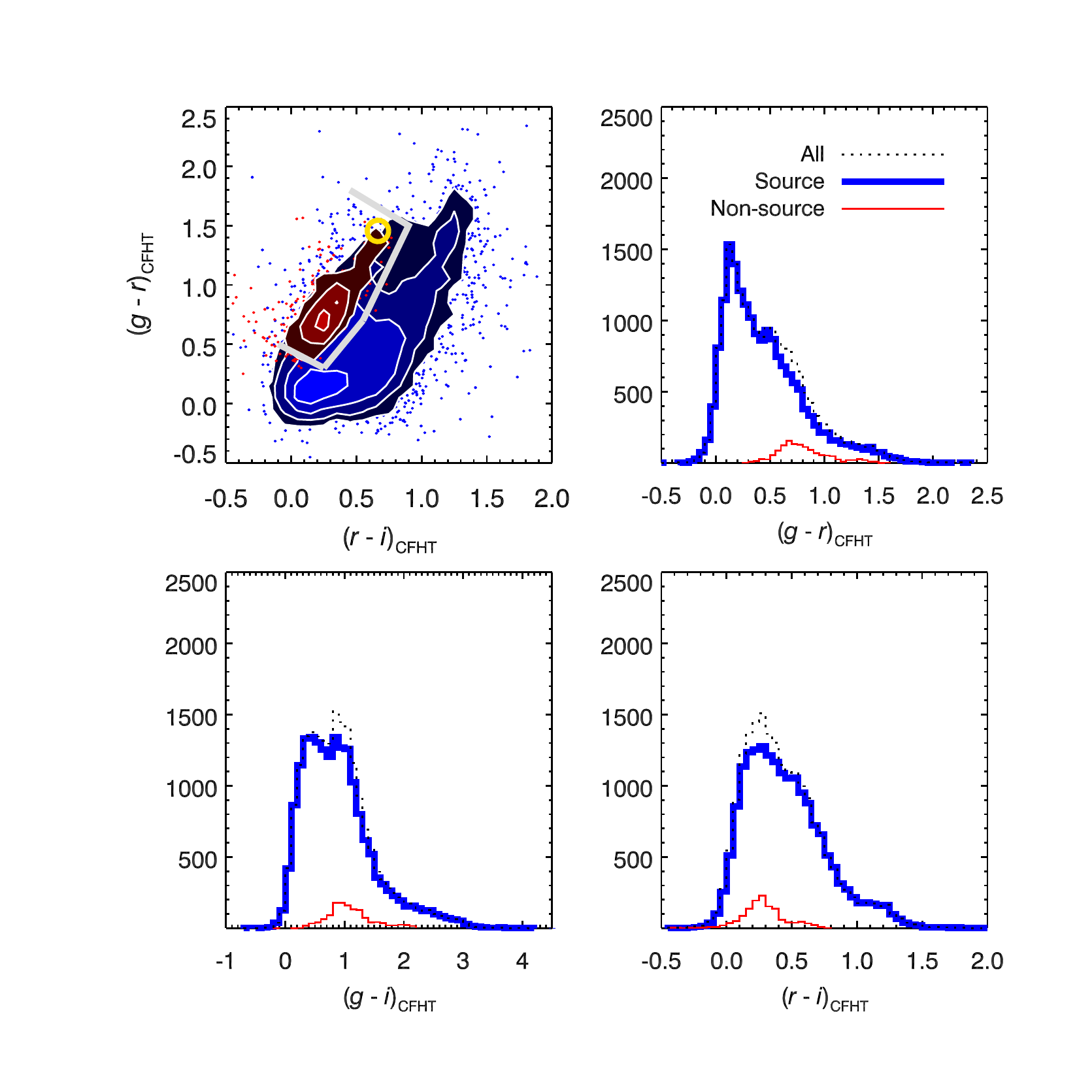}
\caption{Color--color distribution of $19 < \cfhti < 25$ galaxies
  in the CFHTLS-Deep photometric redshift catalog.  {Upper left:}
  the blue region denotes the distribution of source galaxies
  ($|\zphot - \zcluster| > 0.05$), and red denotes non-source galaxies
  ($|\zphot - \zcluster| < 0.05$), both for an example cluster in the
  high redshift bin, \clusterb\ at $z_l = 0.383$.  Contours delineate
  isodensities in color--color bins of size $0.1\,\mathrm{mag}\times
  0.1\,\mathrm{mag}$, and are logarithmically spaced by factors of
  $10^{0.5}$ starting at 10.  The gray polygon delineates
  source/non-source regions.  The yellow circle shows the typical
  color of a luminous red galaxy at the cluster redshift
  \citep{lopes07}. The remaining panels show one-dimensional
  histograms of projections of these colors. These panels illustrate
  that one-dimensional color cuts are a less efficient way of removing
  contaminant galaxies.\label{fig:colorcuts}}
\end{figure*}

We remove cluster galaxies from the shear catalogs using the same
CFHTLS catalogs as a guide.  We first plot the density of galaxies at
$19 < \cfhti < 25$ in the CFHTLS-Deep catalogs in $(g-r,r-i)$
color--color space.  Galaxies with photometric redshifts of
$|z_\mathrm{phot} - z_l| < 0.05$, i.e., near the cluster redshift, are
considered contaminants or ``non-sources'', and all other galaxies are
considered ``sources''.  The densities of these two populations are
shown in Figure \ref{fig:colorcuts}.  We then define a simple polygon
that encloses the majority of the non-source galaxies.

\begin{figure}
\epsscale{1}
\plotone{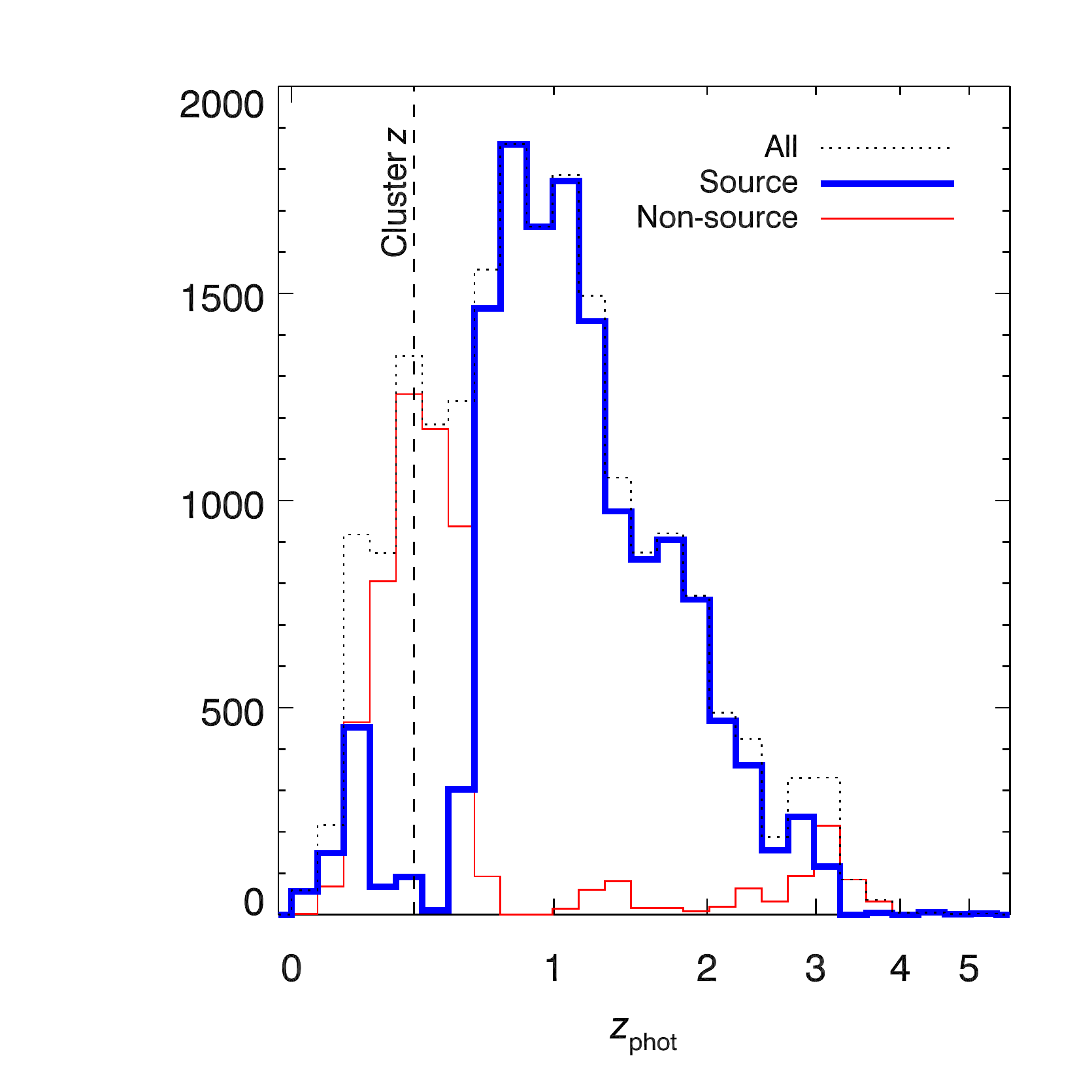}
\caption{CFHTLS Deep field 3 photo-$z$ distribution of $19 <
  \cfhti < 25$ galaxies after we have identified sources and
  non-sources using the color cuts illustrated in Figure
  \ref{fig:colorcuts}. A redshift of $0.383$, corresponding to
  \clusterb, is shown.\label{fig:colorcutsz}}
\end{figure}

Because the clusters are at different redshifts, the location and
shape of this polygon is a function of $z_l$ in general.  Rather than
defining five different polygons, we only define two: one that excises
cluster galaxies for $0.28 < z_l < 0.35$ (two clusters), and one for
$0.35 < z_l < 0.43$ (three clusters).  Figure \ref{fig:colorcutsz} shows
the photometric redshift distribution after removing galaxies within
the high-$z_l$ polygon.  We test the contamination after cuts by using
these polygons to remove galaxies from the CFHTLS-Deep catalogs, and
measuring the fraction of galaxies with photometric redshifts
satisfying $|z_\mathrm{phot} - z_l| < 0.05$.  Under this test, the
fraction of contaminants is $<0.1\%$ for all cluster redshifts
considered here.

\begin{figure}
\epsscale{1}
\plotone{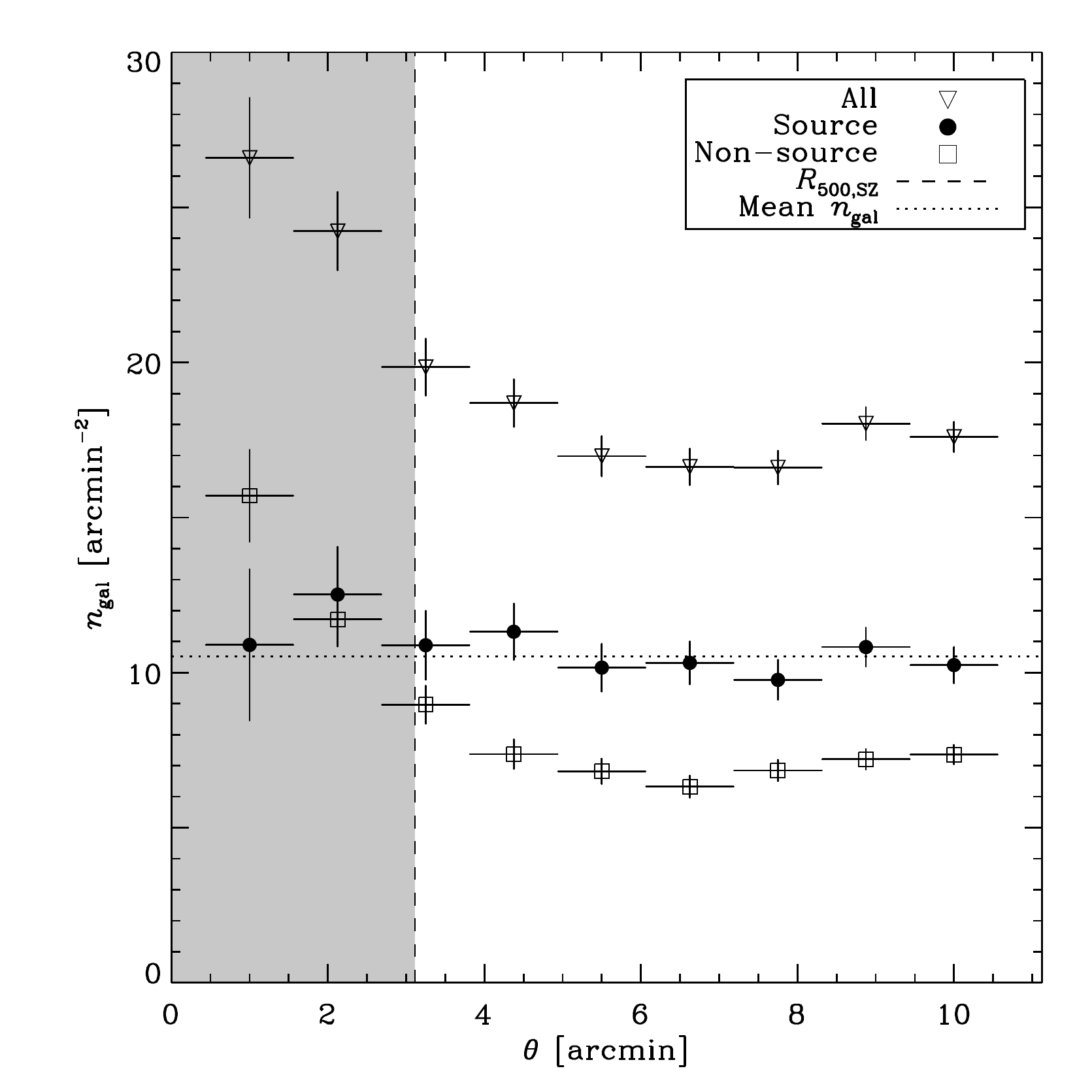}
\caption{Radial distribution of galaxies in the Clay-Megacam
  catalog of cluster \clusterb, before and after making the color cuts
  described in Section \ref{sec:sigmacrit}.  These data are
  representative of all clusters.  The color cuts reduce the
  clustering signal such that the final catalog used in the weak
  lensing analysis is roughly flat.  WL masses are computed outside
  the shaded region, from $\theta_1$ (vertical dashed line) to
  $\theta_2=12\arcmin$.\label{fig:ngal}}
\end{figure}

After transforming the Clay-Megacam catalogs to the CFHT-Megacam
photometric system using color terms \citep{regnault09},
% \begin{align}
%   \cfhtg-\sdssg & = -0.156 (g-r)_{\mathrm{SDSS}} \\
%   \cfhtr-\sdssr & = 0.000 (g-r)_{\mathrm{SDSS}} \\
%   \cfhti-\sdssi & = -0.094 (r-i)_{\mathrm{SDSS}},
% \end{align}
we apply the same magnitude and color cuts to the Clay-Megacam
catalogs.  The radial distribution of catalog galaxies before and
after the color cuts is shown in Figure \ref{fig:ngal} for a
representative cluster.  This procedure removes most of the radially
decreasing trend of galaxy densities such that the final galaxy
surface density is uncorrelated with radius. This constitutes some
empirical evidence that the procedure is removing cluster galaxies.
Some residual surface density trend with radius may be expected due to
WL magnification \citep[e.g.][]{bartelmann01}.  We measure
the slope of $\mathrm{d}\log{N}/\mathrm{d}m$ of selected sources to be $\sim
0.4$ in our data, so the effect would be small.

We estimate $\langle\beta\rangle$ and $\langle\beta^2\rangle$ from the
photo-$z$ distribution in the CFHTLS-Deep catalogs after these
selections are made (Figure \ref{fig:colorcutsz}).  The $\beta$
distribution is expected to vary between fields, both in the CFHTLS
Deep fields and the SPT cluster fields, due to finite galaxy counts
and cosmic variance.  We estimate the uncertainty this induces in the
mean mass ratio results by repeating the entire analysis using each
CFHTLS Deep field individually, and averaging the results.  The
uncertainties on resulting WL to SZ mass ratios are $<2\%$, a level
that is subdominant to the overall statistical uncertainties and other
systematics.  We adopt the mean values of $\langle\beta\rangle$ and
$\langle\beta^2\rangle$ over the four CFHTLS Deep fields in the
baseline analysis.  These are reported in Table \ref{tab:z}.

\begin{deluxetable*}{clccc}
%\tabletypesize{\scriptsize}
%\rotate
\tablecaption{Source galaxy properties.\label{tab:z}}
\tablewidth{0pt}
\tablehead{
\colhead{Cluster Name} & 
\colhead{$z_l$} &
\colhead{$\langle \beta \rangle$} &
\colhead{$\langle \beta^2 \rangle$} &
\colhead{$n_{\mathrm{gal}}$} \\
\colhead{~} &
\colhead{~} &
\colhead{~} &
\colhead{~} &
\colhead{(arcmin$^{-2}$)}
}
\startdata
\clustera & 0.294(1) & 0.64  & 0.43  &           14.8  \\
\clusterb & 0.383(1) & 0.54  & 0.32  &           10.6  \\
\clusterc & 0.40(4)  & 0.53  & 0.31  &           13.3  \\
\clusterd & 0.284(1) & 0.65  & 0.44  & \phantom{1}9.0  \\
\clustere & 0.427(1) & 0.50  & 0.29  &           12.2  
\enddata
\tablecomments{This table summarizes the basic properties of the
  sources used in the weak-lensing mass measurement after making the
  catalog selections described in Section \ref{sec:sigmacrit}.  \\{
    Column 1:} \citetalias{reichardt12} cluster designation. \\{
    Column 2:} cluster redshift. The number in parentheses is the
  uncertainty in the last digit. \\{Column 3:} the mean of $\beta
  = D_{ls}/D_s$ after cuts. \\{Column 4:} the variance of $\beta$
  after cuts.  \\{Column 5:} the source number density after
  cuts.}
\end{deluxetable*}

We perform consistency checks on the WL mass analysis by modifying
this photometric selection procedure in two ways.  First, we test
defining color--color polygons that remove the vast majority of objects
at $z_{\mathrm{phot}} < z_l + 0.1$, i.e., both near and in front of
the cluster.  This significantly widens the area covered by the
polygon, and the effect is that a large number of galaxies that are in
fact behind the cluster are eliminated as well.  This roughly halves
$n_{\mathrm{gal}}$ and does not change $\langle \beta\rangle$
estimates significantly.  The WL-SZ ratios are consistent with unity
and with the baseline result under this test.

Second, we test using limits of $\cfhti = 24.5$ and $24$, which are
brighter than our baseline limit of $\cfhti = 25$.  The effect
is to probe a brighter mean population of galaxies in the catalogs,
which causes $\langle \beta \rangle$ and $n_{\mathrm{gal}}$ to both
take smaller values.  This increases the statistical uncertainty of
the WL masses, but restricts the catalogs to magnitude regimes in
which the CFHTLS-Deep photometric redshift catalogs have been
explicitly tested and verified \citep{coupon09}.  The resulting WL-SZ
aperture mass ratios in both cases are statistically consistent with
the baseline result, and with unity.

\subsection{Shear Measurement Method}
\label{sec:ksb}
 
% Correcting the shear estimators for the convolution of a generally
% anisotropic PSF is key to accurate WL measurements.  
The second key ingredient in estimating cluster masses in WL analyses
is reduced shear.
To estimate reduced shear in the $r'$-band images we employ the
method of \citet*[][hereafter \citetalias{kaiser95b}]{kaiser95b} and
\citet{luppino97}, with the modifications of \citet[][]{hoekstra98}.
This method bases shear measurements on Gaussian-weighted second-order
image moments,
\begin{equation}
  \label{eqn:quad}
  I_{ij} = \int d^2 \vec{x} x_i x_j W(\vec{x}) f(\vec{x}),
\end{equation}
where $W$ is the Gaussian function, $f$ is the object image, and the
origin of the coordinate system has been iteratively determined from
the first-order moment using the same weight.  Variables $x_i$ ($i\in\{1,2\}$)
indicate vertical and horizontal image coordinates. The polarization
is a spin 2 pseudo-vector $e_\alpha$ ($\alpha \in \{1,2\}$), where
\begin{equation}
\label{eqn:polarization}
e_1 = \frac{I_{11} - I_{22}}{I_{11}+I_{22}}\quad\mathrm{and}\quad
e_2 = \frac{2 I_{12}}{I_{11}+I_{22}}.
\end{equation}
The magnitude of the polarization is $e = (e_1^2 + e_2^2)^{1/2}$.

\begin{figure*}
\plottwo{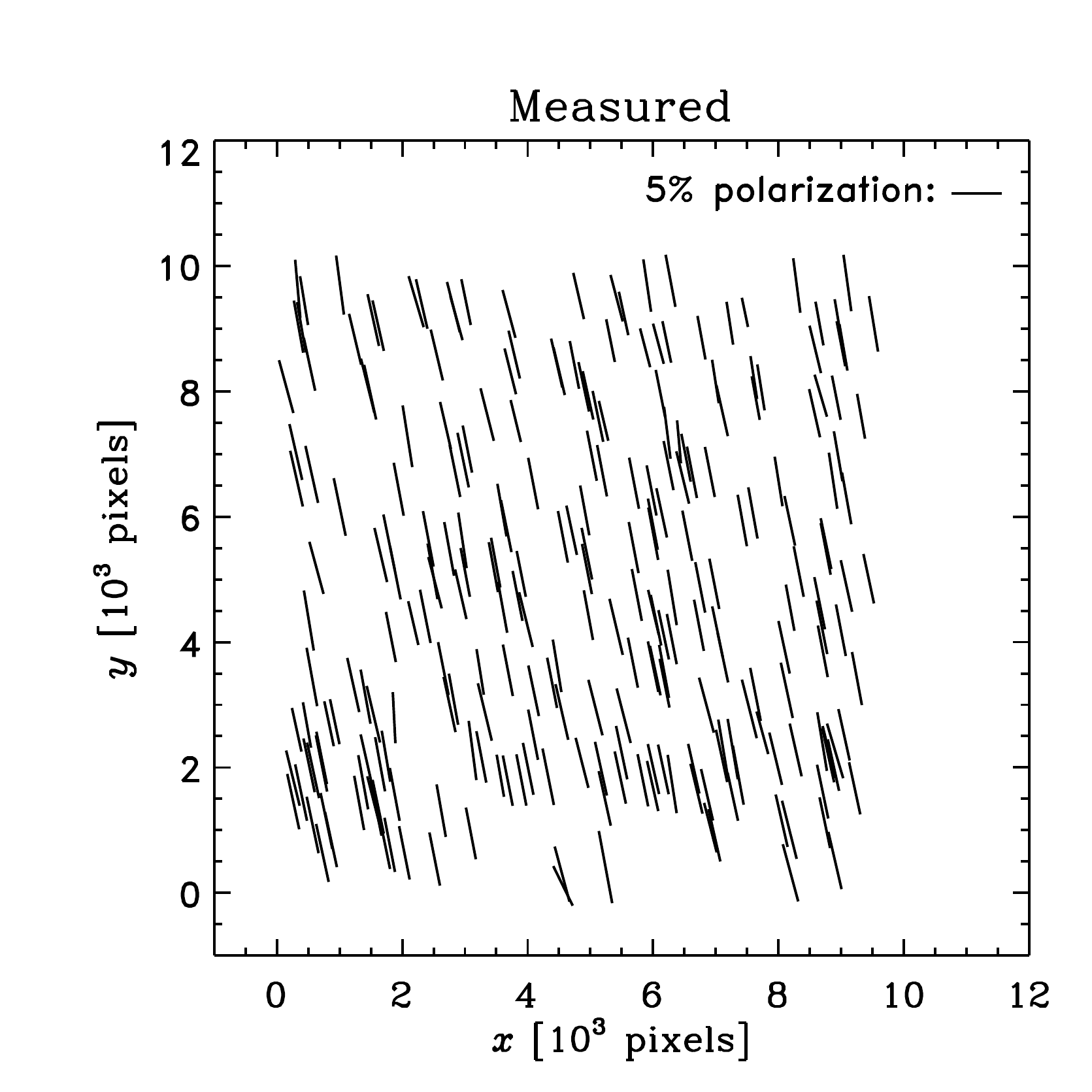}{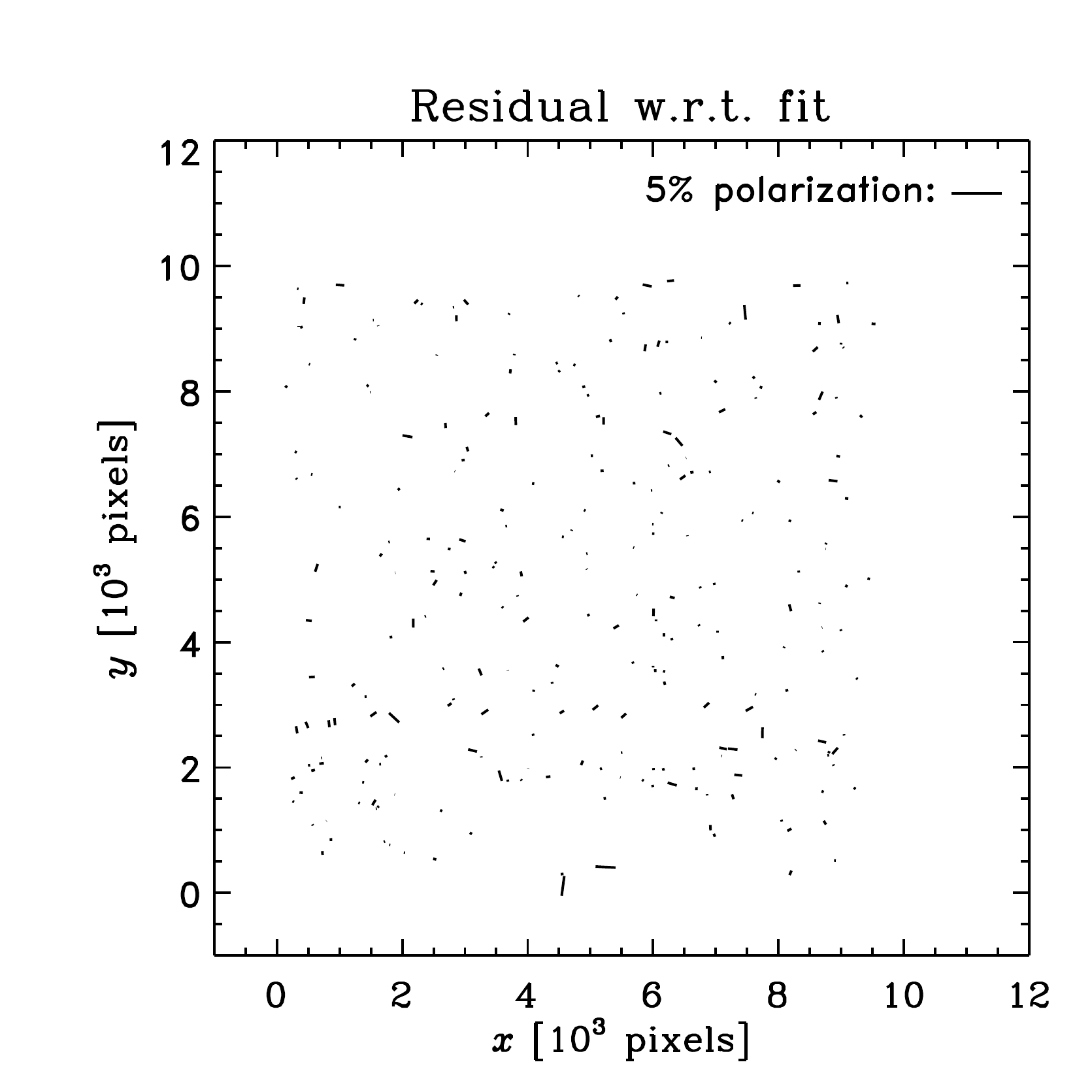}
\caption{PSF polarization plot of the coadded $r$-band image of \clusterb,
  which is representative of the sample.  {Left:} polarization
  magnitude and direction, as measured from the Gaussian-weighted
  image moments of point sources.  {Right:} same as the left
  panel, after a fourth order polynomial in the two spatial dimensions
  is fit to each component of polarization.  Residuals between the
  data and the model are $<0.005$ rms for all clusters.\label{fig:whisker}}
\end{figure*}

The resulting polarizations must be corrected for the effects of the
anisotropic smearing by the PSF. To this end, we fit fourth order
polynomials to the stellar $P^{\mathrm{sh}}_{ii}$ and
$P^{\mathrm{sm}}_{ii}$ (the diagonal entries of the
shear- and smear-polarizability tensors,
see \citetalias{kaiser95b}) and remove outliers.  To the surviving
objects we fit a fourth order polynomial to both $e_1$ and $e_2$. We
do this for a range of weight functions, which is necessary for
computing the pre-seeing shear polarizability $P^\gamma$
\citep[see][]{hoekstra98}.  The rms of the residuals for each
polarization component after the fourth order fit are $< 0.005$ for
all clusters.  These are small compared to the magnitude of signals we
are seeking, which are about $0.01$ to $0.1$.  While residuals of this
size can in principle lead to large shear bias for any given faint
galaxy locally, we expect this effect to be zero on average in the
radial bins because the spatial residual pattern is consistent with
noise.  A representative example PSF polarization plot is shown in
Figure \ref{fig:whisker}. This model is used to correct for PSF
anisotropy using the procedure described in the original works
\citep[\citetalias{kaiser95b};][]{hoekstra98}.

The next step is to account for the seeing, which lowers the observed
shear signal. This is done by rescaling the polarizations to their
``true'' values by $P^\gamma$ \citep{luppino97,hoekstra98}. The
measurements of $P^\gamma$ for individual galaxies are too noisy, and
instead we bin the measurements in galaxy size and use this to compute
the value as a function of size.  The reduced shear $g$ is then
estimated as
\begin{equation}
g_i = e_i/P^\gamma.
\end{equation}
This implementation of the \citetalias{kaiser95b} shear analysis
pipeline is described in more detail in previous works
\citep{hoekstra98,hoekstra00}.  It was subject to blind tests by the
Shear Testing Program \citep[][]{heymans06,massey07}, wherein it
achieved accuracy at the $0.02\gamma$ level under low PSF
anisotropies, such as Clay-Megacam exhibits.  This level of bias
induces mass errors that are subdominant to the statistical and
systematic uncertainties.

\section{Weak-lensing Mass Measurement}
\label{sec:wlmasses}

In this section we discuss how we estimate mass from the shear
catalogs.  We employ two types of WL mass estimators, both of which
make use of the same, azimuthally averaged shear profiles.  We test
the accuracy of
our algorithms using ray-traced $N$-body simulations, and we also
describe the WL convergence field reconstruction procedure.

\subsection{Shear Profiles and Corrections}
\label{sec:profiles}

We compute binned shear profiles using a weighted average in each
radial bin.  The weight for each galaxy is
\begin{equation}
\label{eqn:sigmag}
  w = \frac{1}{\sigma^2_g} =
  \frac{P^{\gamma\,2}}{\sigma_\gamma^2 P^{\gamma\,2} + \langle\Delta
    e^2\rangle},
\end{equation}
where $\sigma_\gamma$ is the scatter in shear due to the intrinsic
ellipticity of the galaxies \citep[set to 0.3; see for
example][]{leauthaud07} and $\langle\Delta e^2\rangle$ is the error
estimate for the polarization \citep{hoekstra00}.  The weighted mean
for $i\in\{+,\times\}$ is then
\begin{equation}
  \langle \gamma_i \rangle = \frac{\sum_n w_n \gamma_{i,n}}{\sum_n w_n},
\end{equation}
and errors on the mean, $\sigma_{\langle\gamma_i\rangle}$, are computed via
\begin{equation}
  \label{eqn:sigmag}
  \frac{1}{\sigma_{\langle\gamma_i\rangle}^2} = \sum_n w_n.
\end{equation}

Two corrections are applied to the binned shear data.  
The first correction accounts for a known error in the binned shear
data due to the averaging operation.  \citet{seitz97} show that
estimating the critical surface density of each cluster using the mean
of the $\beta$ distribution in redshift catalogs induces an error in
the observed reduced-shear, which can be corrected to first order
using
\begin{equation}
  \label{eqn:seitz}
  \frac{\langle g_{\mathrm{obs}}\rangle}{\langle g_{\mathrm{true}}\rangle} = 1 + \left(
  \frac{\langle\beta^2\rangle}{\langle\beta\rangle^2} - 1 \right)\kappa.
\end{equation}
We adopt the model for $\kappa$ described below to make this
correction.

The second correction is to transform the reduced shear $g$ to shear
$\gamma$ by estimating the $\kappa$ in each bin.  We accomplish
this by jointly fitting for the shear and the convergence profiles to
those predicted by \citet*[][hereafter
\citetalias{navarro97}]{navarro97} density profiles assuming
\citet[][hereafter \citetalias{duffy08}]{duffy08}
concentration--mass--redshift scaling, wherein the SZ-derived mass is
used as the input.  The NFW family of density profiles take the form
\begin{equation}
  \rho(r) = \frac{\delta_c \rhocrit}{(rc/r_{200})(1+rc/r_{200})^2},
\end{equation}
where $r$ is the three-dimensional radial distance, $r_{200}$ is the
radius at which the mean NFW overdensity is 200 times $\rhocrit$ at
the cluster redshift, $c$ is concentration, and
\begin{equation}
\delta_c = \frac{200}{3}\frac{c^2}{\ln(1+c) - c/(1+c)}.
\end{equation}
This fitting procedure results in estimates of the $\gamma$ profile
under the assumption of \citetalias{duffy08} concentrations, and it
also yields the convergence profile used to perform the redshift
distribution correction in Equation \ref{eqn:seitz} as well as total
spherical masses in our main results.

% We accomplish
% this by iteratively fitting the reduced shear and the convergence
% to a
% singular isothermal sphere (SIS) model,
% \begin{equation}
%   \kappa = \gamma = \frac{r_{\mathrm{E}}}{2r},
% \end{equation}
% to the tangential shear profiles.  For a cluster whose galaxies are in
% isotropic orbits, the Einstein radius $r_{\mathrm{E}}$ is a function
% only of galaxy line-of-sight velocity dispersion,
% $\sigma_{\mathrm{vel}}$, and of $\beta$:
% \begin{equation}
%   r_{\mathrm{E}} = 4\pi \left( \frac{\sigma_{\mathrm{vel}}}{c} \right)^2 \beta.
% \end{equation}
% We iteratively fit SIS profiles at projected radii
% $>R_{500,\mathrm{SZ}}$ and adopt the final best-fit velocity
% dispersion to transform from $g$ to $\gamma$ at all radii.  We report
% the best-fit velocity dispersion values in Section \ref{sec:results}.
% This correction is only important where $\kappa$ is appreciable, such
% as in the inner regions of clusters.
% aperture masses at $R_{500}$, which we employ here, are
% (filtered) integrals over the shears well outside the clusters' cores, where $\kappa
% \ll 1$, so the correction from reduced shear to shear is expected to
% have a negligible effect on the mass measurements.
% We verify this by testing the effect of conservative velocity dispersion
% errors of $\pm 500\mathrm{km}\,\mathrm{s}^{-1}$ on the final
% mean ratio of WL to SZ masses, and determine that 
% it is subdominant to the statistical uncertainty from
% the full sample of clusters.

Observed shear profiles, after all of these corrections are applied,
are presented in the Appendix. The cross shear is
consistent with zero, as expected for a signal that is not appreciably
contaminated by systematics.

\subsection{Aperture Masses}

% \subsubsection{Aperture Masses}
\label{sec:mapbasic}

One of the advantages of the tangential shear is that it can be used
to directly constrain the projected mass within an aperture.  Aperture
masses are integrals over the shear data multiplied by a 
filter \citep[see][]{fahlman94,schneider96}.  These aperture masses are
parameter-free in the sense that the observable quantity gives
cylindrical mass constraints without reference to analytic density
profiles.
If the filter is compensated, they are also
insensitive to the mass-sheet degeneracy. 
The classical $\zeta$-statistic for circular apertures, which uses a
particular choice of filter, is\footnote{This $\zeta$ statistic is in
  general a function of two angles, $\theta_1$ and $\theta_2$, but we
  fix $\theta_2$ for all analyses and so have simplified the notation.}
\begin{equation}
  \label{eqn:zeta}
\begin{split}
 \zeta(\theta_1) & = \langle \kappa \rangle_{<\theta_1} - \langle \kappa \rangle_{\theta_1 < \theta <\theta_2} \\
 & =\frac{2}{1-\theta_1^2/\theta_2^2}\int_{\theta_1}^{\theta_2}
  \mathrm{d}\theta \, \frac{\langle\gamma_+\rangle(\theta)}{\theta}
\end{split}
\end{equation}
\citep{fahlman94,kaiser95a}.  
The aperture mass is then
\begin{equation}
  M_{\mathrm{ap}}(\theta_1) = \pi (\theta_1 D_l)^2
  \Sigma_{\mathrm{crit}} \zeta(\theta_1). 
\end{equation}
The square of the measurement uncertainty of the aperture mass
statistic takes an analytic form,
\begin{equation}
  \sigma^2_{\mathrm{stat}}(\theta_1) =
  \left( \frac{2}{1-\theta_1^2/\theta_2^2} \right)^2 \int_{\theta_1}^{\theta_2} 
  \mathrm{d}\theta \, \frac{\sigma_{\langle\gamma_+\rangle}^2}{\theta^2},
\end{equation}
where $\sigma_{\langle\gamma_+\rangle}$ is given in Equation
\eqref{eqn:sigmag}. In addition to the measurement uncertainty,
LSS along the line of sight contributes a
random uncertainty to the WL aperture masses, which we add to the
formal statistical WL mass uncertainties in quadrature using the
prescription of \citet{hoekstra11}.  LSS uncertainties are 15\% to
20\% for these clusters.

% To calculate $\zeta(\theta_1)$ and $\sigma(\theta_1)$ in practice, we
% Riemann sum over each galaxy at radial distances $\theta_1 < \theta <
% \theta_2$.  
We fix $\theta_2$ to $12\arcmin$ for all analyses, which is roughly
the maximum radius of the Megacam imaging.  We set $\theta_1$ to
$R_{\mathrm{500,SZ}}/D_l$.

Aperture masses provide lower limits on total cylindrical masses
within $\theta_1$. In the limit of $\theta_2\to\infty$, the aperture
mass converges to true integrated cylindrical mass within $\theta_1$,
but real data extend to finite radius, so $\theta_2$ must be finite.
Direct comparisons of the observed aperture mass to other spherical or
cylindrical mass-observables therefore requires care.

\subsection{Spherical- to Aperture-mass Transformations}
\label{sec:mapcomp}

Some extra computation makes direct comparisons between aperture
masses and spherical masses possible.  This is accomplished by
assuming an analytic three-dimensional profile consistent with a given
spherical mass estimate, projecting it to two dimensions, calculating
the shear assuming some $\Sigma_{\mathrm{crit}}$, and then computing
the aperture mass statistic given this predicted shear.

To illustrate this procedure concretely, say we are given some
spherical mass estimate, $M_{\Delta,\mathrm{obs}}$, derived from
another observable such as the SZ effect.  We wish to compare this to
the WL aperture mass at the same radius,
$M_{\mathrm{ap,WL}}(R_{\Delta,\mathrm{obs}})$, where
$R_{\Delta,\mathrm{obs}} = [3
M_{\Delta,\mathrm{obs}}/(\Delta4\pi\rho_{\mathrm{crit}}(z_l))]^{1/3}$.
We first assume a three-dimensional \citetalias{navarro97} profile,
with total spherical mass within $R_{\Delta,\mathrm{obs}}$ equal to
$M_{\Delta,\mathrm{obs}}$, and concentration $c$ taken from the mean
of clusters of these masses and redshifts according to the
\citetalias{duffy08} $c$--$M$--$z$ scaling relation.  The
corresponding two-dimensional projection of this profile and predicted
shear take an analytic form \citep{brainerd00}.  We use the same
$\Sigma_{\mathrm{crit}}$ as used in the WL analysis.  Then, from this
shear we compute $\zeta$, again using the same filter as used on the
WL data.  Specifically, the outer radius, $\theta_2 = 12\arcmin$, has
been set by the size of the Megacam imaging footprint, and the inner
radius is $\theta_1 = R_{\Delta,\mathrm{obs}}/D_l$.  The result,
$M_{\mathrm{ap,obs}}(R_{\Delta,\mathrm{obs}})$, is the
aperture-equivalent mass of $M_{\Delta,\mathrm{obs}}$, which was
determined by some other method or observable.  It is the direct
analog of the WL aperture mass measured at the same radius,
$M_{\mathrm{ap,WL}}(R_{\Delta,\mathrm{obs}})$, such that the ratio of
the WL to this aperture-equivalent quantity is unity in the absence of
other systematic errors.  We perform this transformation on SZ mass
estimates to test for statistical consistency with the WL aperture
masses.

We propagate spherical mass uncertainties to aperture mass equivalent
by computing the aperture mass of $M_{\Delta,\mathrm{obs}}\pm\sigma_M$
numerically using the same procedure. As illustration, aperture mass
uncertainties are $\sim (0.54,0.67,0.81)$ times the values of their
spherical \citetalias{navarro97} counterparts for $c=(1,3,10)$ at
these clusters' typical $R_{500}$ radii.

\subsection{Calibration Tests with Mock Catalogs}
\label{sec:mocks}

A number of recent works have shown using $N$-body simulations that
WL derived mass estimates are biased at roughly the $-5\%$
to $-10\%$ level \citep{becker11,bahe12,rasia12}.  We perform similar
calibration tests of the WL mass statistics presented in our work
using mock shear catalogs of $220\deg^2$ of sky.  The catalogs are
drawn from an $N$-body dark matter simulation of a standard
$\Lambda$CDM universe.  The dark matter halos are populated with
galaxies using ADDGALS such that they reproduce known luminosity,
color, and clustering relations \citep[][R.\ H.\ Wechsler et al., in
preparation]{wechsler04}.  Shears are assigned to each galaxy by raytracing
through the $N$-body simulation\footnote{The simulated shear catalogs
  were kindly made available to use by R.\ Wechsler, M.\ Busha, and
  M.\ R.\ Becker.}.

We compute the masses of $280$ $N$-body halos at redshifts $0.25 < z <
0.45$ with masses $M_{200}\geq 10^{14}h_{70}^{-1}\Msun$.  To reflect
the choices we have made in analyzing the Clay-Megacam data, we only
use shear profiles between $R_{500}$ and $12\arcmin$, where $R_{500}$
is extrapolated using the \citetalias{duffy08}
concentration--mass--redshift scaling relation with $M_{200}$ determined
from the $N$-body halo finder as the input.  In order to minimize
statistical uncertainty in these tests, we use perfect knowledge of
the shear field at the galaxy locations (i.e., we do not include
intrinsic shape noise) as well as perfect source redshifts, and we
select source galaxies as those at $|z - \zcluster| > 0.05$.  The
resulting bias between measured masses and appropriately transformed
$N$-body masses, averaged over the sample, ranges from $-6\%$ to
$-13\%$, consistent in magnitude and sign with the previous works.
These tests carry a statistical uncertainty of about $2\%$.  The
precise bias values are reported in Section \ref{sec:err}.  We do not
apply these bias corrections to any mass estimates presented in this
work.

% We emphasize that this is a test of the fundamental accuracy of our
% algorithm in a mock universe under the idealized circumstances
% described above, and that effects that appear in data that we have not
% modeled in the mock catalogs may affect the true accuracy of the
% methodology.  We leave more sophisticated tests to future work.

%%% END OLD MOCKS VERIFICATION

% \begin{figure}
% \plotone{map_vs_truth_z_resid_nfw_map.pdf}
% \caption{Verification of the aperture mass code using ray-traced mock
%   catalogs based on an $N$-body simulation.  The upper panel shows the
%   measured aperture masses relative to aperture-equivalent true halo
%   masses.  The median residual in redshift bins is shown with error
%   bars. The lower panel shows the same quantity but scaled to
%   emphasize the variation in the median values.  The gray region
%   delimits $\pm 10\%$ errors.  The overall accuracy of the code is
%   determined from this test to be $(3\pm1)\%$.\label{fig:mocks}}
% \end{figure}

\subsection{Convergence Field Reconstruction}
\label{sec:kaiser}

We reconstruct the convergence field in two dimensions using the
method of 
\citet{kaiser93}.
% and 
% \citet{squires96}.
The purpose is to test the impact of using the peak of the
reconstructed $\kappa $ field as the cluster center when computing
shear profiles.  The convergence is estimated up to a constant as a
sum over galaxies $n$,
\begin{equation}
  \kappa(\vec{x}_0) = -\frac{1}{n_{\mathrm{gal}}\pi} \sum_n \chi_\alpha(\vec{x}_n-\vec{x}_0) 2e_\alpha/P^\gamma,
\end{equation}
where $n_\mathrm{gal}$ is the mean surface density of source galaxies
and $\chi_\alpha = \{x^2-y^2,2xy\}/\theta^4$ and $\theta =
(x^2+y^2)^{1/2}$.  The constant is set such that the mean convergence
across the full field is zero.  The wide field-of-view of the
observations allows us to avoid artifacts from this method which have
caused problems in the past.

We pixelate the convergence maps at $0\farcm4\,\mathrm{pixel}^{-1}$.  The errors
per pixel are independent, and are computed as
$\sigma_{\mathrm{pix}}\approx \sigma_\kappa/\sqrt{n_{\mathrm{gal}}0\farcm4^2}$, where $\sigma_\kappa \sim
\sigma_\gamma \sim 0.3$ and $n_{\mathrm{gal}}$ is measured from the
data and assigned units of $\mathrm{arcmin}^{-2}$.  We then smooth the
map with a Gaussian of size $\sigma_{\mathrm{smooth}}=3\arcmin$.  The peak
in the smoothed convergence field, $\max(\kappa)$, is identified as
the maximum value across the entire smoothed map.  The signal-to-noise
ratio in the smoothed peak value is $\mathrm{S}/\mathrm{N} =
\max(\kappa)/(\sigma_{\mathrm{pix}}/\sqrt{2\pi
  \sigma_{\mathrm{smooth}}/0\farcm4})$.  The uncertainty on the peak position
is then $\mathrm{FWHM}_{\mathrm{smooth}}/(S/N)$, where
$\mathrm{FWHM}_{\mathrm{smooth}}=\sqrt{8\ln{2}}\,\sigma_{\mathrm{smooth}}$.
This is equal to $\sim45\arcsec$ for all five clusters.  Using the
convergence field peaks as the cluster centers gives a mean ratio of
WL-SZ masses that is consistent with unity and with the baseline
result to within the statistical uncertainty.

\section{Results}
\label{sec:results}

We report WL-derived masses, then test the overall accuracy of the SZ
mass determination of \citetalias{reichardt12} by measuring the mean
ratio of equivalent WL and SZ mass estimators. In the baseline
analysis we
\begin{itemize}
\item select source galaxies at $19<\cfhti<25$ with colors exterior to
the polygon shown in Section \ref{sec:sigmacrit},
\item use concentrations of \citetalias{duffy08}, and 
\item use SZ centroid as the cluster centers.
\end{itemize}
We then alter various steps in the analysis in order to test the
robustness of the result, as well as to estimate the magnitude of
various potential sources of error.
Shear profiles and aperture mass profiles are presented in the
Appendix, in addition to optical, SZ, and convergence maps.

\subsection{Masses and Mass Ratios}
\label{sec:ratio}

In this section we use three different methods for estimating mass
from these data.  In all cases, we only use shear data at
$R_{500,\mathrm{SZ}}/D_l < \theta < 12\arcmin$.

\begin{figure*}
\epsscale{1.2}
\plottwo{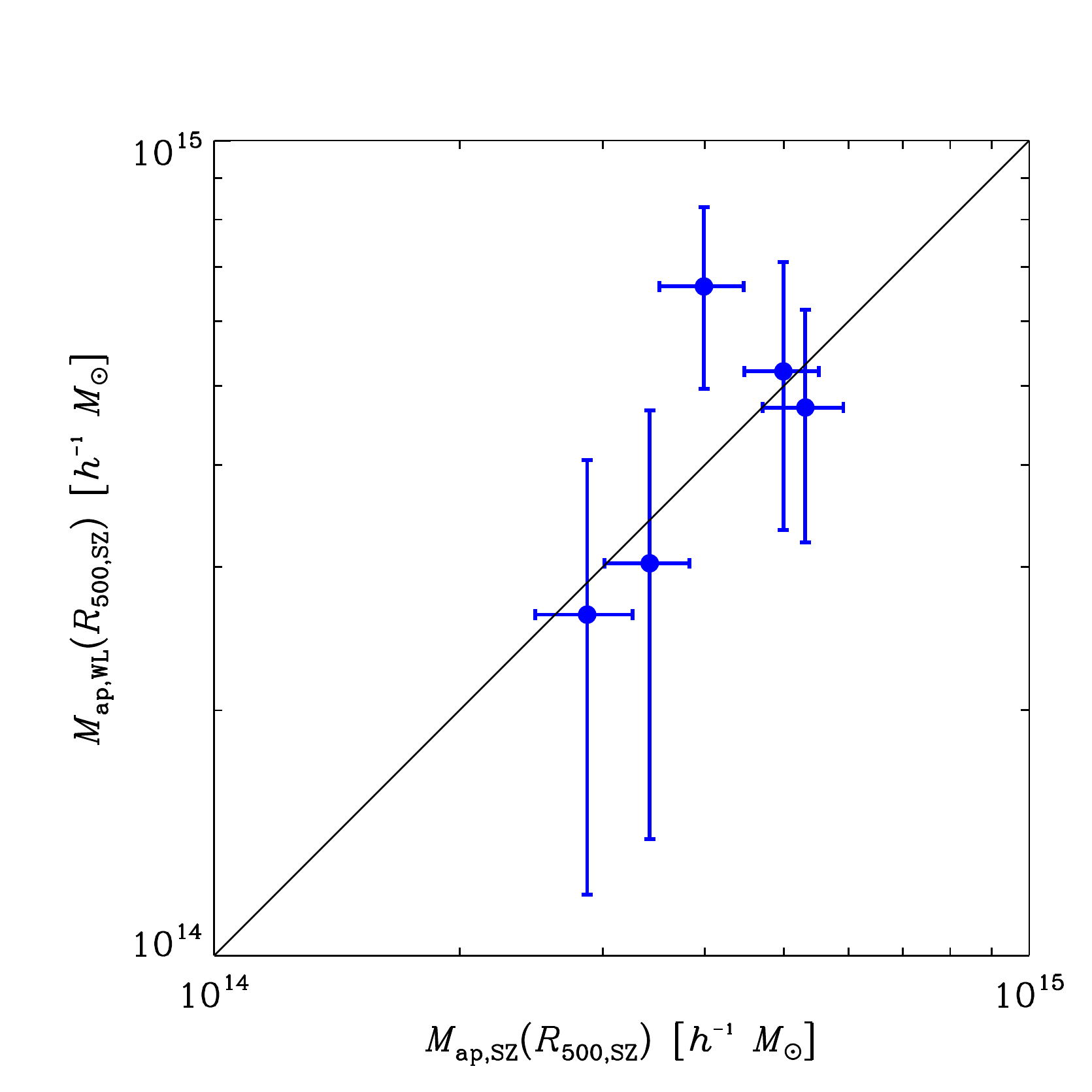}{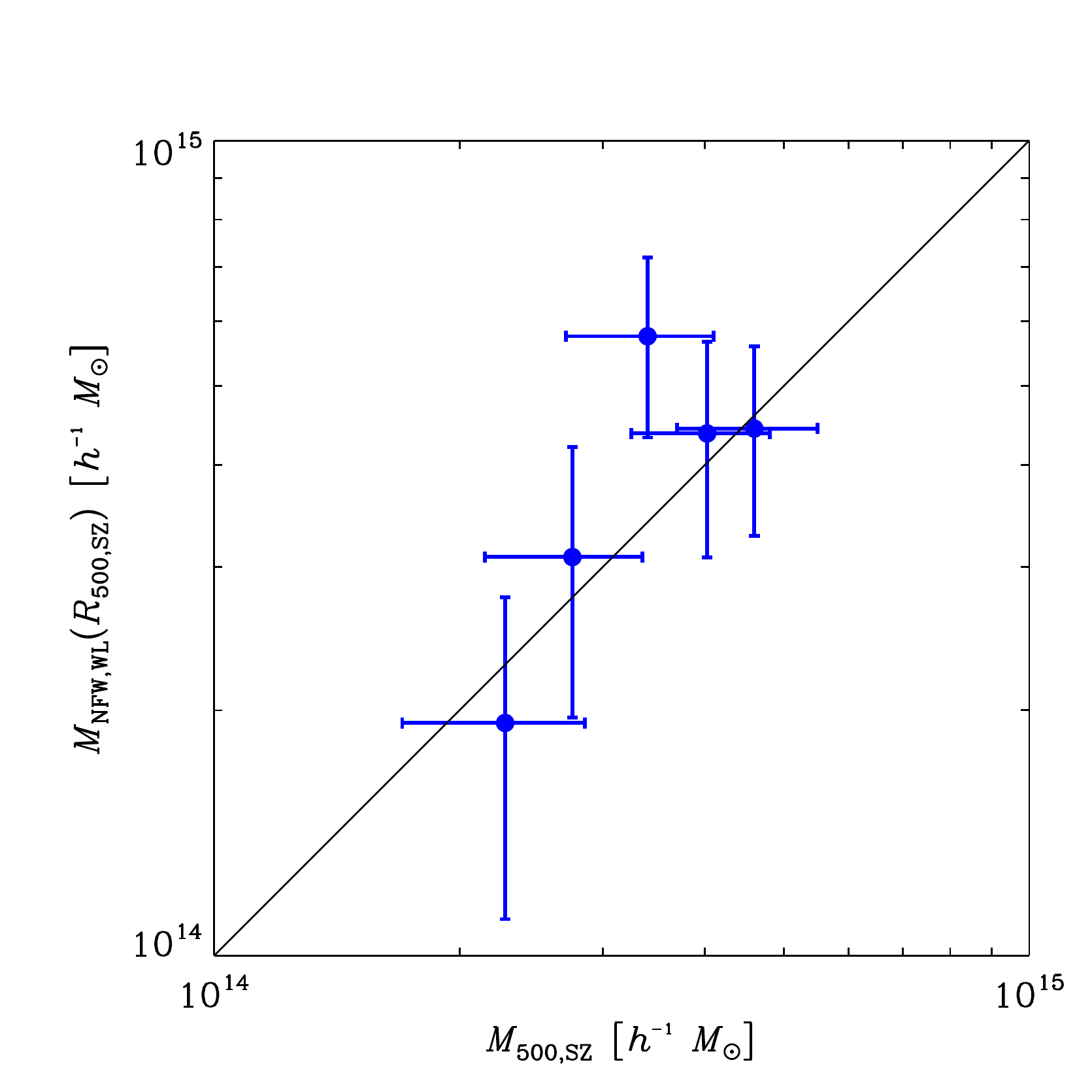}
\caption{{Top:} weak-lensing aperture masses vs.\ inferred aperture
  masses from Sunyaev--Zel'dovich effect data. The one-to-one line is
  also shown. {Bottom:} spherical WL masses vs spherical SZ
  masses.  The WL masses are determined by fitting
  \citetalias{navarro97} profiles to weak lensing shear data at
  $R_{500,\mathrm{SZ}}/D_l < \theta < 12\arcmin$ and evaluating the
  resulting NFW mass profile at $R_{500,\mathrm{SZ}}$.\label{fig:m12}}
\end{figure*}

In the first method, we measure aperture masses with $\theta_1 =
R_{500,\mathrm{SZ}}/D_l$ and $\theta_2 = 12\arcmin$.  Following the
procedure outlined in Section \ref{sec:mapcomp}, we derive the
equivalent estimator from the SZ masses by assigning an NFW profile
consistent with $M_{500,\mathrm{SZ}}$, computing the predicted shear
for such a profile, and then calculating the aperture mass from the
predicted profile over the same radii. These masses are inherently
projected, two-dimensional quantities. We plot the results in the top
panel of Figure \ref{fig:m12}.

In the second method, we estimate the spherical mass by fitting NFW
profiles to binned shear data at $R_{500,\mathrm{SZ}}/D_l < \theta <
12\arcmin$ and computing the total mass of the best fit profile at
$R_{500,\mathrm{SZ}}$.  We label this
$M_{\mathrm{NFW,WL}}(R_{500,\mathrm{SZ}})$.  The results are to be
compared to $M_{500,\mathrm{SZ}}$ directly.  Because of the use of a
radius from an external source, the radius at which the mass is quoted
is {\it not} the radius where $\Delta = 500$ (the overdensity factor
with respect to the cosmological critical density) in the best-fit NFW
model.  These masses are plotted in the bottom panel of Figure
\ref{fig:m12}.

\begin{figure}
\epsscale{1.2}
\plotone{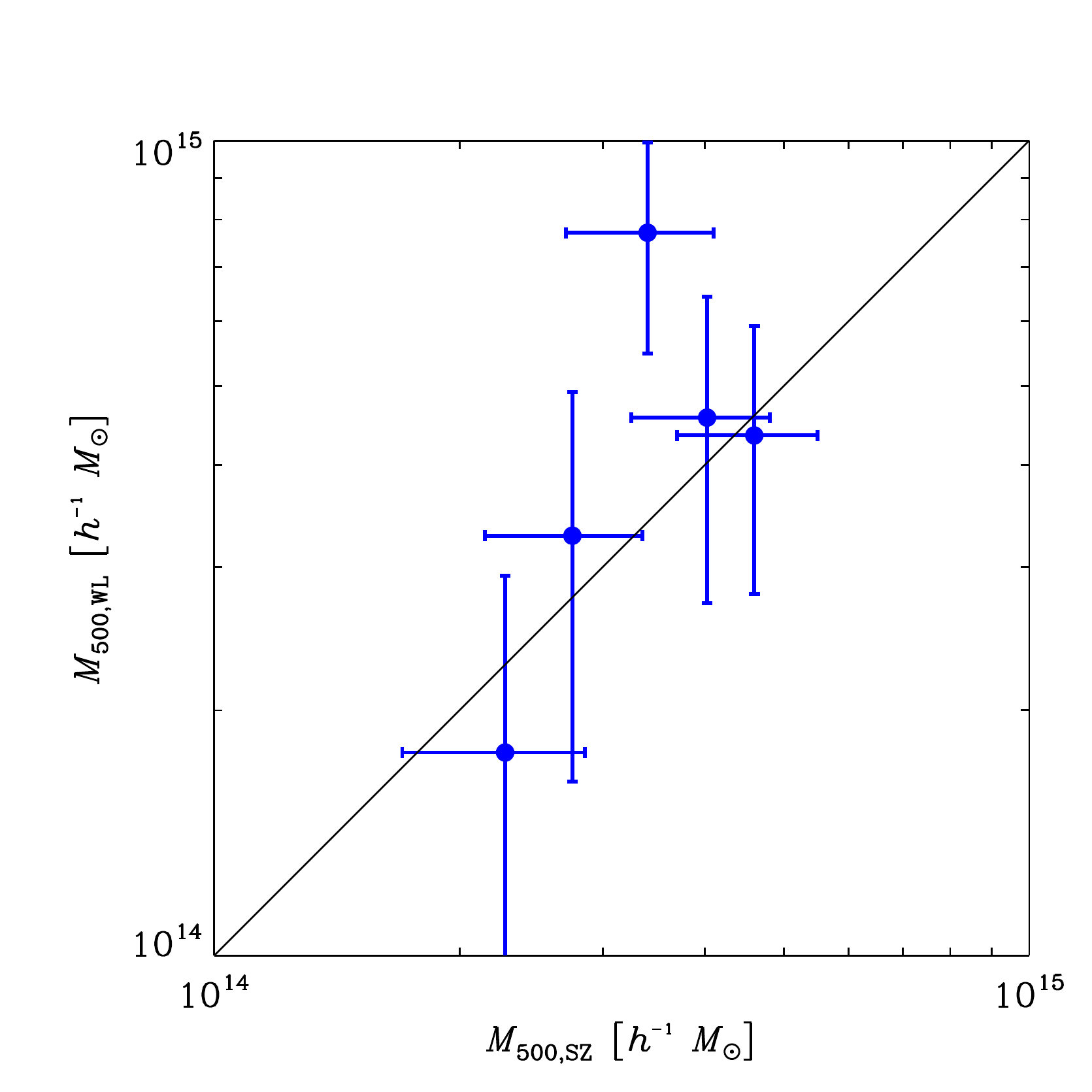}
\caption{Spherical WL masses vs spherical SZ masses.  The WL masses
  are determined by fitting \citetalias{navarro97} profiles to weak
  lensing shear data at $R_{500,\mathrm{SZ}}/D_l < \theta < 12\arcmin$
  and evaluating the resulting NFW mass profile at
  $R_{500,\mathrm{WL}}$ as determined from the best-fit profile
  itself.\label{fig:m3}}
\end{figure}

And in the third method, we use the same best-fit profiles above to
estimate the spherical mass within the radius where $\Delta = 500$ as
determined from the best-fit model itself, i.e.,
$M_{500,\mathrm{WL}}$.
% This is equal to $M_{\mathrm{NFW,WL}}(R_{500,\mathrm{WL}})$ in our
% notation.  
This is also to be compared to $M_{500,\mathrm{SZ}}$ directly.  These
masses are plotted in Figure \ref{fig:m3}.

The expectation of the ratio of WL to SZ masses is unity for all of
these methods in the absence of systematic errors from, for example,
the concentration and cluster centering assumptions. We note that we
use a concentration--mass--redshift scaling relation in all ratio tests:
for the aperture mass comparison, it is only used to transform the
SZ-derived mass to an aperture mass equivalent, whereas for both
spherical mass comparisons it is only used in the WL shear profile
fit.

The mass results and derived quantities are reported in Table
\ref{tab:mass}. 
% We also report the best-fit velocity dispersion in the
% SIS model used to transform reduced shear to shear. We emphasize that
% these velocity dispersion quantities are theoretical predictions
% inferred from the WL data.  
% Only shear at $R_{500,\mathrm{SZ}}/D_l <
% \theta < 12\arcmin$ was used to determine $\sigma_{\mathrm{vel}}$,
% although the correction was applied to all shear data. 
% The shear
% profiles of the SIS and NFW models are in excellent agreement at
% $\theta > R_{500,\mathrm{SZ}}/D_l$ (see figures in Appendix
% \ref{app:figures}).

\begin{deluxetable*}{cc|cc|ccc}
%\tabletypesize{\scriptsize}
%\rotate
\tablecaption{Cluster mass results\label{tab:mass}}
\tablewidth{0pt}
\tablehead{
\colhead{Cluster Name} & 
\colhead{$R_{500,\mathrm{SZ}}$} &
\colhead{$M_{\mathrm{ap,WL}}(R_{500,\mathrm{SZ}})$} &
\colhead{$M_{\mathrm{ap,SZ}}(R_{500,\mathrm{SZ}})$} &
\colhead{$M_{\mathrm{NFW,WL}}(R_{500,\mathrm{SZ}})$} &
\colhead{$M_{500,\mathrm{WL}}$} &
\colhead{$M_{500,\mathrm{SZ}}$} \\
\colhead{~} & 
\colhead{($h^{-1}\mathrm{Mpc}$)} &
\colhead{($10^{14}h^{-1}\Msun$)} &
\colhead{($10^{14}h^{-1}\Msun$)} &
\colhead{($10^{14}h^{-1}\Msun$)} &
\colhead{($10^{14}h^{-1}\Msun$)} &
\colhead{($10^{14}h^{-1}\Msun$)} 
}
\startdata
\clustera & 0.84 & $4.70\pm1.49$ & $5.31\pm0.60$ & $4.43\pm1.16$ & $4.35\pm1.57$ & $4.60\pm0.90$ \\%[2mm]
\clusterb & 0.69 & $3.03\pm1.64$ & $3.42\pm0.41$ & $3.08\pm1.12$ & $3.27\pm1.64$ & $2.75\pm0.60$ \\%[2mm]
\clusterc & 0.64 & $2.62\pm1.43$ & $2.87\pm0.39$ & $1.93\pm0.82$ & $1.77\pm1.15$ & $2.28\pm0.57$ \\%[2mm]
\clusterd & 0.77 & $6.62\pm1.66$ & $3.99\pm0.47$ & $5.75\pm1.43$ & $7.71\pm2.23$ & $3.40\pm0.70$ \\%[2mm]
\clustere & 0.77 & $5.21\pm1.88$ & $4.99\pm0.52$ & $4.37\pm1.29$ & $4.57\pm1.87$ & $4.03\pm0.78$
\enddata
\tablecomments{Results from the WL measurements, as well SZ
  data and derived quantities. We have divided two dimensional
  aperture masses from spherical masses.  \\{Column 1:}
  \citetalias{reichardt12} cluster designation. \\{Column 2:}
  cluster radius determined from the SPT SZ mass estimate in Column 7.
  \\{Column 3:} WL aperture mass at $R_{500,\mathrm{SZ}}$. \\{
    Column 4:} SZ aperture mass at $R_{500,\mathrm{SZ}}$. \\{
    Column 5:} WL mass at $R_{500,\mathrm{SZ}}$ from
  \citetalias{navarro97} profile fitted to shear data at
  $R_{500,\mathrm{SZ}}/D_l < \theta < 12\arcmin$.  \\{Column 6:}
  WL mass at $R_{500,\mathrm{WL}}$ from \citetalias{navarro97} profile
  fitted to shear data at $R_{500,\mathrm{SZ}}/D_l < \theta <
  12\arcmin$. \\{Column 7:} SZ mass estimate from
  \citetalias{reichardt12}.}
\end{deluxetable*}

The mean ratios of the three WL to SZ mass statistics are summarized
in Table \ref{tab:ratios}. In all cases we report weighted mean
values, where, for each cluster $n$, weights $1/\sigma_n^2$ are a
combination of the WL aperture mass statistical uncertainty (including
the estimated LSS contribution, which is between 15\% and 20\% for
these clusters) and the total SZ mass uncertainties from
\citetalias{reichardt12} propagated to the derived quantity when
necessary. The uncertainty on the mean is computed via $1/\sigma^2 =
\sum_n 1/\sigma_n^2$.

\begin{deluxetable*}{lcccccc}
%\tabletypesize{\scriptsize}
%\rotate
\tablecaption{Mass ratio results and consistency tests\label{tab:ratios}}
\tablewidth{0pt}
\tablehead{
\colhead{Procedure} & 
\multicolumn{2}{c}{Aperture Masses} &
\multicolumn{2}{c}{Spherical Masses at $R_{500,\mathrm{SZ}}$} &
\multicolumn{2}{c}{Spherical Masses at $R_{500,\mathrm{WL}}$} \\
\cmidrule(rl){2-3}\cmidrule(rl){4-5}\cmidrule(rl){6-7}
\colhead{~} & 
\colhead{Mean Ratio} &
\colhead{Total Scatter} &
\colhead{Mean Ratio} &
\colhead{Total Scatter} &
\colhead{Mean Ratio} &
\colhead{Total Scatter} \\
\colhead{~} &
\colhead{~} &
\colhead{(\%)} &
\colhead{~} &
\colhead{[\%]} &
\colhead{~} &
\colhead{(\%)}
}
\startdata
Baseline results & $1.04\pm0.18$ & 33 & $1.07\pm0.18$ & 33 & $1.10\pm0.24$ & 58 \\   
$\cfhti < 24.5$  & $1.05\pm0.20$ & 37 & $1.05\pm0.19$ & 37 & $1.08\pm0.25$ & 65 \\
$\cfhti < 24.0$  & $1.07\pm0.23$ & 34 & $1.07\pm0.21$ & 35 & $1.12\pm0.29$ & 60 \\
Conservative color cuts & $1.07\pm0.19$ & 35 & $1.12\pm0.18$ & 35 & $1.16\pm0.23$ & 58
\enddata
\tablecomments{WL-SZ mass ratio results. The baseline analysis employs
  a magnitude limit of $\cfhti = 25$ and color cuts described in
  Section \ref{sec:sigmacrit}. \\{Column 1:} procedure
  used to compute mean ratio. \\{Column 2:} weighted mean of the
  ratio of WL to SZ aperture masses, $\langle
  M_{\mathrm{ap,WL}}(R_{500,\mathrm{SZ}})/M_{\mathrm{ap,SZ}}(R_{500,\mathrm{SZ}})
  \rangle$. \\{Column 3:} the unweighted standard deviation of the
  WL to SZ aperture mass ratio. \\{Column 4:} weighted mean of the
  ratio of WL spherical mass (evaluated at $R_{500,\mathrm{SZ}}$) to
  SZ spherical mass, $\langle
  M_{NFW,\mathrm{WL}}(R_{500,\mathrm{SZ}})/M_{500,\mathrm{SZ}}
  \rangle$. \\{Column 5:} the unweighted standard deviation of the
  Column 4 ratio statistic.  \\{Column 6:} weighted mean of the
  ratio of WL spherical mass (evaluated at $R_{500,\mathrm{WL}}$) to
  SZ spherical mass, $\langle M_{500,\mathrm{WL}}/M_{500,\mathrm{SZ}}
  \rangle$. \\{Column 7:} the unweighted standard deviation of the
  Column 6 ratio statistic.}
\end{deluxetable*}

We note that this method does not take into account any correlated
uncertainty between clusters. The SZ mass estimates from
\citetalias{reichardt12} have been marginalized over cosmological and
scaling relation parameters, and this results in a $\sim 10\%$
systematic uncertainty that is highly correlated between clusters. We
have checked for the effect of these correlations. Briefly, we adopt a
Gaussian likelihood for both the SZ and WL mass statistics and
introduce nuisance parameters representing the true mass of each
cluster.  We introduce an additional free parameter representing an
overall scaling of the SZ mass estimates. We then explore the
resulting likelihood surface using an MCMC.  If we use a diagonal
covariance matrix with the uncertainties given in Table
\ref{tab:mass}, we recover the nominal mean ratios of WL to SZ mass
statistics of Table \ref{tab:ratios} at the maximum likelihood points
in the chains. If we use the covariance matrix for the five clusters
as estimated from the \citetalias{reichardt12} cosmological
chains, we find that the increase in the uncertainty on the mean
ratios is small ($\sim 10\%$ of the baseline ratio uncertainty values)
when compared to when we set these correlations to zero. As discussed
in Section \ref{sec:err}, WL systematic uncertainties are estimated to
be significantly smaller than the statistical components, so we only
use a diagonal WL covariance matrix.  Thus, we ignore the correlated
component to the SZ and WL cluster mass estimate uncertainties for the
remainder of this work.

The total scatter reported in Table \ref{tab:ratios} is the unweighted
standard deviation of the respective ratio data. We do not estimate
intrinsic scatter because this requires an estimate of the level of
correlation in intrinsic scatter between the WL and SZ
mass-observables.  Estimating these quantities is beyond the scope of
this work (see Section \ref{sec:scatter} for further discussion).

We have performed additional consistency tests by using brighter
$\cfhti$ magnitude limits, and by adopting more conservative color
cuts. The brighter magnitude limits are meant to probe the CFHTLS-Deep
catalogs at magnitudes where the photo-$z$ accuracy has been
explicitly verified and stated \citep{coupon09}. Results of these
tests are also reported in Table \ref{tab:ratios}. For all tests we
have performed, mean ratios are statistically consistent with unity
and with the baseline result to within $1\sigma$.

\subsection{Systematic Error Analysis}
\label{sec:err}

We have explored various potential sources of systematic error, as
discussed throughout the text, and we report the effects on the mean
WL-SZ mass ratios in Table \ref{tab:err} for the dominant or most
interesting cases.  Changes in mean ratios are quoted relative to the
baseline results of the aperture mass and spherical mass comparisons
(Table \ref{tab:ratios}).

\begin{deluxetable*}{lccc}
%\tabletypesize{\scriptsize}
%\rotate
\tablecaption{Possible sources of systematic error.\label{tab:err}}
\tablewidth{0pt}
\tablehead{
\colhead{Perturbation} & 
\multicolumn{3}{c}{Change in Mean Ratio} \\
\cmidrule(rl){2-4}
\colhead{~} &
\colhead{Aperture Masses} &
\colhead{Spherical Masses at $R_{500,\mathrm{SZ}}$} &
\colhead{Spherical Masses at $R_{500,\mathrm{WL}}$}
}
\startdata
\cutinhead{Concentration assumption}
\citet{duffy08} (baseline) & $+0.00$ & $+0.00$ & $+0.00$  \\
\citet{maccio08} & $+0.03$ & $+0.04$ & $+0.06$  \\
\citet{neto07}  & $+0.02$ & $+0.03$ & $+0.04$  \\
%\citet{dolag04} & $+0.05$ & $+0.08$  \\
\citet{prada11} 50th percentile & $+0.04$ & $+0.06$ &  $+0.08$  \\
\citet{prada11} 90th percentile & $+0.06$ & $+0.09$ &  $+0.11$  \\
\cutinhead{Cluster center assumption} 
Use BCG centers & $+0.00$ & $-0.01$ &  $-0.02$  \\
Use $\kappa$ map centers & $-0.14$ & $-0.12$ &  $-0.17$ \\
\cutinhead{Other} 
Calibration to $N$-body simulation & $+0.06$ & $+0.09$ &  $+0.13$ 
% $i$-band zeropoint wrong by $\pm0.05\,\mathrm{mag}$ & $0.98\pm0.14$ & $+0.02$ 
%aperture mass code bias & 3\%  \\
%Shear bias of $0.02\gamma_+$ & 2\% 
% SIS $\sigma_{\mathrm{vel}}=500$ & $0.17\pm0.76$ \\
% SIS $\sigma_{\mathrm{vel}}=1500$ & $0.05\pm0.71$
\enddata
\tablecomments{Exploration of possible sources of uncertainty in the
  WL-SZ mass ratios with respect to the baseline results shown in
  Table \ref{tab:ratios}. \\{Column 1:} brief description of the
  perturbation performed.  \\{Column 2:} change in weighted mean,
  $\langle
  M_{\mathrm{ap,WL}}(R_{500,\mathrm{SZ}})/M_{\mathrm{ap,SZ}}(R_{500,\mathrm{SZ}})
  \rangle$. \\{Column 3:} change in weighted mean, $\langle
  M_{\mathrm{NFW,WL}}(R_{500,\mathrm{SZ}})/M_{500,\mathrm{SZ}}
  \rangle$. \\{Column 4:} change in weighted mean, $\langle
  M_{500,\mathrm{WL}}/M_{500,\mathrm{SZ}} \rangle$. }
\end{deluxetable*}

Our first tests are to measure the effect that the assumed
concentration scaling relation has on the mean ratios.  Our baseline
analysis uses that of \citetalias{duffy08}, which gives $3.0 \leq c
\leq 3.2$ for these five clusters. The work of \citet{maccio08} give
$4.2 \leq c \leq 4.5$ and \citet{neto07} give $3.8 \leq c \leq
4.1$. \citet{prada11} find concentration values higher than these
other works, with 50th percentile values of $c\approx 5.0$ and
90th percentile values of $c\approx 7.1$. The latter represents the
most extreme concentrations we might expect for this sample, while the
\citetalias{duffy08} gives the lowest expected values.  This range of
concentrations change the aperture mass ratios by 2\% to
6\%. Spherical mass ratios using NFW fits to WL data evaluated at
$R_{\mathrm{500,SZ}}$ change by 3\% to 9\%, and those evaluated at
$R_{\mathrm{500,WL}}$ change by 4\% to 11\%.  This level of
uncertainty is less than our statistical uncertainty on the mass ratio
for the five clusters, but it is the dominant source of systematic
uncertainty in this work.

While we use the SZ positions as the cluster centers in the baseline
analyses, we also test using two other definitions of the cluster
center: (1) the BCG and (2) the peak in the WL-reconstructed $\kappa$
field. We use the BCGs selected of \citet{song12}, which are typically
$\lesssim 30\arcsec$ from the SZ centroid, with the exception of
\clusterd, which is nearly $2\arcmin$ away. Mean ratio results using
the BCG as the cluster center are in near perfect agreement with those
using the SZ positions. However, adoption of the $\kappa$ peak
location as the cluster center reduces the mean ratios of aperture
masses and spherical masses by significantly greater amounts. The
$\kappa$ peak locations have statistical uncertainties of $\sim
30\arcsec$ to $60\arcsec$, which are the largest uncertainties of the
three center choices we have considered. While the mean ratios using
$\kappa$ peak locations are statistically consistent with unity and
with the baseline results, this represents the largest deviation from
the baseline. This is probably because of noise in centering due to
the large statistical uncertainties in determining convergence field
peak locations; indeed, the universal reduction in the ratio values
using these centers provides some evidence that the convergence peaks
are poorer estimators of cluster centers than the SZ and BCG positions
given the data, as centering errors suppress shear profiles,
particularly in the inner regions \citep{george12}.

Tests using $N$-body simulations (Section \ref{sec:mocks}) result in
the calibration bias levels reported in Table \ref{tab:err}.  The
statistical uncertainties of these calibration bias estimates are each
$2\%$.  We have not applied these bias corrections to any individual
mass estimates nor to any mean mass ratio statistics presented in this
work.  We note that such bias corrections increase tension with the
mean mass ratio expectation of unity by up to $1\sigma$ at most.

% In addition to these, the shear measurement method is estimated to be
% accurate to $0.02\gamma$ in shear, or $\sim 2\%$ in mass (Section
% \ref{sec:ksb}).

The effects of other known sources of systematic uncertainty are
smaller than those of concentration. The shear measurement method is
estimated to be accurate to $0.02\gamma$ in shear (Section
\ref{sec:shear}), which translates to $\leq2\%$ in the mean aperture
mass ratio and $\leq3\%$ in mean spherical mass ratios. 
% The
% fundamental accuracy of our methodology has been verified to the $2\%$
% level under idealized conditions as well (Section \ref{sec:mocks}).

The source redshift distribution inferred from the four different
CFHTLS Deep fields (Section \ref{sec:sigmacrit}) is expected to vary
due to finite galaxy counts and cosmic variance. The source redshift
distribution in the SPT cluster fields we have observed is also
expected to vary across fields in the same way. We estimate the
uncertainty this induces in the mean ratio results by repeating the
entire analysis using each CFHTLS Deep field individually and
averaging the results. Under the assumption of randomness, the mean
aperture mass ratio carries an uncertainty from finite galaxy counts
and cosmic variance of $\pm 1.1\%$, while the mean spherical mass
ratios are affected at $\pm 1.0\%$ (using $R_{\mathrm{500,SZ}}$) and
$\pm 1.6\%$ (using $R_{\mathrm{500,WL}}$). This uncertainty can in
principle be reduced further by using standard photo-$z$ catalogs of
additional, disparate fields, or by measuring photo-$z$'s directly in
the SPT cluster fields, but this is not necessary given the data, as
it is significantly subdominant to other uncertainties.

Bias and uncertainty from assumed cosmological parameters are also
subdominant. We have assumed cosmological parameter values that are
slightly different from the results of \citetalias{reichardt12}
(combining the 100 high purity SPT detected clusters with the
CMB+BAO+$H_0$+SNe data). To estimate the bias and scatter this
induces, we have tested marginalizing the WL to SZ mass ratios over
the \citetalias{reichardt12} $\Lambda$CDM chains that combine
CMB+BAO+$H_0$+SNe+SPT$_{\mathrm{CL}}$ data. The results are that both
the bias and statistical uncertainty due to cosmology are sub-percent
in the mass ratio statistics, and are therefore significantly below
those expected from other sources of potential error.

In summary, no potential sources of systematic error that we have
explored contributes at or above the level of statistical accuracy for
this sample of clusters, suggesting they are all subdominant.
Uncertainty in the assumed NFW profile concentration and calibration
biases from tests using $N$-body simulations are likely the
dominant systematic uncertainties.

\section{Discussion}
\label{sec:discussion}

In this section we discuss the level and possible origin of scatter in
WL and SZ mass statistics, as well as individual cluster systems that
exhibit interesting features in the data.

\subsection{The Effects of Choice of Radius on Mass Ratios}
\label{sec:scatter}

We have tested three methods of comparing WL- and SZ-derived masses,
and find that WL mass statistics evaluated at a fixed radius of
$R_{\mathrm{500,SZ}}$ exhibit less total scatter than spherical WL
masses evaluated at $R_{\mathrm{500,WL}}$ (Table \ref{tab:ratios}).
This is primarily because $R_{\mathrm{500,WL}}$, effectively
determined from the best-fit profiles to the WL data, has a greater
uncertainty than $R_{\mathrm{500,SZ}}$.  Fixing the radius to some
externally specified value that has greater precision has the benefit
of reducing this source of uncertainty.

On the other hand, masses measured at externally specified radii that
are themselves determined from a mass proxy (such as
$R_{\mathrm{500,SZ}}$) must suffer from correlated scatter in addition
to any that is intrinsic between the observables. For illustration, if
the $M_{\mathrm{500,SZ}}$ estimate of \citetalias{reichardt12} is 20\%
larger than the true $M_{\mathrm{500}}$, then the
$R_{\mathrm{500,SZ}}$ estimate scatters up by $7\%$, and as a result,
the WL aperture mass and NFW mass evaluated at $R_{\mathrm{500,SZ}}$
also scatter up by some amount, under the assumption of monotonically
increasing mass profiles. This means that scatter in SZ-derived masses
induces additional, positively correlated scatter between WL and SZ
masses due to the use of $R_{\mathrm{500,SZ}}$, in addition to any
intrinsic correlation.

This motivates fixing the radius to some value in arcminutes or in
Mpc, or more generally to a value derived from any data that are
minimally correlated with the mass-observables or have negligible
uncertainty. We have tested measuring aperture masses setting
$\theta_1 = 0.8h^{-1}\mathrm{Mpc}/D_l$ and $\theta_2 = 12\arcmin$ for
all clusters, where the inner radius is roughly equal to the median
$R_{\mathrm{500,SZ}}$ for the ensemble. The equivalent statistic for
the SZ-derived mass is also computed. The weighted mean ratio result
is $\langle
M_{\mathrm{ap,WL}}(0.8h^{-1}\mathrm{Mpc})/M_{\mathrm{ap,SZ}}(0.8h^{-1}\mathrm{Mpc})\rangle
=1.02\pm0.18$, with total scatter of 35\%. Compared to the baseline
spherical mass ratio with 34\% scatter in Table \ref{tab:ratios}, this
increased scatter could be consistent with a $\sim10\%$ correlated
scatter induced by estimating both the WL and SZ spherical masses at
$R_{\mathrm{500,SZ}}$.

We have also tested fitting NFW profiles to shear data at
$0.8h^{-1}\mathrm{Mpc}/D_l <\theta < 12\arcmin$, and evaluating the
total mass at $0.8h^{-1}\mathrm{Mpc}$.  To compute the equivalent SZ
mass statistic, we evaluate the NFW profile that is consistent with
$M_{\mathrm{500,SZ}}$ also at $0.8h^{-1}\mathrm{Mpc}$.  The weighted
mean result is $\langle
M_{\mathrm{NFW,WL}}(0.8h^{-1}\mathrm{Mpc})/M_{\mathrm{NFW,SZ}}(0.8h^{-1}\mathrm{Mpc})
\rangle =1.12\pm0.19$, with total scatter of 36\%.  The spherical mass
ratio result at $R_{\mathrm{500,SZ}}$ in Table \ref{tab:ratios}
exhibits 15\% less total scatter (in quadrature), which again may be
due in part to the added correlation induced by using
$R_{\mathrm{500,SZ}}$.

% We have confirmed in $N$-body simulations, as with the other ratio
% statistics (Section \ref{sec:mocks}), that these fixed aperture tests
% are also unbiased in the absence of systematic errors in, for example,
% concentration and cluster centering.

This preliminary analysis is suggestive that minimum variance mass
comparisons may be achievable by fixing the radius of analysis in Mpc,
or some other quantity that has negligible uncertainty or is
uncorrelated with mass. We note that these choice of radii would have
non-trivial complications for cluster cosmological analyses, which
usually compare the measured cluster abundance to predictions where
cluster masses are typically defined as $M_{\Delta}$. However, a
rigorous comparison of different mass measures and identification of
minimum variance estimators requires quantifying correlations in
scatter. We leave this to future work.

\subsection{Disturbed Systems and Line-of-sight Structure}
\label{sec:unrelaxed}

The primary goal of this work is to provide cosmologically unbiased tests of the
scaling of the SZ observable with total mass. 
As described in Section
\ref{sec:mocks}, we have performed calibration tests on mock
catalogs based on ray-traced $N$-body simulations for all halos above
a given mass, as identified by the $N$-body cluster finder, without
regard to whether the halos have undergone recent merger activity or
contain significant structures along the line of sight. This
approximately mimics the SZ selection, which is roughly uniform in
mass at $z\gtrsim 0.3$. The verification exercise indicates no signs
of bias to $\sim 2\%$ for both the aperture mass and spherical mass
ratio tests, under the simple assumptions adopted in that verification
study. Similarly, 
we must include any SPT-detected clusters that may
exhibit merger activity or contain known structures in the line of
sight, as cutting them out risks introducing additional bias in the
mean mass
ratio tests.

Nonetheless, identifying disturbed systems and significant line of
sight structures is of some interest. For example, we note that the
inner regions of the shear profiles ($<R_{500,\mathrm{SZ}}$) of
\clustera\ and \clusterd\ show disagreement with the NFW profiles
fitted at $R_{500,\mathrm{SZ}}/D_l < \theta < 12\arcmin$, and these
two cluster systems also show the greatest disagreement between SZ
significance centroid, BCG, and $\kappa$ peak locations. We explore
possible explanations here; however, ultimately we conduct WL tests in
this work in the same way we have tested our procedure in $N$-body
simulations, and we therefore do not alter the WL-SZ mass comparison
methodology based on the results of these explorations.

\subsubsection{\clustera}

\citet{andersson10} note that \clustera\ exhibits north--south
elongation in X-ray data obtained with {\it Chandra}. \citet{zhang08}
also note ellipticity in {\it XMM-Newton} X-ray data for this
cluster. Both the distribution of red cluster-galaxies and the
convergence-map morphology show north--south elongation as well, in
rough agreement with the X-ray emission structure and SPT SZ
significance map. In addition, the convergence peak disagrees with the
SZ significance centroid and the BCG locations at about $3\sigma$. This
suggests that this cluster is unrelaxed. It has been shown that
core-excised X-ray mass-observables are nearly universal for relaxed
and unrelaxed clusters \citep[e.g.,][]{mantz10,vikhlinin09b}. The SPT
SZ observable is similarly insensitive to the details of the core gas
activity, so the SZ-derived mass should not be greatly affected.  We
note that the WL- and SZ-derived masses are in excellent agreement,
which may be due to our use of shear profile data only outside the
cluster core.
% It is also interesting to note that the cluster
% morphology does seem to affect the X-ray mass estimate in
% \citet{andersson10}, which is $\sim2$ times larger than either the WL
% or SZ mass estimate.  This level of offset in the X-ray mass estimate
% is approximately what would be expected if the merger was entirely
% along the plane-of-the-sky (Andersson, private communication).

% We test including the inner portions of the shear profile (which, by
% construction, affects the \citetalias{navarro97} fit and not the
% aperture mass). This gives $M_{\mathrm{500,WL}}=(1.96\pm1.12)\times
% 10^{14}h^{-1}\Msun$, which is $1.8\sigma$ discrepant from
% $M_{\mathrm{500,SZ}}$. This disagreement is statistically
% insignificant. While this disagreement is statistically
% insignificant, this cluster system qualitatively illustrates our
% motivation for adopting the aperture mass statistic to test SZ
% masses, as well for fitting NFW profiles only at large radii.

\citet{mcinnes09} measured the WL mass of \clustera\ using data from
the Blanco Cosmology Survey.  The measured masses agree at the $\sim
1\sigma$ level after analytically accounting for the different
concentration values used, as well as for the different overdensity
definitions.

\subsubsection{\clusterd}

The data of the \clusterd\ field show a number of interesting
features.  First, the two brightest cluster galaxies (both with
spectroscopic redshifts consistent with $0.28$) have relatively large
offsets from the SZ centroid at $\sim 1\arcmin$ each, are of similar
luminosity, and are also consistent with two peaks seen in the WL
convergence map (see Appendix \ref{app:figures}), which suggests
possible merging activity.  Second, the shear profile is suppressed
within $R_{500,\mathrm{SZ}}$, similar to \clustera.  And third, there
is evidence in the literature for a structure in the foreground of the
higher redshift, spectroscopically confirmed SZ cluster.  In this
section we explore the possible implications of a foreground
interloper and a plane-of-sky merger on the data.

The evidence for a foreground cluster or group is strong.  As
discussed in detail in \citet{song12}, \clusterd\ is listed by
\citetalias{reichardt12} as having an SZ location that is consistent
with the optically identified cluster A3685 (location
$115\arcsec\pm150\arcsec$ from the SPT SZ position) and the X-ray
detected cluster RXC J2032.1-5627 (location $87\arcsec$ from the SPT
SZ position). A3685 is assigned $z=0.062$ from just one galaxy
\citep{fetisova81,struble99}, but it is more likely to be $z=0.14$
because five galaxies near its location are identified at this
redshift in the spectroscopic galaxy catalog of \citet{guzzo09}.
While RXC J2032.1-5627 is identified at $z=0.14$ from the
\citet{guzzo09} catalog, that catalog also contains six galaxies
consistent with the redshift we identify for \clusterd, which is
derived from 31 galaxies from our own observations. \citet{song12}
conclude that both the X-ray and the SZ signals are likely
predominantly arising from a massive $z=0.28$ cluster, but there is
also a cluster at $z=0.14$ (possibly the object identified originally
as A3685) spatially consistent with the SZ detection.

We explore the possible effects of a foreground interloper and a
plane-of-sky merger on the SZ data by simulating halos using analytic
profiles of \citet{arnaud10}, and on the WL data by simulating NFW
halos.  In our first set of simulations, we model an $M_{200} =
10^{14} h^{-1}\Msun$ ($M_{500} = 0.75\times 10^{14} h^{-1}\Msun$)
cluster at $z=0.14$ and an $M_{500} = 4.30\times 10^{14} h^{-1}\Msun$
cluster at $z=0.28$.  The chosen mass of the interloper is motivated
by the designation of Richness Class 0 to A3685 by
\citet{abell89}, which indicates identification of 30--49 clustered
galaxies and is the smallest possible richness on that scale.  The
velocity dispersion of the five galaxies at this redshift in the
\citet{guzzo09} catalog ($470\pm190\mathrm{km}\,\mathrm{s}^{-1}$)
suggests a mass of
$M_{200}=(0.8^{+1.4}_{-0.6})\times10^{14}h^{-1}\Msun$ using the
scaling relation of \citet{evrard08}, which is also consistent.  To
reflect the spatial offset of the convergence peaks seen in the data,
we separate the cluster centers by $\sim 4\arcmin$ by centering the
clusters at the two $\kappa$ peak locations.  We confirm from these
simulations that a foreground object would boost WL masses by roughly
50\%, while the SZ observable would be boosted by only $\lesssim 2\%$.
This configuration also reproduces the double peak in the convergence
field and the suppression of the shear profiles within $R_{500}$.  The
SZ centroid in this case would be centered on the gas of the more massive
system, which is inconsistent with our data.  To reproduce the WL-SZ
mass ratios we see in the data, the foreground cluster would have to
be $M_{200} = 3\times 10^{14} h^{-1}\Msun$ ($M_{500} = 2.2\times
10^{14} h^{-1}\Msun$), which is different from the dynamical mass at
$1.8\sigma$; however, taking the data together, this larger mass is no
more or less favored than the central dynamical mass value, as the WL
and dynamical masses and the convergence peak locations carry large
uncertainties.

In our second set of simulations, we model a plane-of-sky merger by
placing at $z=0.28$ two $M_{200} =3\times 10^{14} h^{-1}\Msun$
($M_{500} = 2.2\times 10^{14} h^{-1}\Msun$) clusters at the $\kappa$
peak locations, which roughly reproduces a single $M_{500} =
4.30\times 10^{14} h^{-1}\Msun$ cluster if naively summed.  This
configuration also reproduces the double peak in the convergence field
and the suppression of the shear profiles within $R_{500}$.  However,
the mass estimates in this case agree well with the SZ derived mass,
in contrast to the foreground contaminant simulation and the real
data.  We would expect the SZ centroid to fall in between the two merging
clusters of similar mass, which is consistent with the data.

In summary, we cannot fully differentiate between the two scenarios of
a foreground cluster or a merger, or quantify their exact effect on
the WL and SZ mass estimates, though both scenarios could explain the
suppressed shear profile within $R_{500,\mathrm{SZ}}$.  The
spectroscopic data show strong evidence for a lower mass interloping
cluster at $z=0.14$, which could lead to large WL-SZ mass ratios as
seen in the data; however, there is significant uncertainty on the
interloper's mass making this interpretation inconclusive. The merging
scenario would be more consistent with the observed offset in the WL
peaks from the SZ data and their coincidence with the BCGs at
$z=0.28$, but a merger alone would not explain the large WL-SZ mass
ratio. Therefore, the data may be pointing to a combination of a
plane-of-sky merger and a foreground interloper.  In this section we
have quantified the impact of either scenario on the WL and SZ mass
estimates.  Regardless, either scenario does not impact the goal of
this work, which is to provide an unbiased test of the accuracy of the
masses of an ensemble of SZ-selected clusters. We therefore do not
treat this cluster differently from the others in the mean ratio
tests.

% In summary, the spectroscopic data show strong evidence for a lower
% mass cluster or group at $z=0.14$ in the line of sight of the massive
% $z=0.28$ cluster.  Our simulations show that this scenario could in
% principle be consistent with the larger WL to SZ mass ratios we see in
% our results, as well as the double convergence peaks and the shear
% profile that is suppressed at radii $<R_{500,\mathrm{SZ}}$.  Some of
% the data also suggest merging activity at the $0.28$ cluster
% redshift. Our simulations show the WL and SZ data could be consistent
% with a combination of a merger and a foreground interloper.
% Nonetheless, these simulations are not exhaustive and carry large
% uncertainties; as the goal of this work is to provide an unbiased a
% test of the accuracy of the masses of an ensemble of SZ-selected
% clusters as possible, we do not treat this cluster differently from
% the others in the mean ratio tests.

\subsection{Other Works Comparing WL and SZ-derived Masses}

\citetalias{hurleywalker12} measured the mass of galaxy clusters with
WL data from the CFHT and SZ data from the Arcminute
Microkelvin Imager (AMI).  The four clusters for which both WL and SZ
masses were successfully estimated exhibited a weighted-mean ratio of
WL-to-SZ masses of $0.86\pm 0.14$, with total scatter of $14\%$.
While this agrees with unity and with our baseline results at
$\lesssim 1\sigma$, such a direct comparison between their work and
ours is not straightforward given the different overdensity used
($\Delta = 200$), their use of a joint Bayesian analysis of the two
data sets, and other differences in analysis techniques.

\citetalias{planck12} showed that the amplitude of SZ signal derived from
{\it Planck} data versus WL masses derived from Local Cluster
Sub-structure Survey data obtained with the Subaru telescope
disagrees at $>1\sigma$ with results calibrated to X-ray data assuming
hydrostatic equilibrium, but in the opposite sense than that predicted
by non-thermal pressure support.  That work proposes that systematic
errors in the WL analysis could give rise to the discrepancy seen, and
that further, careful study of systematics is needed to determine
whether the difference is astrophysical in origin.  The discrepancy is
also seen in the scaling relation result of \citet{marrone11}, which
used the same WL data but used SZ data from the SZ
Array (SZA).

\section{Conclusion}
\label{sec:conclusion}

We have observed five galaxy clusters with the Megacam camera on the
$6.5\mathrm{m}$ Magellan Clay telescope, with the goal of measuring
total masses with weak gravitational lensing and empirically testing
the accuracy of the SZ and X-ray--based mass estimates of
\citet[][\citetalias{reichardt12}]{reichardt12}. Shear is estimated in
deep $r'$-band data, and additional $g'$- and $i'$-band data are used to
calibrate photometry using the stellar color--color locus, as well as
to remove cluster galaxies using color cuts. The source redshift
distribution is estimated from CFHTLS-Deep photometric redshift
catalogs (which do not overlap our fields) under the same photometric
selections.

We adopt three measures of total mass derivable from WL data: aperture
masses, which are inherently two-dimensional quantities; spherical
masses derived from NFW fits to shear data and evaluated at
$R_{500,\mathrm{SZ}}$ determined from the SZ data; and the same NFW
fits evaluated at $R_{500,\mathrm{WL}}$ as determined from the
best-fit profiles themselves. In all cases, only shear profiles from
$R_{\mathrm{500,SZ}}$ to $12\arcmin$ are used. To make one-to-one
comparisons with the WL derived aperture masses, we compute the
aperture mass equivalent statistic of $M_{\mathrm{500,SZ}}$, which
requires the assignment of an NFW profile using
concentration--mass--redshift scaling relations from the
literature. For the spherical mass comparison, a $c$--$M$--$z$
relation is also adopted to fit models to the shear profiles, and the
resulting WL mass is compared directly to $M_{\mathrm{500,SZ}}$.
Calibration tests are performed on mock catalogs based on ray-traced
$N$-body simulations.  These show evidence for bias at levels
consistent with a number of previous works, under the assumption of
perfect knowledge of cluster centers, galaxy redshifts, and galaxy
shears.  Under all methods, the mean ratio of WL to SZ masses is
consistent with expectation of unity to within the statistical
uncertainty: the mean aperture mass ratio is $1.04\pm0.18$, the mean
spherical mass ratio using $R_{\mathrm{500,SZ}}$ is $1.07\pm0.18$, and
the mean spherical mass ratio using $R_{\mathrm{500,WL}}$ is
$1.10\pm0.24$. Consistency checks are performed, wherein we make more
conservative selections on both the Clay-Megacam and CFHTLS-Deep
photometric catalogs, and resulting mean ratios remain statistically
consistent with unity and with the baseline results in all cases.

Possible sources of systematic error are explored. The dominant sources
are most likely the assumption about concentration and calibration to
$N$-body simulations.  Different concentration scaling relations change mean
aperture mass ratios by 2\% to 6\%, and a few percent more in the mean
NFW mass ratios, depending on the 
relation adopted. Reducing this source of error will be key to
unbiased and precise mass constraints, and this requires obtaining
knowledge of the concentration of the cluster population that SPT
detects.  The bias levels with respect to $N$-body simulations is
consistent with levels seen in other works, at $-6\%$ to $-13\%$.

The assumed cluster center is another potential source of
uncertainty. We show that using the BCG position give nearly exact
agreement with baseline results in which the SZ position is used.
While using the reconstructed convergence field peak gives the
greatest deviations from the baseline result, this is probably because
of noise in centering due to the large statistical uncertainties in
determining convergence field peak locations.  Other sources of
systematic error, including shear bias, assumed cosmological
parameters, and statistical uncertainty on the source redshift
distribution, are subdominant to these.

% While aperture mass ratios appear to show less total scatter than
% spherical mass ratios, they may also have greater correlated scatter.
% This arises from the adoption of an aperture that is a function of the
% mass we are testing. Spherical masses show more total scatter, but are
% expected to carry less correlated scatter. Measuring these
% correlations is left to future work. 
For two of the five cluster systems in this work, we identify signs of
possible merging activity and structures in the line-of-sight. We
discuss these systems and simulate the effects of the proposed
contaminants.  The simulations show that both plane-of-sky merging
activity and line-of-sight structures can induce multiple peaks in the
convergence field and suppress shear profiles within
$R_{500,\mathrm{SZ}}$, but have different effects on the WL derived
masses.  The tests are not conclusive about the actual physical
configuration of the clusters.  The goal of this work is to provide as
unbiased a test of the accuracy of the \citetalias{reichardt12} mass
estimates of SZ-selected clusters as possible, so we do not treat
these clusters differently in our analysis.

In conclusion, we find statistical consistency between masses derived
from WL data and those derived from SZ and X-ray data at the
$\sim20\%$ level.  This represents first steps toward improved galaxy
cluster mass estimates in the SPT survey.  Improving the calibration
of mass-observables is critical for exploiting the full statistical
power of the SPT $2500\deg^2$ survey data set in cosmological cluster
abundance studies.

% This is the first result in an ongoing campaign to calibrate the
% normalization of the SZ mass-observable using weak lensing.  We have
% observed with Clay-Megacam more than three times the number of cluster
% presented here, and the analysis of this full data set is underway.
% We have also obtained imaging of over two dozen clusters at high
% redshift with the Advanced Camera for Surveys aboard the Hubble Space
% Telescope, which allows for weak lensing analyses of clusters at
% higher redshifts than are otherwise accessible from the ground.  These
% clusters have X-ray and multi-object spectroscopy observations as
% well.  Together these represent major steps toward achieving the
% calibration goal of the SPT cluster survey over the full redshift
% range it is sensitive to.

%\phantom{\citet{h

\acknowledgments

{\it Facilities:}
% \facility{Blanco (NEWFIRM)},
% \facility{Blanco (MOSAIC)},
% \facility{CXO (ACIS)},
% \facility{Gemini-S (GMOS)},
\facility{Magellan:Baade (IMACS)},
% \facility{Magellan:Clay (LDSS3)},
\facility{Magellan:Clay (Megacam)},
% \facility{Spitzer (IRAC)},
% \facility{South Pole Telescope},
% \facility{VLT:Antu (FORS2)}

We thank M.\ Holman for trading Megacam observing time for a
proof-of-concept study, which enabled the first ever detection of weak
gravitational shear using the Clay-Megacam system (\clustera). We
extend thanks to P.\ Protopapas for observing assistance, as well as
the entire staff of Las Campanas Observatory and the Megacam
instrument scientists.

We gratefully acknowledge Risa Wechsler, Michael Busha, and Matt
Becker for providing us simulated shear catalogs. The simulation used
to produce the catalogs is one of the Carmen simulations, a 1 Gpc
simulation run by M.\ Busha as part of the Las Damas project.

We also thank
D.\ Johnston, E.\ Rozo, J.\ Dietrich, J.\ Rhodes, D.\
Clowe, P.\ Melchior, P.\ Shechter, and S.\ Dodelson for useful
discussions.

We gratefully acknowledge the anonymous referee for providing useful
comments that improved this manuscript.

The South Pole Telescope program is supported by the National Science
Foundation through grant ANT-0638937.  Partial support is also
provided by the NSF Physics Frontier Center grant PHY-0114422 to the
Kavli Institute of Cosmological Physics at the University of Chicago,
the Kavli Foundation, and the Gordon and Betty Moore Foundation.  HH
acknowledges support from Marie Curie IRG grant 230924 and the
Netherlands Organisation for Scientific Research (NWO) grant number
639.042.814.  The Munich group acknowledges support from the
Excellence Cluster Universe and the DFG research program TR33 The Dark
Universe.  Galaxy cluster research at Harvard is supported by NSF
grant AST-1009012, and research at SAO is supported in part by NSF
grants AST-1009649 and MRI-0723073.  The McGill group acknowledges
funding from the National Sciences and Engineering Research Council of
Canada, Canada Research Chairs program, and the Canadian Institute for
Advanced Research. R.\ J.\ F.\ is supported by a Clay fellowship.

This paper used data products produced by the OIR Telescope Data
Center, supported by the Smithsonian Astrophysical Observatory.
 
This work is based in part on data products produced at the Canadian
Astronomy Data Centre and Terapix as part of the Canada-France-Hawaii
Telescope Legacy Survey, a collaborative project of the National
Research Council of Canada and the French Centre National de la
Recherche Scientifique.

% We used the cosmology calculator tools of \citet{wright06}.
%\phantom{\citep{hurleywalker12}}

\bibliography{../../BIBTEX/oir.bib}

\begin{thebibliography}{80}
\expandafter\ifx\csname natexlab\endcsname\relax\def\natexlab#1{#1}\fi

\bibitem[{{Abell} {et~al.}(1989){Abell}, {Corwin}, \& {Olowin}}]{abell89}
{Abell}, G.~O., {Corwin}, Jr., H.~G., \& {Olowin}, R.~P. 1989, \apjs, 70, 1

\bibitem[{{Albrecht} {et~al.}(2006){Albrecht}, {Bernstein}, {Cahn}, {Freedman},
  {Hewitt}, {Hu}, {Huth}, {Kamionkowski}, {Kolb}, {Knox}, {Mather}, {Staggs},
  \& {Suntzeff}}]{albrecht06}
{Albrecht}, A., {et~al.} 2006, ArXiv e-prints, astro-ph/0609591

\bibitem[{{AMI Consortium: Hurley-Walker} {et~al.}(2012){AMI Consortium:
  Hurley-Walker}, {Bridle}, {Cypriano}, {Davies}, {Erben}, {Feroz}, {Franzen},
  {Grainge}, {Hobson}, {Lasenby}, {Marshall}, {Olamaie}, {Pooley},
  {Rodr{\'{\i}}guez-Gonz{\'a}lvez}, {Saunders}, {Scaife}, {Schammel}, {Scott},
  {Shimwell}, {Titterington}, {Waldram}, \& {Zwart}}]{hurleywalker12}
{AMI Consortium: Hurley-Walker}, N., {et~al.} 2012, \mnras, 419, 2921

\bibitem[{{Andersson} {et~al.}(2011){Andersson}, {Benson}, {Ade}, {Aird},
  {Armstrong}, {Bautz}, {Bleem}, {Brodwin}, {Carlstrom}, {Chang}, {Crawford},
  {Crites}, {de Haan}, {Desai}, {Dobbs}, {Dudley}, {Foley}, {Forman},
  {Garmire}, {George}, {Gladders}, {Halverson}, {High}, {Holder}, {Holzapfel},
  {Hrubes}, {Jones}, {Joy}, {Keisler}, {Knox}, {Lee}, {Leitch}, {Lueker},
  {Marrone}, {McMahon}, {Mehl}, {Meyer}, {Mohr}, {Montroy}, {Murray}, {Padin},
  {Plagge}, {Pryke}, {Reichardt}, {Rest}, {Ruel}, {Ruhl}, {Schaffer}, {Shaw},
  {Shirokoff}, {Song}, {Spieler}, {Stalder}, {Staniszewski}, {Stark}, {Stubbs},
  {Vanderlinde}, {Vieira}, {Vikhlinin}, {Williamson}, {Yang}, {Zahn}, \&
  {Zenteno}}]{andersson10}
{Andersson}, K., {et~al.} 2011, \apj, 738, 48

\bibitem[{{Arnaud} {et~al.}(2010){Arnaud}, {Pratt}, {Piffaretti},
  {B{\"o}hringer}, {Croston}, \& {Pointecouteau}}]{arnaud10}
{Arnaud}, M., {Pratt}, G.~W., {Piffaretti}, R., {B{\"o}hringer}, H., {Croston},
  J.~H., \& {Pointecouteau}, E. 2010, \aap, 517, A92+

\bibitem[{{Bah{\'e}} {et~al.}(2012){Bah{\'e}}, {McCarthy}, \& {King}}]{bahe12}
{Bah{\'e}}, Y.~M., {McCarthy}, I.~G., \& {King}, L.~J. 2012, \mnras, 421, 1073

\bibitem[{{Bartelmann} \& {Schneider}(2001)}]{bartelmann01}
{Bartelmann}, M., \& {Schneider}, P. 2001, \physrep, 340, 291

\bibitem[{{Becker} \& {Kravtsov}(2011)}]{becker11}
{Becker}, M.~R., \& {Kravtsov}, A.~V. 2011, \apj, 740, 25

\bibitem[{{Beers} {et~al.}(1990){Beers}, {Flynn}, \& {Gebhardt}}]{beers90}
{Beers}, T.~C., {Flynn}, K., \& {Gebhardt}, K. 1990, \aj, 100, 32

\bibitem[{{Benson} {et~al.}(2011){Benson}, {de Haan}, {Dudley}, {Reichardt},
  {Aird}, {Andersson}, {Armstrong}, {Bautz}, {Bayliss}, {Bazin}, {Bleem},
  {Brodwin}, {Carlstrom}, {Chang}, {Cho}, {Clocchiatti}, {Crawford}, {Crites},
  {Desai}, {Dobbs}, {Foley}, {Forman}, {George}, {Gladders}, {Halverson},
  {High}, {Holder}, {Holzapfel}, {Hoover}, {Hrubes}, {Jones}, {Joy}, {Keisler},
  {Knox}, {Lee}, {Leitch}, {Liu}, {Lueker}, {Luong-Van}, {Mantz}, {Marrone},
  {McDonald}, {McMahon}, {Mehl}, {Meyer}, {Mocanu}, {Mohr}, {Montroy},
  {Murray}, {Natoli}, {Padin}, {Plagge}, {Pryke}, {Rest}, {Ruel}, {Ruhl},
  {Saliwanchik}, {Saro}, {Schaffer}, {Shaw}, {Shirokoff}, {Song}, {Spieler},
  {Stalder}, {Staniszewski}, {Stark}, {Story}, {Stubbs}, {Suhada}, {van
  Engelen}, {Vanderlinde}, {Vieira}, {Vikhlinin}, {Williamson}, {Zahn}, \&
  {Zenteno}}]{benson11}
{Benson}, B.~A., {et~al.} 2011, ArXiv e-prints, 1112.5435

\bibitem[{{Bertin} \& {Arnouts}(1996)}]{bertin96}
{Bertin}, E., \& {Arnouts}, S. 1996, \aaps, 117, 393

\bibitem[{{Bertin} {et~al.}(2002){Bertin}, {Mellier}, {Radovich}, {Missonnier},
  {Didelon}, \& {Morin}}]{bertin02}
{Bertin}, E., {Mellier}, Y., {Radovich}, M., {Missonnier}, G., {Didelon}, P.,
  \& {Morin}, B. 2002, in Astronomical Society of the Pacific Conference
  Series, Vol. 281, Astronomical Data Analysis Software and Systems XI, ed.
  {D.~A.~Bohlender, D.~Durand, \& T.~H.~Handley}, 228--+

\bibitem[{{Carlstrom} {et~al.}(2011){Carlstrom}, {Ade}, {Aird}, {Benson},
  {Bleem}, {Busetti}, {Chang}, {Chauvin}, {Cho}, {Crawford}, {Crites}, {Dobbs},
  {Halverson}, {Heimsath}, {Holzapfel}, {Hrubes}, {Joy}, {Keisler}, {Lanting},
  {Lee}, {Leitch}, {Leong}, {Lu}, {Lueker}, {Luong-van}, {McMahon}, {Mehl},
  {Meyer}, {Mohr}, {Montroy}, {Padin}, {Plagge}, {Pryke}, {Ruhl}, {Schaffer},
  {Schwan}, {Shirokoff}, {Spieler}, {Staniszewski}, {Stark}, {Tucker},
  {Vanderlinde}, {Vieira}, \& {Williamson}}]{carlstrom11}
{Carlstrom}, J.~E., {et~al.} 2011, \pasp, 123, 568

\bibitem[{{Carlstrom} {et~al.}(2002){Carlstrom}, {Holder}, \&
  {Reese}}]{carlstrom02}
{Carlstrom}, J.~E., {Holder}, G.~P., \& {Reese}, E.~D. 2002, \araa, 40, 643

\bibitem[{{Coupon} {et~al.}(2009){Coupon}, {Ilbert}, {Kilbinger}, {McCracken},
  {Mellier}, {Arnouts}, {Bertin}, {Hudelot}, {Schultheis}, {Le F{\`e}vre}, {Le
  Brun}, {Guzzo}, {Bardelli}, {Zucca}, {Bolzonella}, {Garilli}, {Zamorani},
  {Zanichelli}, {Tresse}, \& {Aussel}}]{coupon09}
{Coupon}, J., {et~al.} 2009, \aap, 500, 981

\bibitem[{{Dressler} {et~al.}(2003){Dressler}, {Sutin}, \&
  {Bigelow}}]{dressler03}
{Dressler}, A.~M., {Sutin}, B.~M., \& {Bigelow}, B.~C. 2003, in Society of
  Photo-Optical Instrumentation Engineers (SPIE) Conference Series, Vol. 4834,
  Society of Photo-Optical Instrumentation Engineers (SPIE) Conference Series,
  ed. P.~{Guhathakurta}, 255--263

\bibitem[{{Duffy} {et~al.}(2008){Duffy}, {Schaye}, {Kay}, \& {Dalla
  Vecchia}}]{duffy08}
{Duffy}, A.~R., {Schaye}, J., {Kay}, S.~T., \& {Dalla Vecchia}, C. 2008,
  \mnras, 390, L64

\bibitem[{{Evrard} {et~al.}(2008){Evrard}, {Bialek}, {Busha}, {White}, {Habib},
  {Heitmann}, {Warren}, {Rasia}, {Tormen}, {Moscardini}, {Power}, {Jenkins},
  {Gao}, {Frenk}, {Springel}, {White}, \& {Diemand}}]{evrard08}
{Evrard}, A.~E., {et~al.} 2008, \apj, 672, 122

\bibitem[{{Fahlman} {et~al.}(1994){Fahlman}, {Kaiser}, {Squires}, \&
  {Woods}}]{fahlman94}
{Fahlman}, G., {Kaiser}, N., {Squires}, G., \& {Woods}, D. 1994, \apj, 437, 56

\bibitem[{{Fetisova}(1981)}]{fetisova81}
{Fetisova}, T.~S. 1981, \sovast, 25, 647

\bibitem[{{George} {et~al.}(2012){George}, {Leauthaud}, {Bundy}, {Finoguenov},
  {Ma}, {Rykoff}, {Tinker}, {Wechsler}, {Massey}, \& {Mei}}]{george12}
{George}, M.~R., {et~al.} 2012, ArXiv e-prints, 1205.4262

\bibitem[{{Guzzo} {et~al.}(2009){Guzzo}, {Schuecker}, {B{\"o}hringer},
  {Collins}, {Ortiz-Gil}, {de Grandi}, {Edge}, {Neumann}, {Schindler},
  {Altucci}, \& {Shaver}}]{guzzo09}
{Guzzo}, L., {et~al.} 2009, \aap, 499, 357

\bibitem[{{Haiman} {et~al.}(2001){Haiman}, {Mohr}, \& {Holder}}]{haiman01}
{Haiman}, Z., {Mohr}, J.~J., \& {Holder}, G.~P. 2001, \apj, 553, 545

\bibitem[{{Heymans} {et~al.}(2006){Heymans}, {Van Waerbeke}, {Bacon}, {Berge},
  {Bernstein}, {Bertin}, {Bridle}, {Brown}, {Clowe}, {Dahle}, {Erben}, {Gray},
  {Hetterscheidt}, {Hoekstra}, {Hudelot}, {Jarvis}, {Kuijken}, {Margoniner},
  {Massey}, {Mellier}, {Nakajima}, {Refregier}, {Rhodes}, {Schrabback}, \&
  {Wittman}}]{heymans06}
{Heymans}, C., {et~al.} 2006, \mnras, 368, 1323

\bibitem[{{High} {et~al.}(2010){High}, {Stalder}, {Song}, {Ade}, {Aird},
  {Allam}, {Armstrong}, {Barkhouse}, {Benson}, {Bertin}, {Bhattacharya},
  {Bleem}, {Brodwin}, {Buckley-Geer}, {Carlstrom}, {Challis}, {Chang},
  {Crawford}, {Crites}, {de Haan}, {Desai}, {Dobbs}, {Dudley}, {Foley},
  {George}, {Gladders}, {Halverson}, {Hamuy}, {Hansen}, {Holder}, {Holzapfel},
  {Hrubes}, {Joy}, {Keisler}, {Lee}, {Leitch}, {Lin}, {Lin}, {Loehr}, {Lueker},
  {Marrone}, {McMahon}, {Mehl}, {Meyer}, {Mohr}, {Montroy}, {Morell}, {Ngeow},
  {Padin}, {Plagge}, {Pryke}, {Reichardt}, {Rest}, {Ruel}, {Ruhl}, {Schaffer},
  {Shaw}, {Shirokoff}, {Smith}, {Spieler}, {Staniszewski}, {Stark}, {Stubbs},
  {Tucker}, {Vanderlinde}, {Vieira}, {Williamson}, {Wood-Vasey}, {Yang},
  {Zahn}, \& {Zenteno}}]{high10}
{High}, F.~W., {et~al.} 2010, \apj, 723, 1736

\bibitem[{{High} {et~al.}(2009){High}, {Stubbs}, {Rest}, {Stalder}, \&
  {Challis}}]{high09}
{High}, F.~W., {Stubbs}, C.~W., {Rest}, A., {Stalder}, B., \& {Challis}, P.
  2009, \aj, 138, 110

\bibitem[{{Hoekstra}(2007)}]{hoekstra07}
{Hoekstra}, H. 2007, \mnras, 379, 317

\bibitem[{{Hoekstra} {et~al.}(2000){Hoekstra}, {Franx}, \&
  {Kuijken}}]{hoekstra00}
{Hoekstra}, H., {Franx}, M., \& {Kuijken}, K. 2000, \apj, 532, 88

\bibitem[{{Hoekstra} {et~al.}(1998){Hoekstra}, {Franx}, {Kuijken}, \&
  {Squires}}]{hoekstra98}
{Hoekstra}, H., {Franx}, M., {Kuijken}, K., \& {Squires}, G. 1998, \apj, 504,
  636

\bibitem[{{Hoekstra} {et~al.}(2011){Hoekstra}, {Hartlap}, {Hilbert}, \& {van
  Uitert}}]{hoekstra11}
{Hoekstra}, H., {Hartlap}, J., {Hilbert}, S., \& {van Uitert}, E. 2011, \mnras,
  412, 2095

\bibitem[{{Holder} {et~al.}(2001){Holder}, {Haiman}, \& {Mohr}}]{holder01b}
{Holder}, G., {Haiman}, Z., \& {Mohr}, J.~J. 2001, \apjl, 560, L111

\bibitem[{{Israel} {et~al.}(2010){Israel}, {Erben}, {Reiprich}, {Vikhlinin},
  {Hildebrandt}, {Hudson}, {McLeod}, {Sarazin}, {Schneider}, \&
  {Zhang}}]{israel10}
{Israel}, H., {et~al.} 2010, \aap, 520, A58

\bibitem[{{Israel} {et~al.}(2011){Israel}, {Erben}, {Reiprich}, {Vikhlinin},
  {Sarazin}, \& {Schneider}}]{israel11}
{Israel}, H., {Erben}, T., {Reiprich}, T.~H., {Vikhlinin}, A., {Sarazin},
  C.~L., \& {Schneider}, P. 2011, ArXiv e-prints, 1112.4444

\bibitem[{{Ivezi{\'c}} {et~al.}(2007){Ivezi{\'c}}, {Smith}, {Miknaitis}, {Lin},
  {Tucker}, {Lupton}, {Gunn}, {Knapp}, {Strauss}, {Sesar}, {Doi}, {Tanaka},
  {Fukugita}, {Holtzman}, {Kent}, {Yanny}, {Schlegel}, {Finkbeiner},
  {Padmanabhan}, {Rockosi}, {Juri{\'c}}, {Bond}, {Lee}, {Stoughton}, {Jester},
  {Harris}, {Harding}, {Morrison}, {Brinkmann}, {Schneider}, \&
  {York}}]{ivezic07}
{Ivezi{\'c}}, {\v Z}., {et~al.} 2007, \aj, 134, 973

\bibitem[{{Ivison} {et~al.}(2007){Ivison}, {Greve}, {Dunlop}, {Peacock},
  {Egami}, {Smail}, {Ibar}, {van Kampen}, {Aretxaga}, {Babbedge}, {Biggs},
  {Blain}, {Chapman}, {Clements}, {Coppin}, {Farrah}, {Halpern}, {Hughes},
  {Jarvis}, {Jenness}, {Jones}, {Mortier}, {Oliver}, {Papovich},
  {P{\'e}rez-Gonz{\'a}lez}, {Pope}, {Rawlings}, {Rieke}, {Rowan-Robinson},
  {Savage}, {Scott}, {Seigar}, {Serjeant}, {Simpson}, {Stevens}, {Vaccari},
  {Wagg}, \& {Willott}}]{ivison07}
{Ivison}, R.~J., {et~al.} 2007, \mnras, 380, 199

\bibitem[{{Kaiser}(1995)}]{kaiser95a}
{Kaiser}, N. 1995, \apjl, 439, L1

\bibitem[{{Kaiser} \& {Squires}(1993)}]{kaiser93}
{Kaiser}, N., \& {Squires}, G. 1993, \apj, 404, 441

\bibitem[{{Kaiser} {et~al.}(1995){Kaiser}, {Squires}, \&
  {Broadhurst}}]{kaiser95b}
{Kaiser}, N., {Squires}, G., \& {Broadhurst}, T. 1995, \apj, 449, 460

\bibitem[{{Komatsu} {et~al.}(2011){Komatsu}, {Smith}, {Dunkley}, {Bennett},
  {Gold}, {Hinshaw}, {Jarosik}, {Larson}, {Nolta}, {Page}, {Spergel},
  {Halpern}, {Hill}, {Kogut}, {Limon}, {Meyer}, {Odegard}, {Tucker}, {Weiland},
  {Wollack}, \& {Wright}}]{komatsu11}
{Komatsu}, E., {et~al.} 2011, \apjs, 192, 18

\bibitem[{{Kravtsov} {et~al.}(2006){Kravtsov}, {Vikhlinin}, \&
  {Nagai}}]{kravtsov06a}
{Kravtsov}, A.~V., {Vikhlinin}, A., \& {Nagai}, D. 2006, \apj, 650, 128

\bibitem[{{Kurtz} \& {Mink}(1998)}]{kurtz98}
{Kurtz}, M.~J., \& {Mink}, D.~J. 1998, \pasp, 110, 934

\bibitem[{{Leauthaud} {et~al.}(2007){Leauthaud}, {Massey}, {Kneib}, {Rhodes},
  {Johnston}, {Capak}, {Heymans}, {Ellis}, {Koekemoer}, {Le F{\`e}vre},
  {Mellier}, {R{\'e}fr{\'e}gier}, {Robin}, {Scoville}, {Tasca}, {Taylor}, \&
  {Van Waerbeke}}]{leauthaud07}
{Leauthaud}, A., {et~al.} 2007, \apjs, 172, 219

\bibitem[{{Leccardi} \& {Molendi}(2008)}]{leccardi08}
{Leccardi}, A., \& {Molendi}, S. 2008, \aap, 486, 359

\bibitem[{{Lopes}(2007)}]{lopes07}
{Lopes}, P.~A.~A. 2007, \mnras, 380, 1608

\bibitem[{{Luppino} \& {Kaiser}(1997)}]{luppino97}
{Luppino}, G.~A., \& {Kaiser}, N. 1997, \apj, 475, 20

\bibitem[{{Macci{\`o}} {et~al.}(2008){Macci{\`o}}, {Dutton}, \& {van den
  Bosch}}]{maccio08}
{Macci{\`o}}, A.~V., {Dutton}, A.~A., \& {van den Bosch}, F.~C. 2008, \mnras,
  391, 1940

\bibitem[{{Mahdavi} {et~al.}(2008){Mahdavi}, {Hoekstra}, {Babul}, \&
  {Henry}}]{mahdavi08}
{Mahdavi}, A., {Hoekstra}, H., {Babul}, A., \& {Henry}, J.~P. 2008, \mnras,
  384, 1567

\bibitem[{{Mantz} {et~al.}(2010{\natexlab{a}}){Mantz}, {Allen}, {Ebeling},
  {Rapetti}, \& {Drlica-Wagner}}]{mantz10}
{Mantz}, A., {Allen}, S.~W., {Ebeling}, H., {Rapetti}, D., \& {Drlica-Wagner},
  A. 2010{\natexlab{a}}, \mnras, 406, 1773

\bibitem[{{Mantz} {et~al.}(2010{\natexlab{b}}){Mantz}, {Allen}, {Rapetti}, \&
  {Ebeling}}]{mantz10b}
{Mantz}, A., {Allen}, S.~W., {Rapetti}, D., \& {Ebeling}, H.
  2010{\natexlab{b}}, \mnras, 406, 1759

\bibitem[{{Marrone} {et~al.}(2012){Marrone}, {Smith}, {Okabe}, {Bonamente},
  {Carlstrom}, {Culverhouse}, {Gralla}, {Greer}, {Hasler}, {Hawkins},
  {Hennessy}, {Joy}, {Lamb}, {Leitch}, {Martino}, {Mazzotta}, {Miller},
  {Mroczkowski}, {Muchovej}, {Plagge}, {Pryke}, {Sanderson}, {Takada}, {Woody},
  \& {Zhang}}]{marrone11}
{Marrone}, D.~P., {et~al.} 2012, \apj, 754, 119

\bibitem[{{Massey} {et~al.}(2007){Massey}, {Heymans}, {Berg{\'e}}, {Bernstein},
  {Bridle}, {Clowe}, {Dahle}, {Ellis}, {Erben}, {Hetterscheidt}, {High},
  {Hirata}, {Hoekstra}, {Hudelot}, {Jarvis}, {Johnston}, {Kuijken},
  {Margoniner}, {Mandelbaum}, {Mellier}, {Nakajima}, {Paulin-Henriksson},
  {Peeples}, {Roat}, {Refregier}, {Rhodes}, {Schrabback}, {Schirmer}, {Seljak},
  {Semboloni}, \& {van Waerbeke}}]{massey07}
{Massey}, R., {et~al.} 2007, \mnras, 376, 13

\bibitem[{{McInnes} {et~al.}(2009){McInnes}, {Menanteau}, {Heavens}, {Hughes},
  {Jimenez}, {Massey}, {Simon}, \& {Taylor}}]{mcinnes09}
{McInnes}, R.~N., {Menanteau}, F., {Heavens}, A.~F., {Hughes}, J.~P.,
  {Jimenez}, R., {Massey}, R., {Simon}, P., \& {Taylor}, A. 2009, \mnras, 399,
  L84

\bibitem[{{McLeod} {et~al.}(1998){McLeod}, {Gauron}, {Geary}, {Ordway}, \&
  {Roll}}]{mcleod98}
{McLeod}, B.~A., {Gauron}, T.~M., {Geary}, J.~C., {Ordway}, M.~P., \& {Roll},
  J.~B. 1998, in Society of Photo-Optical Instrumentation Engineers (SPIE)
  Conference Series, Vol. 3355, Society of Photo-Optical Instrumentation
  Engineers (SPIE) Conference Series, ed. {S.~D'Odorico}, 477--486

\bibitem[{{Melin} {et~al.}(2006){Melin}, {Bartlett}, \&
  {Delabrouille}}]{melin06}
{Melin}, J.-B., {Bartlett}, J.~G., \& {Delabrouille}, J. 2006, \aap, 459, 341

\bibitem[{{Mellier}(1999)}]{mellier99}
{Mellier}, Y. 1999, \araa, 37, 127

\bibitem[{{Miralda-Escude}(1995)}]{miralda-escude95a}
{Miralda-Escude}, J. 1995, \apj, 438, 514

\bibitem[{{Nagai} {et~al.}(2007){Nagai}, {Kravtsov}, \& {Vikhlinin}}]{nagai07}
{Nagai}, D., {Kravtsov}, A.~V., \& {Vikhlinin}, A. 2007, \apj, 668, 1

\bibitem[{{Navarro} {et~al.}(1997){Navarro}, {Frenk}, \& {White}}]{navarro97}
{Navarro}, J.~F., {Frenk}, C.~S., \& {White}, S.~D.~M. 1997, \apj, 490, 493

\bibitem[{{Neto} {et~al.}(2007){Neto}, {Gao}, {Bett}, {Cole}, {Navarro},
  {Frenk}, {White}, {Springel}, \& {Jenkins}}]{neto07}
{Neto}, A.~F., {et~al.} 2007, \mnras, 381, 1450

\bibitem[{{Osip} {et~al.}(2008){Osip}, {Floyd}, \& {Covarrubias}}]{osip08}
{Osip}, D.~J., {Floyd}, D., \& {Covarrubias}, R. 2008, in Society of
  Photo-Optical Instrumentation Engineers (SPIE) Conference Series, Vol. 7014,
  Society of Photo-Optical Instrumentation Engineers (SPIE) Conference Series

\bibitem[{{Planck Collaboration: Aghanim} {et~al.}(2012){Planck Collaboration:
  Aghanim}, {Arnaud}, {Ashdown}, {Atrio-Barandela}, {Aumont}, {Balbi},
  {Banday}, {Barreiro}, {Bartlett}, {Battaner}, {Battye}, {Bernard},
  {Bersanelli}, {Bhatia}, {Bikmaev}, {B{\"o}hringer}, {Bonaldi}, {Bond},
  {Borgani}, {Borrill}, {Bourdin}, {Brown}, {Bucher}, {Burenin}, {Burigana},
  {Butler}, {Cabella}, {Cardoso}, {Carvalho}, {Chamballu}, {Chiang}, {Chon},
  {Clements}, {Colafrancesco}, {Cuttaia}, {Da Silva}, {Dahle}, {Davis}, {de
  Bernardis}, {de Gasperis}, {Delabrouille}, {D{\'e}mocl{\`e}s}, {D{\'e}sert},
  {Diego}, {Dolag}, {Dole}, {Donzelli}, {Douspis}, {Dupac}, {Efstathiou},
  {En{\ss}lin}, {Eriksen}, {Finelli}, {Flores-Cacho}, {Forni}, {Frailis},
  {Franceschi}, {Frommert}, {Ganga}, {G{\'e}nova-Santos}, {Giard},
  {Giraud-H{\'e}raud}, {Gonz{\'a}lez-Nuevo}, {G{\'o}rski}, {Gruppuso},
  {Hansen}, {Harrison}, {Hern{\'a}ndez-Monteagudo}, {Herranz}, {Hildebrandt},
  {Hivon}, {Hobson}, {Huffenberger}, {Hurier}, {Jagemann}, {Juvela},
  {Keih{\"a}nen}, {Khamitov}, {Kneissl}, {Knoche}, {Kunz}, {Kurki-Suonio},
  {Lagache}, {Lamarre}, {Lasenby}, {Lawrence}, {Le Jeune}, {Leach}, {Leonardi},
  {Liddle}, {Lilje}, {Linden-V{\o}rnle}, {L{\'o}pez-Caniego}, {Luzzi},
  {Mac{\'{\i}}as-P{\'e}rez}, {Maino}, {Mandolesi}, {Marleau}, {Marshall},
  {Mart{\'{\i}}nez-Gonz{\'a}lez}, {Masi}, {Matarrese}, {Matthai}, {Mazzotta},
  {Melchiorri}, {Melin}, {Mendes}, {Miville-Desch{\^e}nes}, {Montier},
  {Morgante}, {Munshi}, {Natoli}, {Noviello}, {Osborne}, {Pajot}, {Paoletti},
  {Pearson}, {Perdereau}, {Perrotta}, {Piacentini}, {Piat}, {Pierpaoli},
  {Piffaretti}, {Platania}, {Pointecouteau}, {Polenta}, {Ponthieu}, {Popa},
  {Poutanen}, {Pratt}, {Puget}, {Rachen}, {Rebolo}, {Reinecke}, {Remazeilles},
  {Renault}, {Ricciardi}, {Ristorcelli}, {Rocha}, {Rosset}, {Rossetti},
  {Rubi{\~n}o-Mart{\'{\i}}n}, {Rusholme}, {Savini}, {Starck}, {Stivoli},
  {Stolyarov}, {Sunyaev}, {Sutton}, {Suur-Uski}, {Tauber}, {Terenzi},
  {Toffolatti}, {Tomasi}, {Tristram}, {Valenziano}, {Van Tent}, {Vielva},
  {Villa}, {Vittorio}, {Weller}, {White}, {Zacchei}, \& {Zonca}}]{planck12}
{Planck Collaboration: Aghanim}, N., {et~al.} 2012, ArXiv e-prints, 1204.2743

\bibitem[{{Prada} {et~al.}(2011){Prada}, {Klypin}, {Cuesta}, {Betancort-Rijo},
  \& {Primack}}]{prada11}
{Prada}, F., {Klypin}, A.~A., {Cuesta}, A.~J., {Betancort-Rijo}, J.~E., \&
  {Primack}, J. 2011, ArXiv e-prints, 1104.5130

\bibitem[{{Rasia} {et~al.}(2012){Rasia}, {Meneghetti}, {Martino}, {Borgani},
  {Bonafede}, {Dolag}, {Ettori}, {Fabjan}, {Giocoli}, {Mazzotta}, {Merten},
  {Radovich}, \& {Tornatore}}]{rasia12}
{Rasia}, E., {et~al.} 2012, New Journal of Physics, 14, 055018

\bibitem[{{Regnault} {et~al.}(2009){Regnault}, {Conley}, {Guy}, {Sullivan},
  {Cuillandre}, {Astier}, {Balland}, {Basa}, {Carlberg}, {Fouchez}, {Hardin},
  {Hook}, {Howell}, {Pain}, {Perrett}, \& {Pritchet}}]{regnault09}
{Regnault}, N., {et~al.} 2009, \aap, 506, 999

\bibitem[{{Reichardt} {et~al.}(2012){Reichardt}, {Stalder}, {Bleem}, {Montroy},
  {Aird}, {Andersson}, {Armstrong}, {Ashby}, {Bautz}, {Bayliss}, {Bazin},
  {Benson}, {Brodwin}, {Carlstrom}, {Chang}, {Cho}, {Clocchiatti}, {Crawford},
  {Crites}, {de Haan}, {Desai}, {Dobbs}, {Dudley}, {Foley}, {Forman}, {George},
  {Gladders}, {Gonzalez}, {Halverson}, {Harrington}, {High}, {Holder},
  {Holzapfel}, {Hoover}, {Hrubes}, {Jones}, {Joy}, {Keisler}, {Knox}, {Lee},
  {Leitch}, {Liu}, {Lueker}, {Luong-Van}, {Mantz}, {Marrone}, {McDonald},
  {McMahon}, {Mehl}, {Meyer}, {Mocanu}, {Mohr}, {Murray}, {Natoli}, {Padin},
  {Plagge}, {Pryke}, {Rest}, {Ruel}, {Ruhl}, {Saliwanchik}, {Saro}, {Sayre},
  {Schaffer}, {Shaw}, {Shirokoff}, {Song}, {Spieler}, {Staniszewski}, {Stark},
  {Story}, {Stubbs}, {Suhada}, {van Engelen}, {Vanderlinde}, {Vieira},
  {Vikhlinin}, {Williamson}, {Zahn}, \& {Zenteno}}]{reichardt12}
{Reichardt}, C.~L., {et~al.} 2012, ArXiv e-prints, 1203.5775

\bibitem[{{Rozo} {et~al.}(2010){Rozo}, {Wechsler}, {Rykoff}, {Annis}, {Becker},
  {Evrard}, {Frieman}, {Hansen}, {Hao}, {Johnston}, {Koester}, {McKay},
  {Sheldon}, \& {Weinberg}}]{rozo10}
{Rozo}, E., {et~al.} 2010, \apj, 708, 645

\bibitem[{{Schneider}(1996)}]{schneider96}
{Schneider}, P. 1996, \mnras, 283, 837

\bibitem[{{Seitz} \& {Schneider}(1997)}]{seitz97}
{Seitz}, C., \& {Schneider}, P. 1997, \aap, 318, 687

\bibitem[{{Skrutskie} {et~al.}(2006){Skrutskie}, {Cutri}, {Stiening},
  {Weinberg}, {Schneider}, {Carpenter}, {Beichman}, {Capps}, {Chester},
  {Elias}, {Huchra}, {Liebert}, {Lonsdale}, {Monet}, {Price}, {Seitzer},
  {Jarrett}, {Kirkpatrick}, {Gizis}, {Howard}, {Evans}, {Fowler}, {Fullmer},
  {Hurt}, {Light}, {Kopan}, {Marsh}, {McCallon}, {Tam}, {Van Dyk}, \&
  {Wheelock}}]{skrutskie06}
{Skrutskie}, M.~F., {et~al.} 2006, \aj, 131, 1163

\bibitem[{{Song} {et~al.}(2012){Song}, {Zenteno}, {Stalder}, {Desai}, {Bleem},
  {Aird}, {Armstrong}, {Ashby}, {Bayliss}, {Bazin}, {Benson}, {Bertin},
  {Brodwin}, {Carlstrom}, {Chang}, {Cho}, {Clocchiatti}, {Crawford}, {Crites},
  {de Haan}, {Dobbs}, {Dudley}, {Foley}, {George}, {Gettings}, {Gladders},
  {Gonzalez}, {Halverson}, {Harrington}, {High}, {Holder}, {Holzapfel},
  {Hoover}, {Hrubes}, {Joy}, {Keisler}, {Knox}, {Lee}, {Leitch}, {Liu},
  {Lueker}, {Luong-Van}, {Marrone}, {McDonald}, {McMahon}, {Mehl}, {Meyer},
  {Mocanu}, {Mohr}, {Montroy}, {Natoli}, {Nurgaliev}, {Padin}, {Plagge},
  {Pryke}, {Reichardt}, {Rest}, {Ruel}, {Ruhl}, {Saliwanchik}, {Saro}, {Sayre},
  {Schaffer}, {Shaw}, {Shirokoff}, {Suhada}, {Spieler}, {Stanford},
  {Staniszewski}, {Stark}, {Story}, {Stubbs}, {van Engelen}, {Vanderlinde},
  {Vieira}, {Williamson}, \& {Zahn}}]{song12}
{Song}, J., {et~al.} 2012, ArXiv e-prints, 1207.4369

\bibitem[{{Story} {et~al.}(2011){Story}, {Aird}, {Andersson}, {Armstrong},
  {Bazin}, {Benson}, {Bleem}, {Bonamente}, {Brodwin}, {Carlstrom}, {Chang},
  {Clocchiatti}, {Crawford}, {Crites}, {de Haan}, {Desai}, {Dobbs}, {Dudley},
  {Foley}, {George}, {Gladders}, {Gonzalez}, {Halverson}, {High}, {Holder},
  {Holzapfel}, {Hoover}, {Hrubes}, {Joy}, {Keisler}, {Knox}, {Lee}, {Leitch},
  {Lueker}, {Luong-Van}, {Marrone}, {McMahon}, {Mehl}, {Meyer}, {Mohr},
  {Montroy}, {Padin}, {Plagge}, {Pryke}, {Reichardt}, {Rest}, {Ruel}, {Ruhl},
  {Saliwanchik}, {Saro}, {Schaffer}, {Shaw}, {Shirokoff}, {Song}, {Spieler},
  {Stalder}, {Staniszewski}, {Stark}, {Stubbs}, {Vanderlinde}, {Vieira},
  {Williamson}, \& {Zenteno}}]{story11}
{Story}, K., {et~al.} 2011, \apjl, 735, L36

\bibitem[{{Struble} \& {Rood}(1999)}]{struble99}
{Struble}, M.~F., \& {Rood}, H.~J. 1999, \apjs, 125, 35

\bibitem[{{van Dokkum}(2001)}]{vandokkum01}
{van Dokkum}, P.~G. 2001, \pasp, 113, 1420

\bibitem[{{Vikhlinin} {et~al.}(2009{\natexlab{a}}){Vikhlinin}, {Burenin},
  {Ebeling}, {Forman}, {Hornstrup}, {Jones}, {Kravtsov}, {Murray}, {Nagai},
  {Quintana}, \& {Voevodkin}}]{vikhlinin09b}
{Vikhlinin}, A., {et~al.} 2009{\natexlab{a}}, \apj, 692, 1033

\bibitem[{{Vikhlinin} {et~al.}(2009{\natexlab{b}}){Vikhlinin}, {Kravtsov},
  {Burenin}, {Ebeling}, {Forman}, {Hornstrup}, {Jones}, {Murray}, {Nagai},
  {Quintana}, \& {Voevodkin}}]{vikhlinin09}
------. 2009{\natexlab{b}}, \apj, 692, 1060

\bibitem[{{Wang} \& {Steinhardt}(1998)}]{wang98}
{Wang}, L., \& {Steinhardt}, P.~J. 1998, \apj, 508, 483

\bibitem[{Wechsler(2004)}]{wechsler04}
Wechsler, R.~H. 2004, Clusters of Galaxies: Probes of Cosmological Structure
  and Galaxy Evolution

\bibitem[{{Weller} \& {Battye}(2003)}]{weller03}
{Weller}, J., \& {Battye}, R.~A. 2003, New Astronomy Review, 47, 775

\bibitem[{{Wright} \& {Brainerd}(2000)}]{brainerd00}
{Wright}, C.~O., \& {Brainerd}, T.~G. 2000, \apj, 534, 34

\bibitem[{{Zhang} {et~al.}(2008){Zhang}, {Finoguenov}, {Bohringer}, Kneib,
  Smith, Kneissl, Okabe, \& Dahle}]{zhang08}
{Zhang}, Y.-Y., {Finoguenov}, A., {Bohringer}, H., Kneib, J., Smith, G.,
  Kneissl, R., Okabe, N., \& Dahle, H. 2008, Astronomy and Astrophysics, 482,
  451

\end{thebibliography}
%\bibliography{megacam_wl}

\appendix

\section{Additional Figures}
\label{app:figures}

\begin{figure*}
\epsscale{1.2}
\plotone{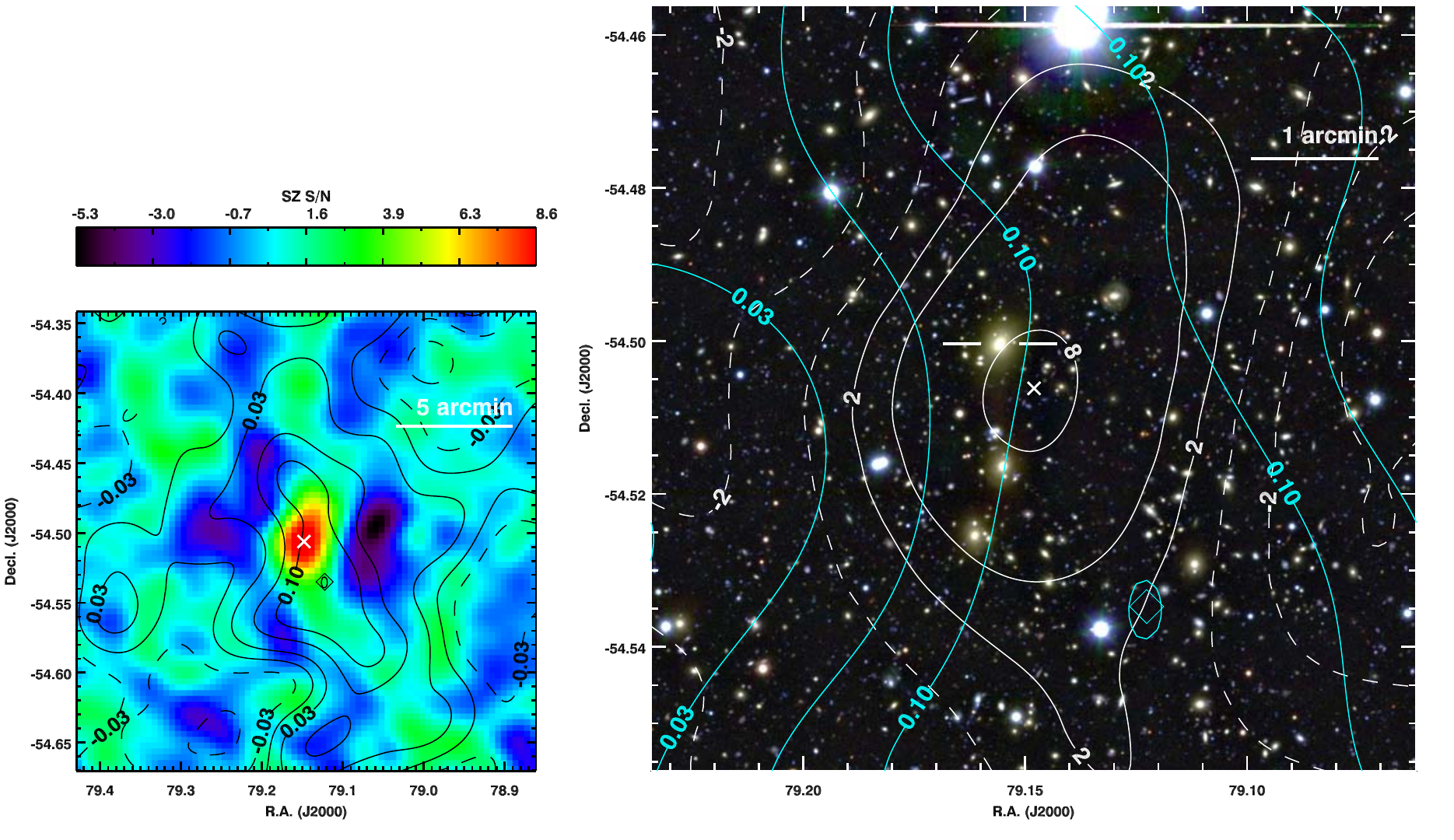}
\caption{SZ, optical, and $\kappa$ data for \clustera. See the Appendix for a description.\label{fig:ppa}}
\end{figure*}

\begin{figure*}
\epsscale{1.2}
\plotone{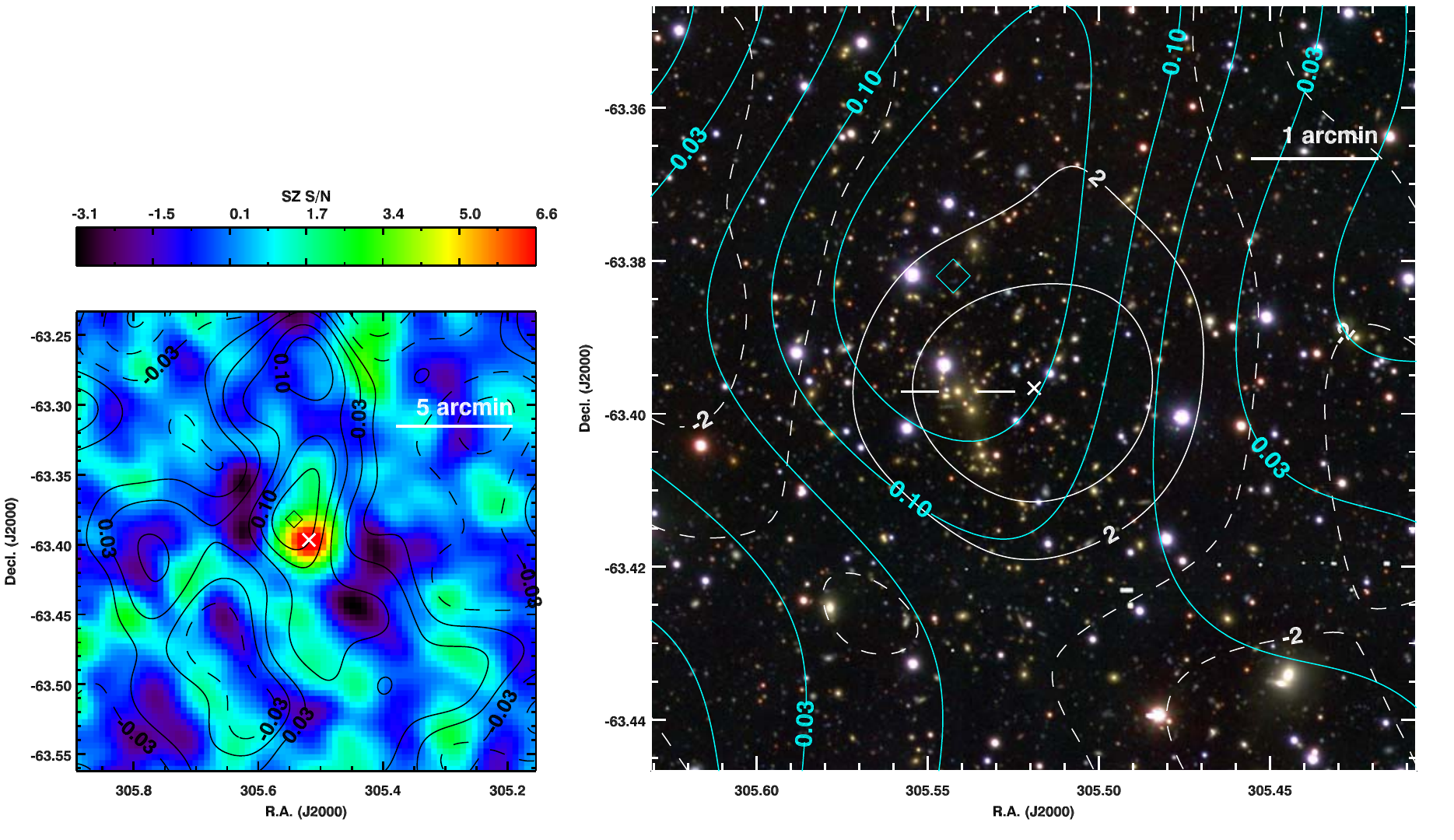}
\caption{SZ, optical, and $\kappa$ data for \clusterb.  See the Appendix for a description.\label{fig:ppb}}
\end{figure*}

\begin{figure*}
\epsscale{1.2}
\plotone{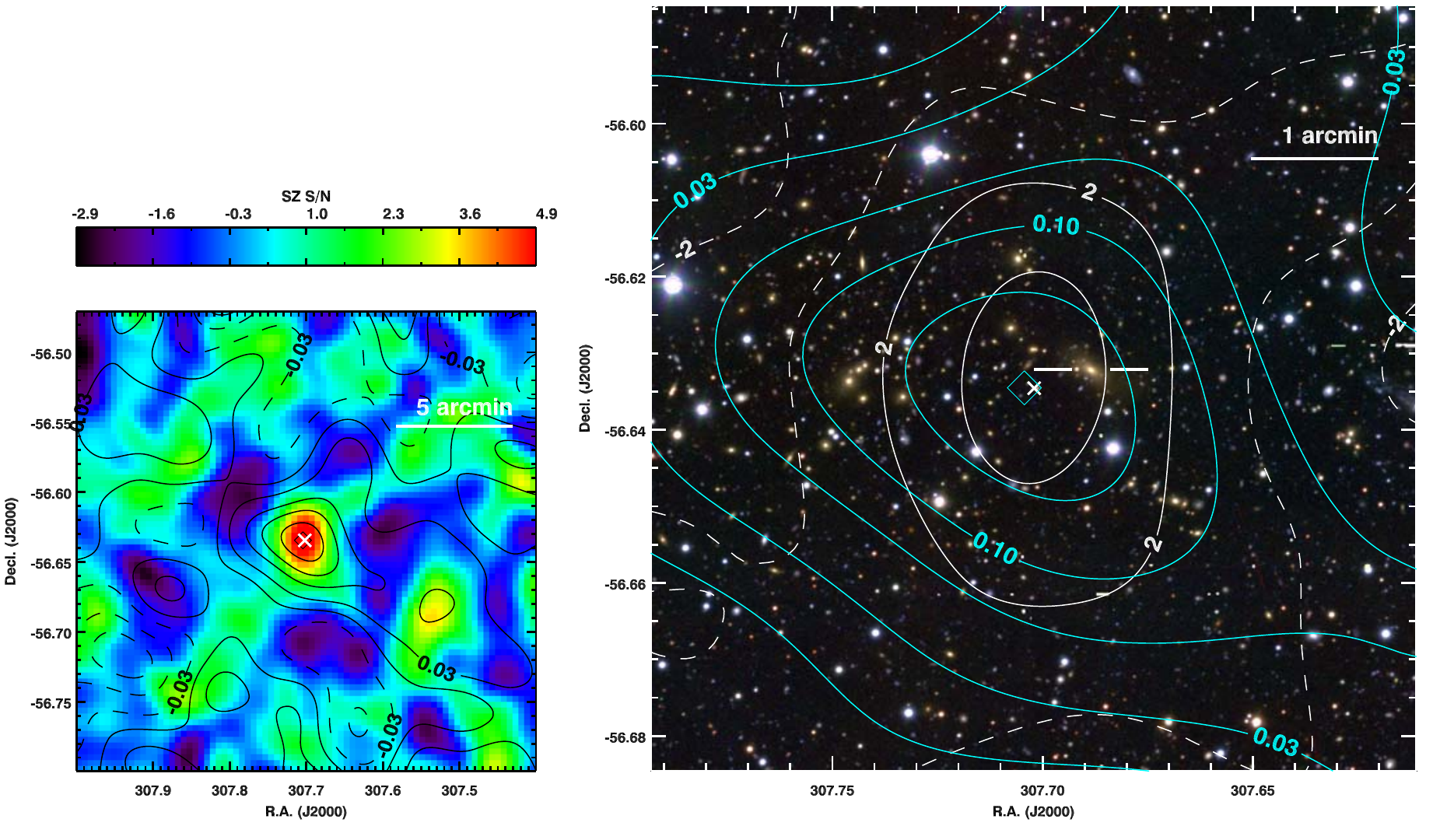}
\caption{SZ, optical, and $\kappa$ data for \clusterc.  See the Appendix for a description.\label{fig:ppc}}
\end{figure*}

\begin{figure*}
\epsscale{1.2}
\plotone{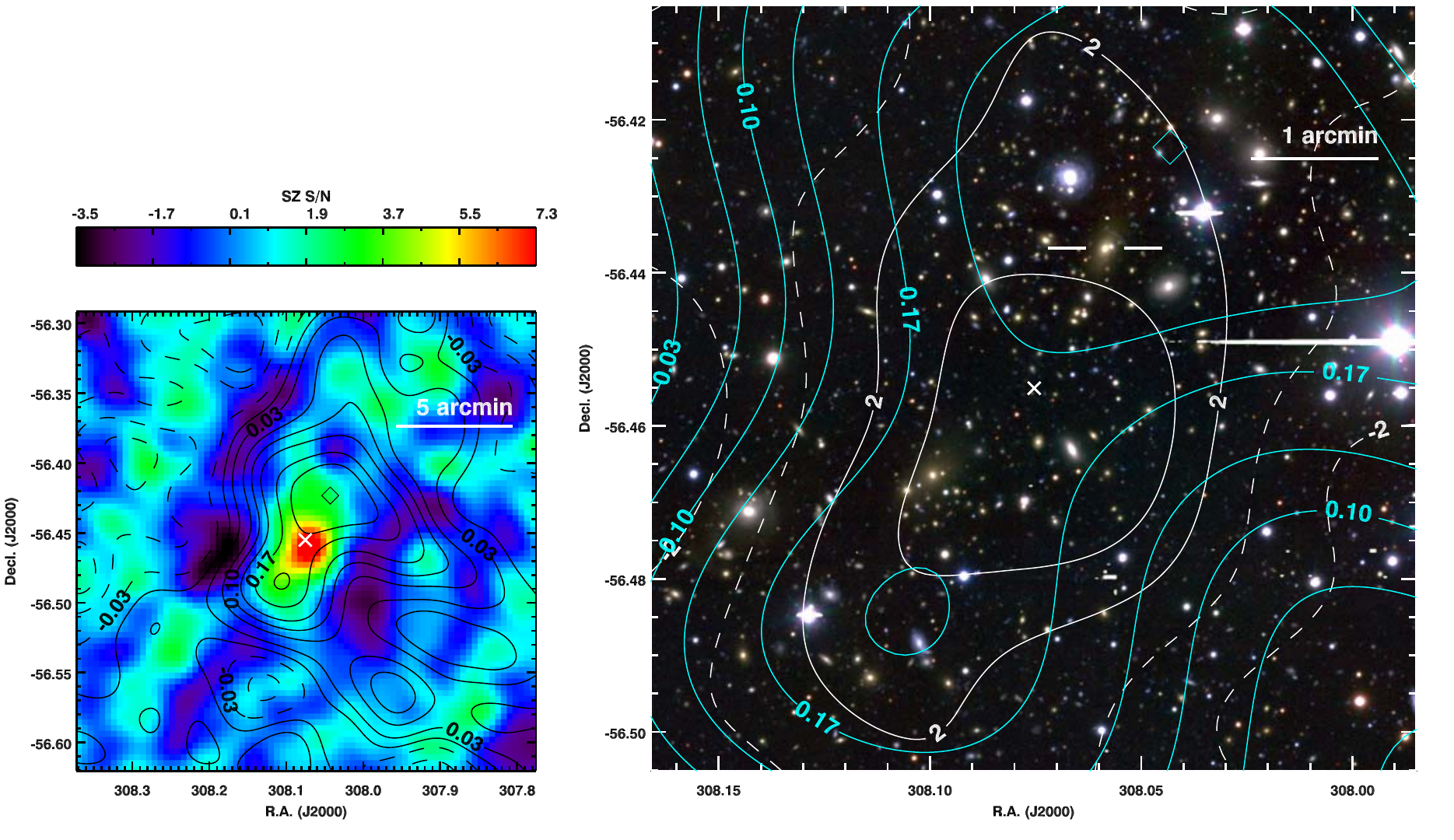}
\caption{SZ, optical, and $\kappa$ data for \clusterd.  See the Appendix for a description.\label{fig:ppd}}
\end{figure*}

\begin{figure*}
\epsscale{1.2}
\plotone{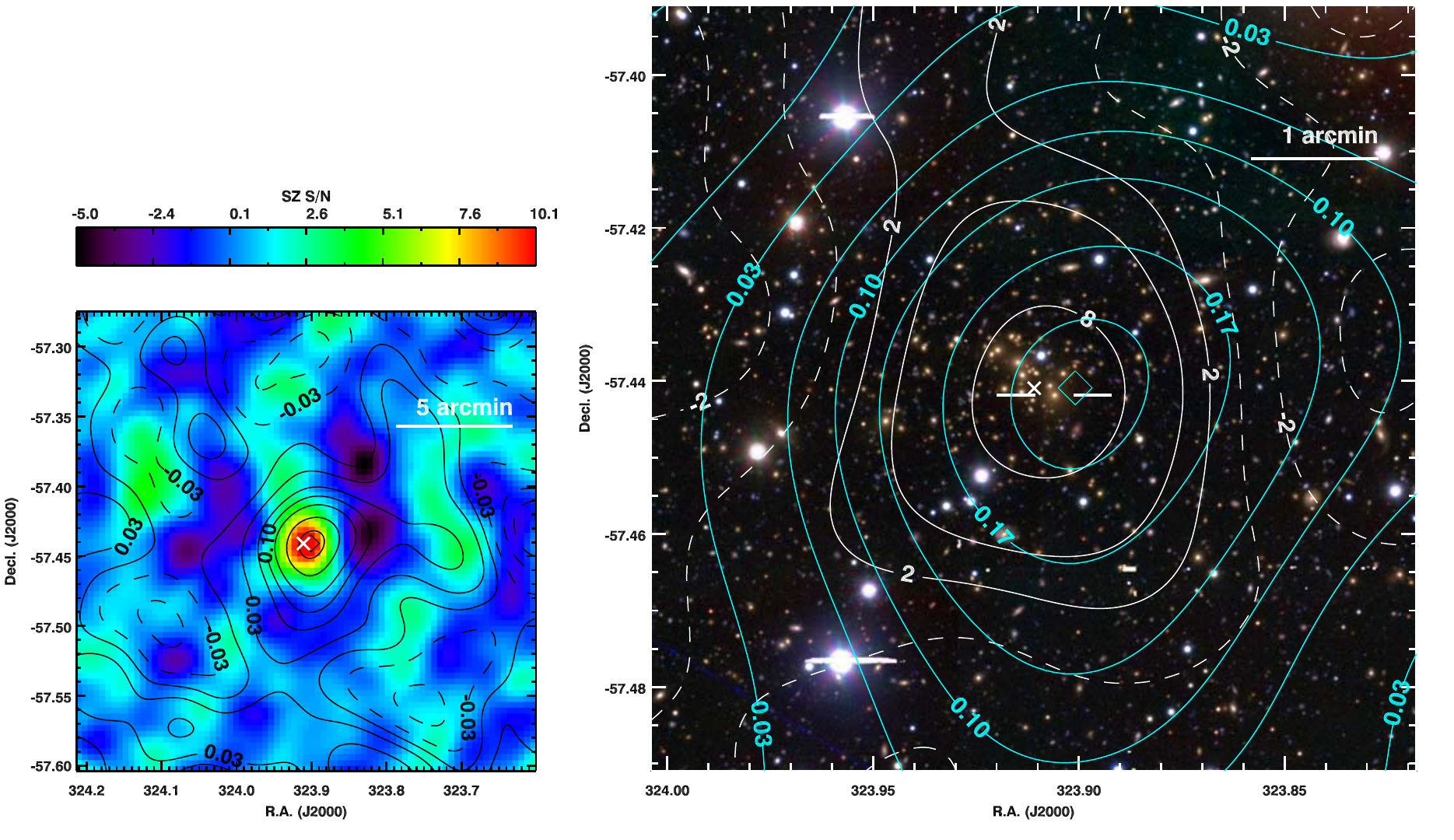}
\caption{SZ, optical, and $\kappa$ data for \clustere.  See the Appendix for a description.\label{fig:ppe}}
\end{figure*}

The left panels of Figures \ref{fig:ppa}, \ref{fig:ppb},
\ref{fig:ppc}, \ref{fig:ppd}, and \ref{fig:ppe} show the SPT SZ
detection significance maps.  The SZ images subtend $20\arcmin$ on a
side. The mapping between color and SZ significance $\xi$ spans the
full range of SZ pixel values in the region of sky shown. The negative
lobes are due to the filtering of the time-ordered data and the maps.
Black contours correspond to the $\kappa$ reconstructed from the WL
shear catalogs and denote values spaced by $\Delta \kappa =
1/30$. The diamond symbols indicate the peak of the reconstructed
$\kappa$ maps over the full field.  Contours are dashed where values
are negative, and solid where values are positive.  The white X
symbols indicate the centroid from the SZ significance map.

The right panels of these figures show the Clay-Megacam images.  A
false-color composite is presented, with $irg$ mapped to the RGB
channels, respectively.  The same $\kappa$ contours from the left
panels are shown in cyan.  White contours correspond to the SZ
significance data from the left panels, spaced at significance values
of $(-8,-4,-2,0,2,4,8)$.  The horizontal white lines enclose the BCG
of each cluster.

\begin{figure*}
\epsscale{1.15}
\plottwo{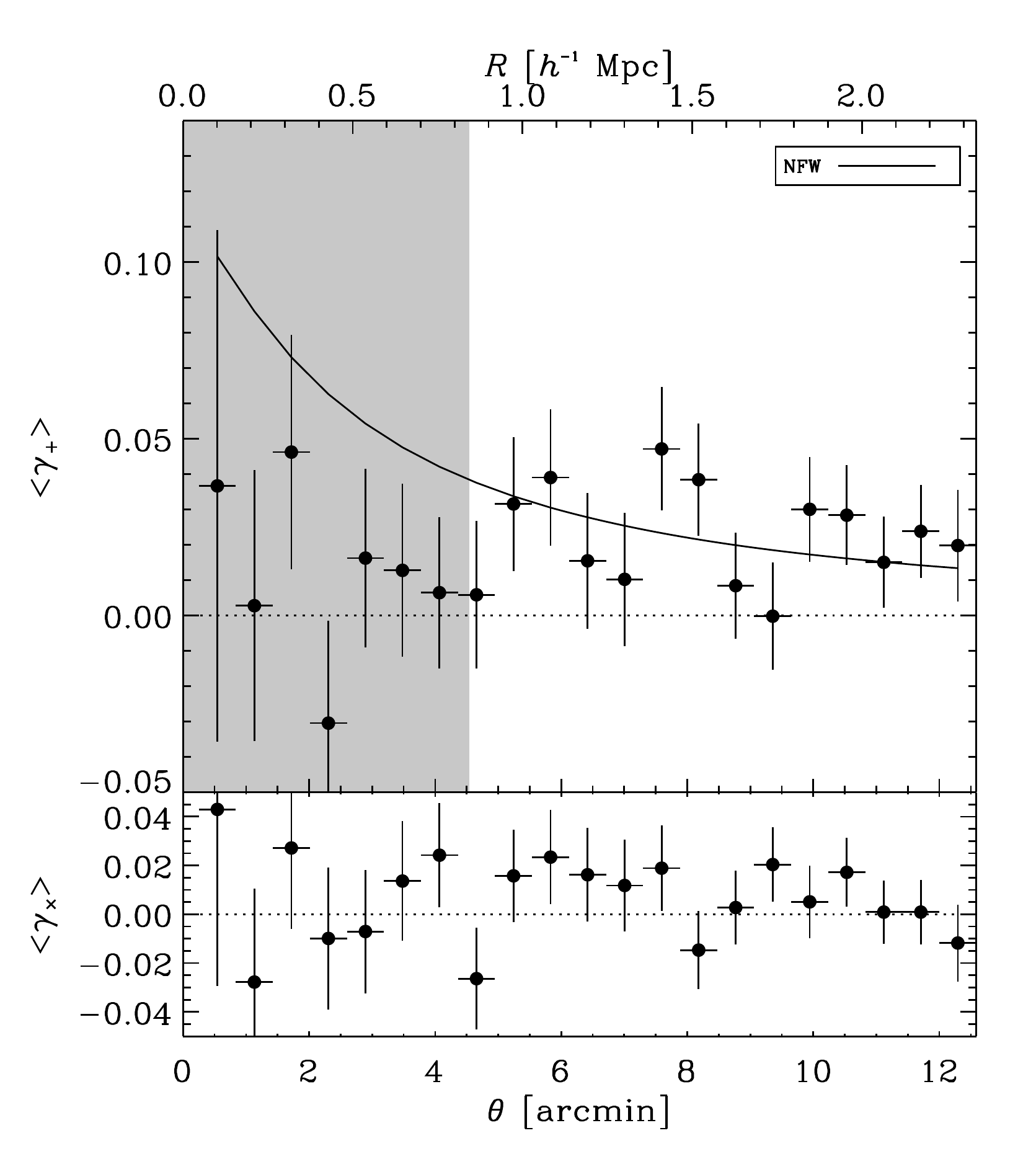}{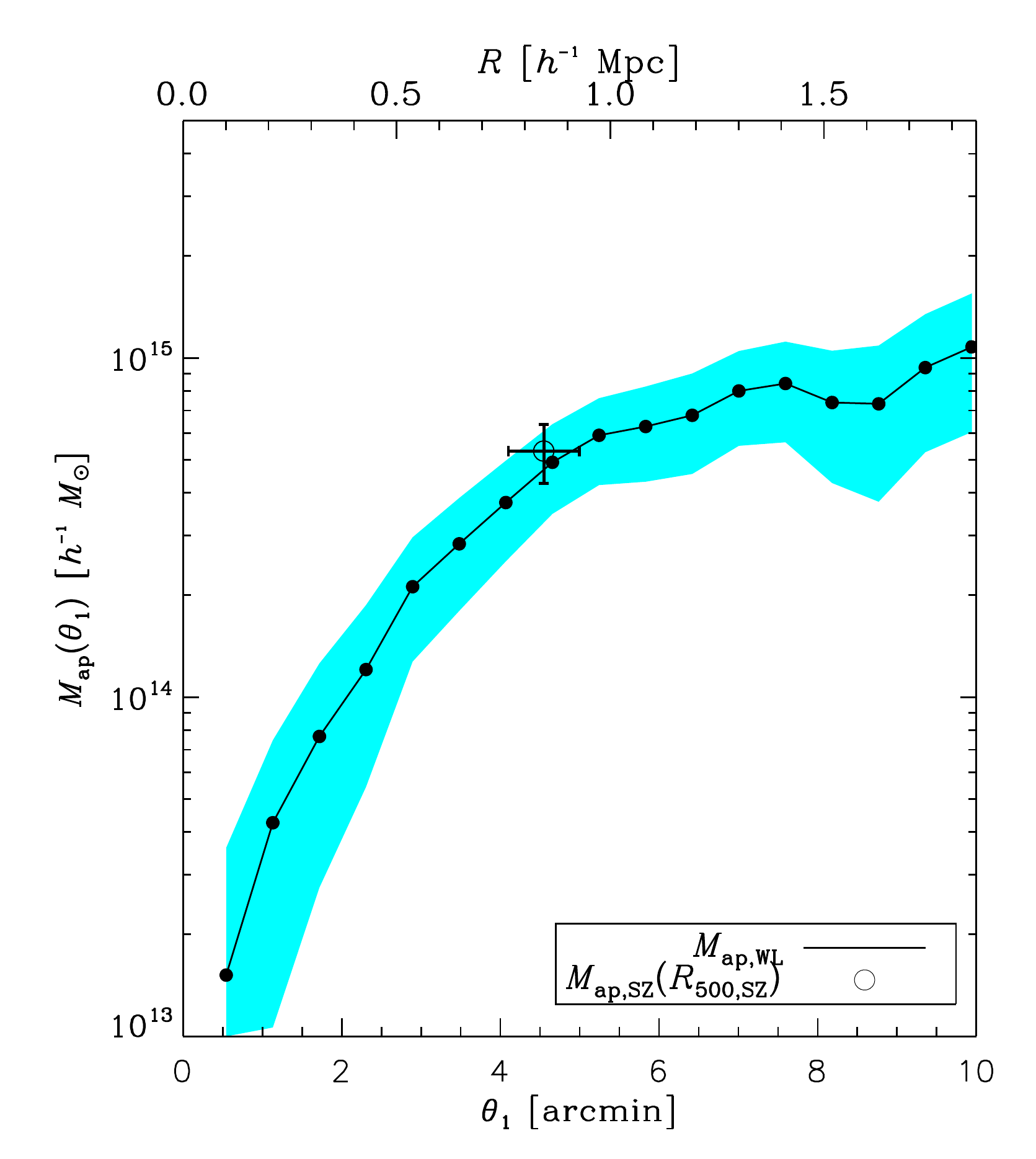}
\caption{Shear and aperture mass profiles of \clustera.  See the Appendix for a description.\label{fig:profilea}}
\end{figure*}

\begin{figure*}
\epsscale{1.15}
\plottwo{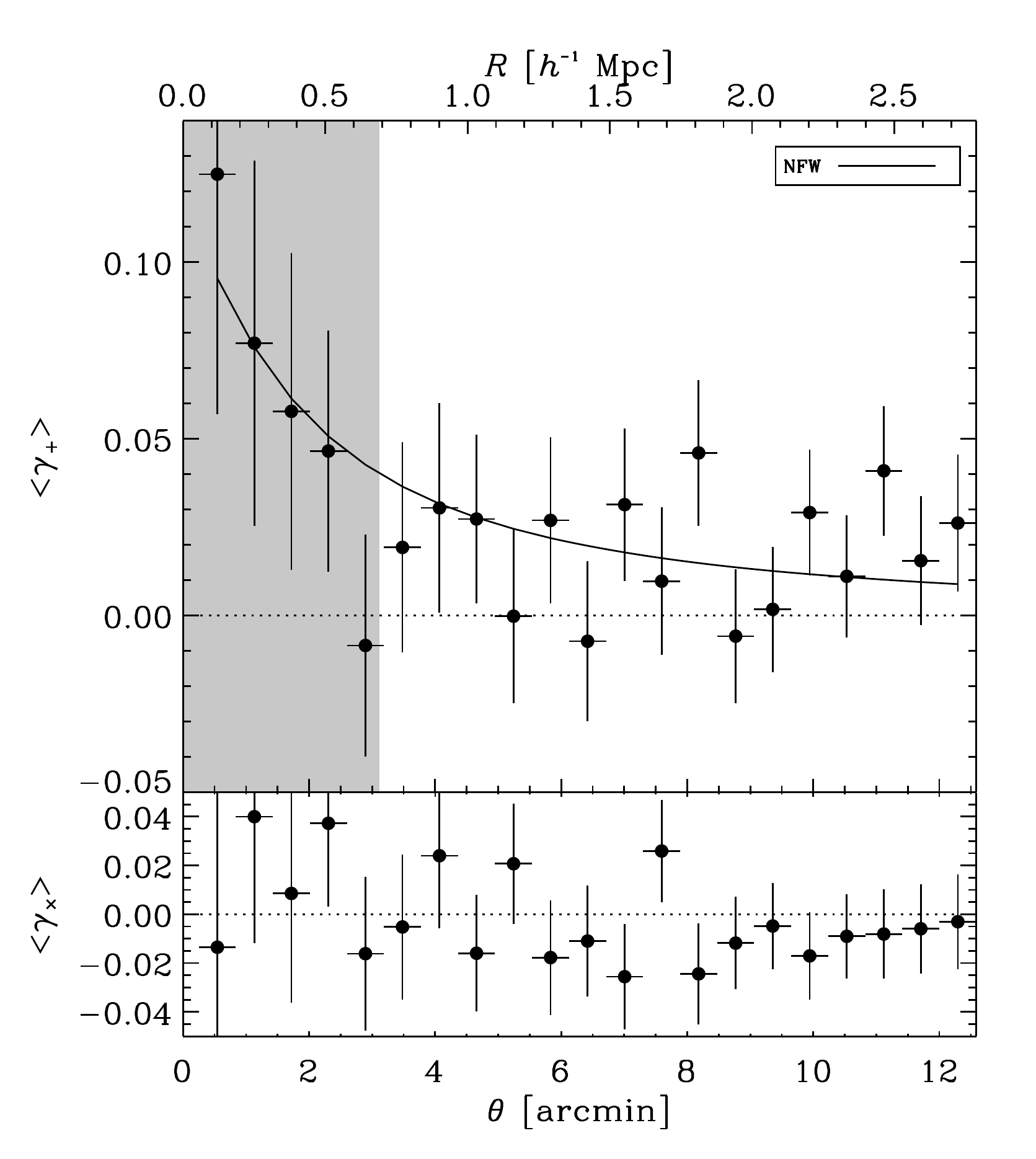}{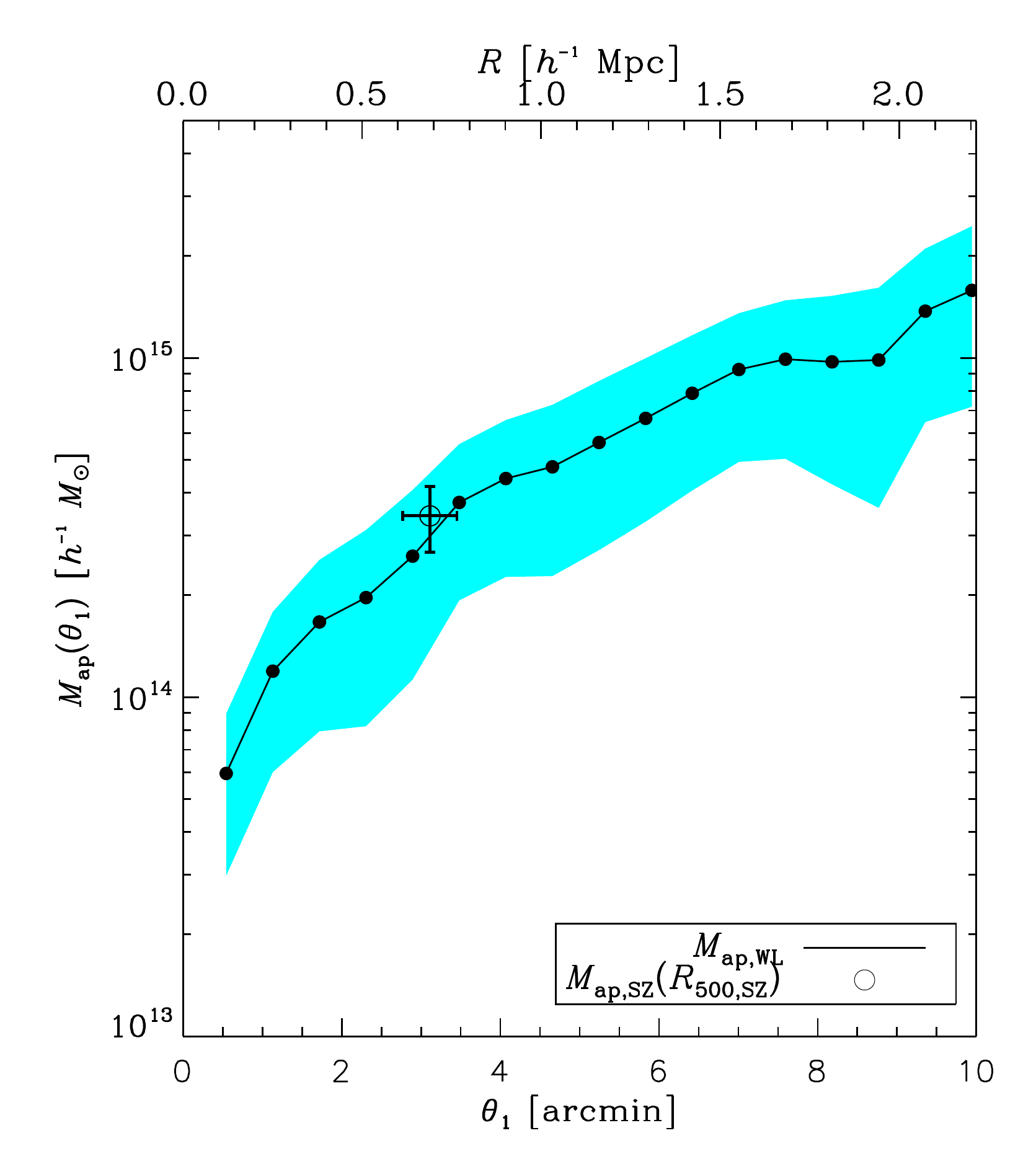}
\caption{Shear and aperture mass profiles of \clusterb.  See the Appendix for a description.\label{fig:profileb}}
\end{figure*}

\begin{figure*}
\epsscale{1.15}
\plottwo{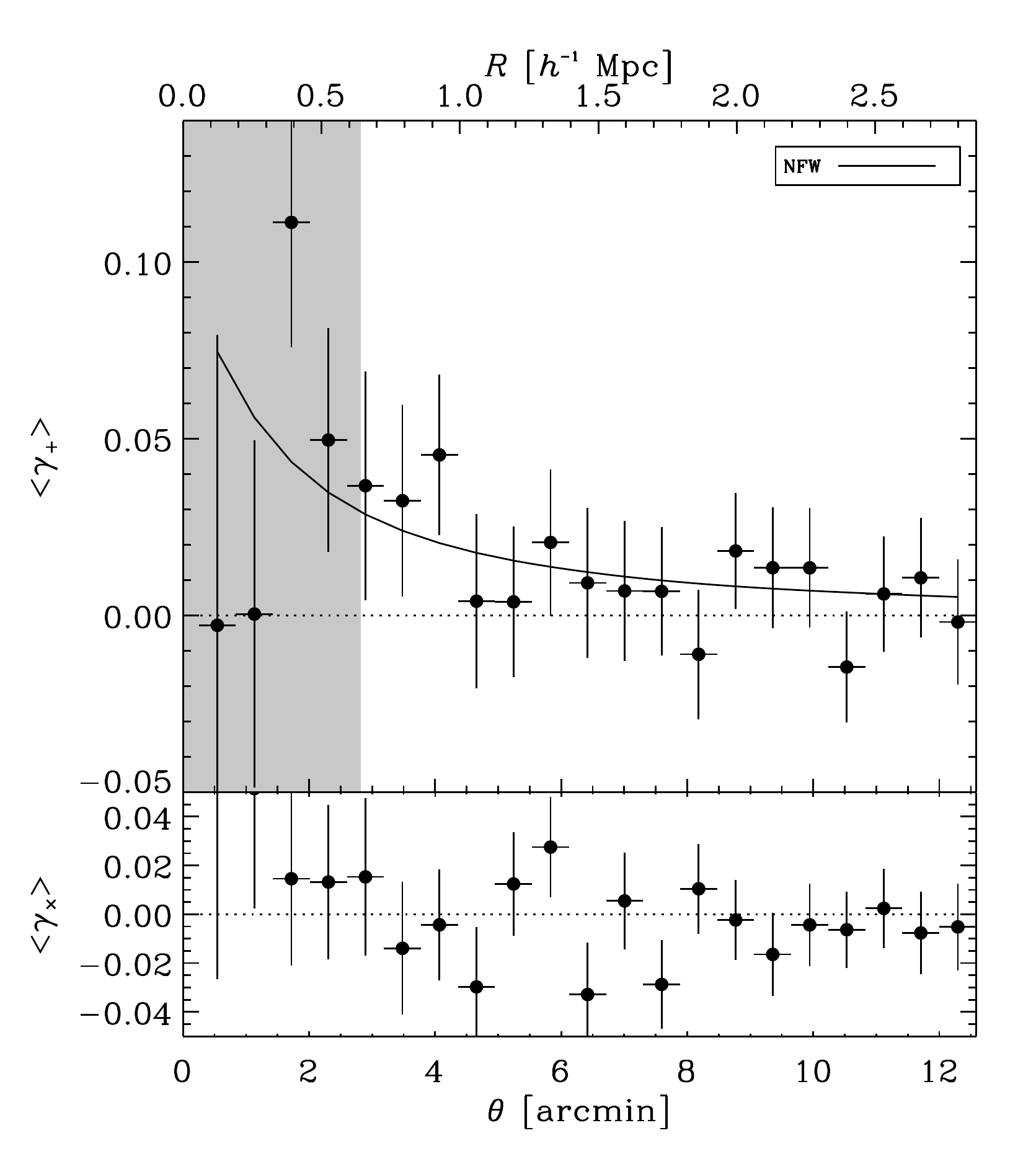}{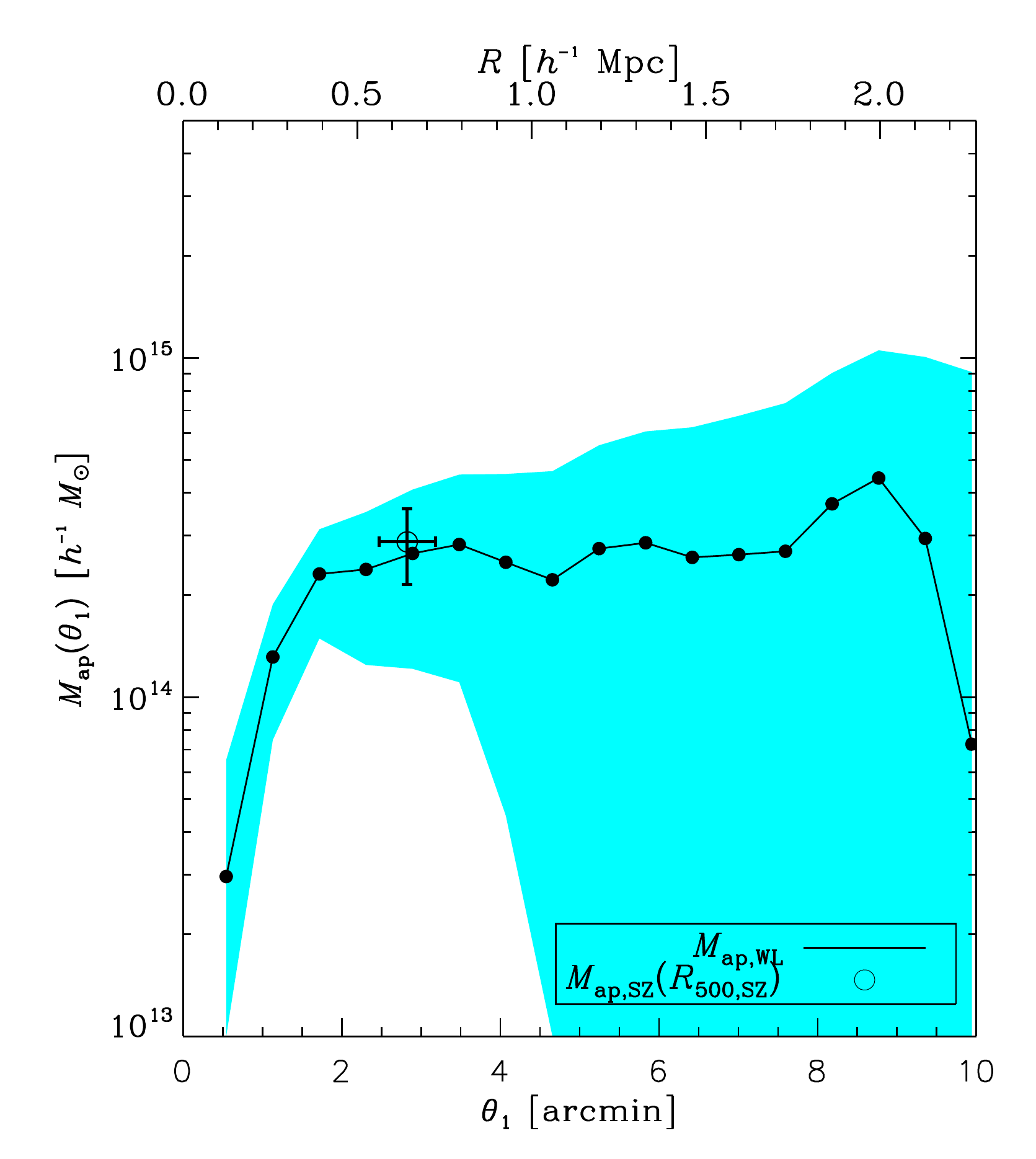}
\caption{Shear and aperture mass profiles of \clusterc.  See the Appendix for a description.\label{fig:profilec}}
\end{figure*}

\begin{figure*}
\epsscale{1.15}
\plottwo{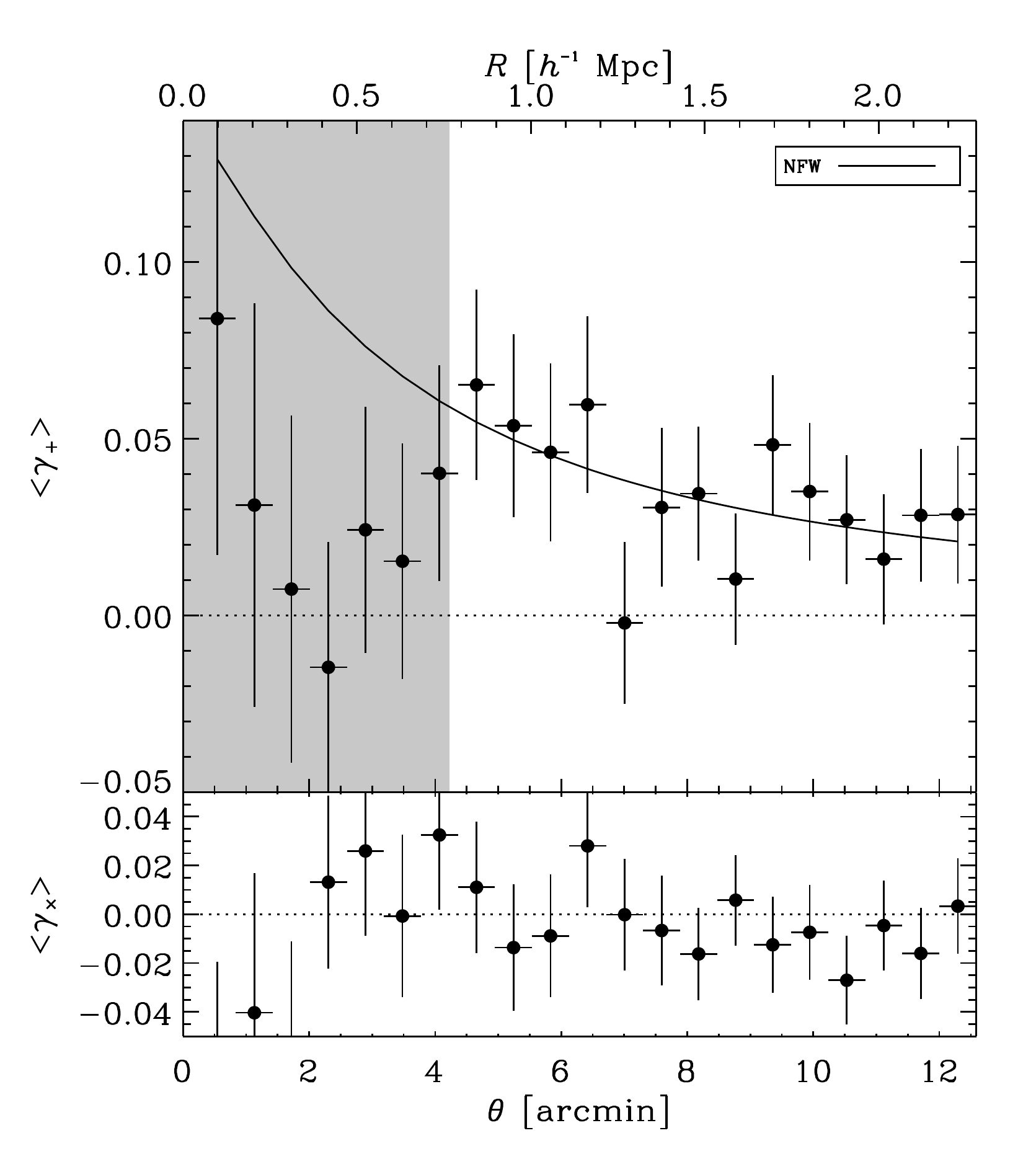}{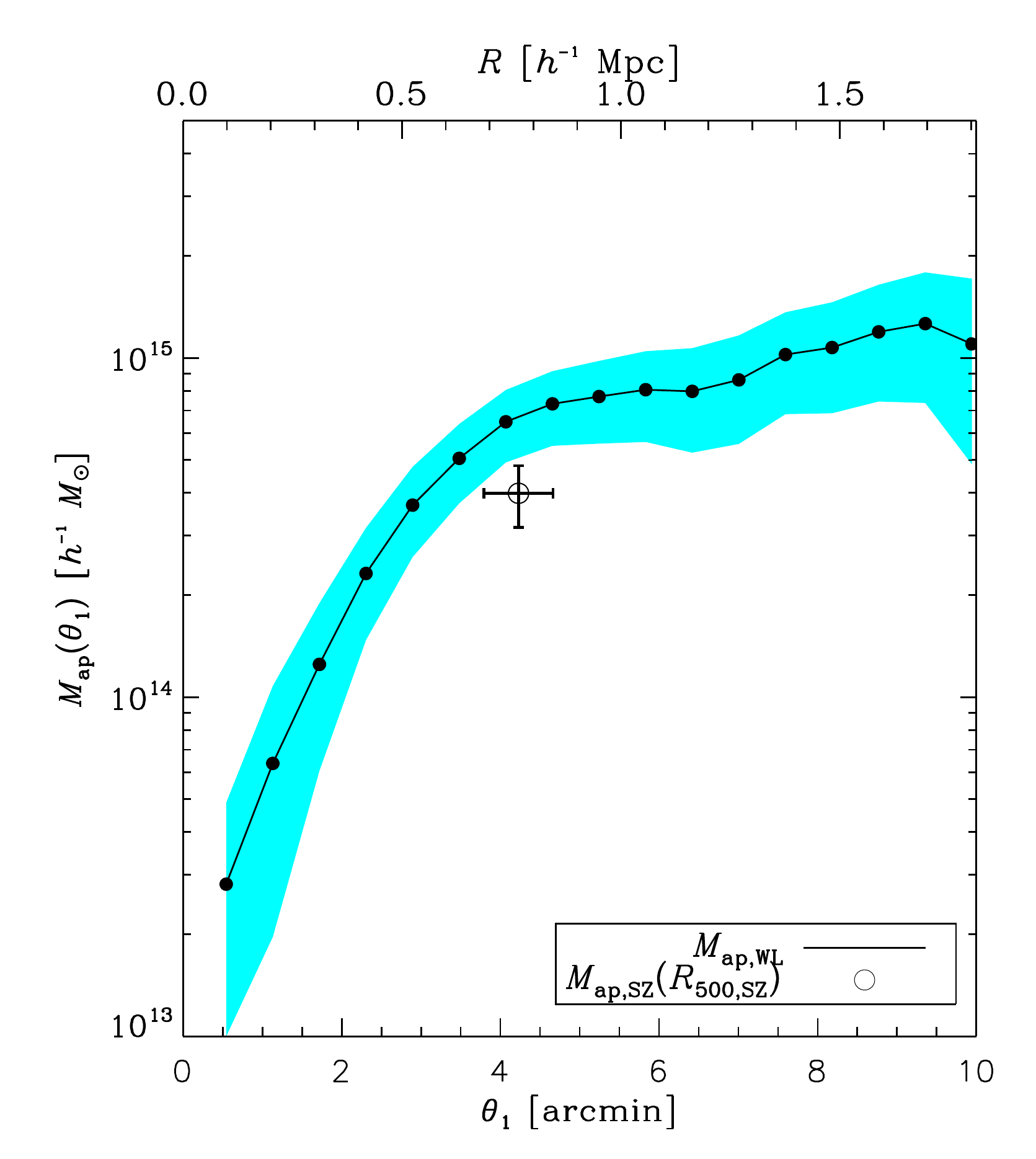}
\caption{Shear and aperture mass profiles of \clusterd.  See the Appendix for a description.\label{fig:profiled}}
\end{figure*}

\begin{figure*}
\epsscale{1.15}
\plottwo{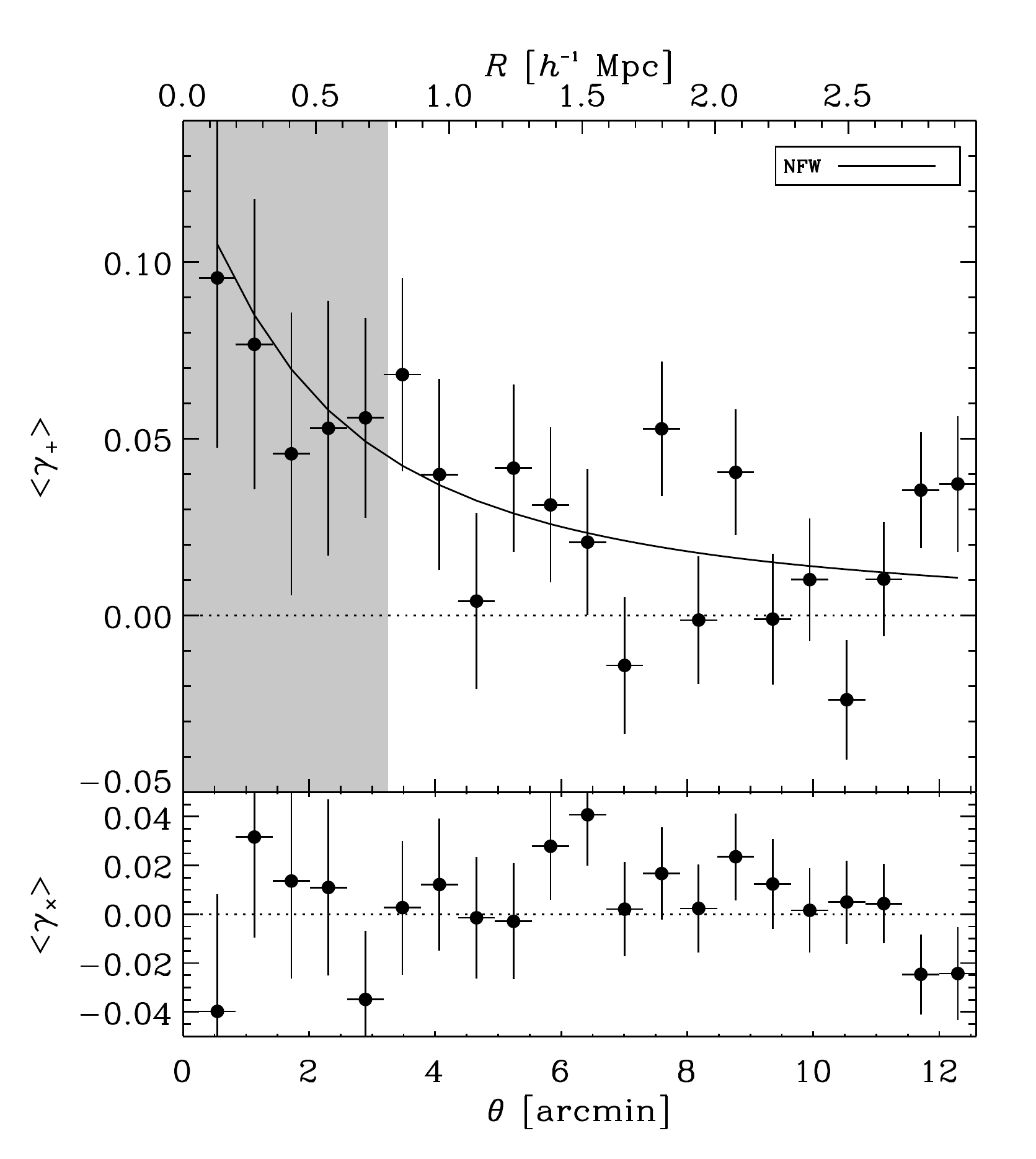}{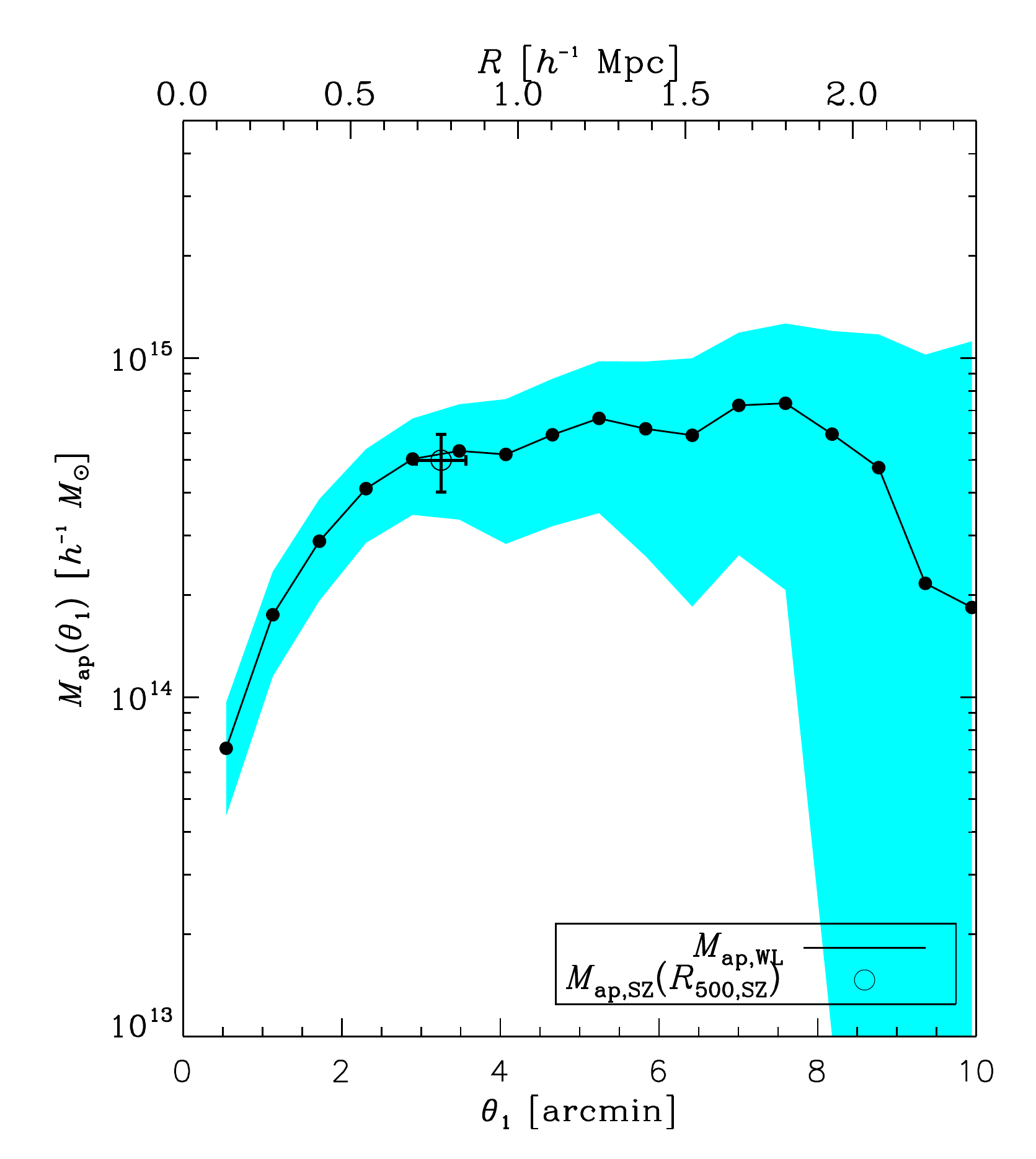}
\caption{Shear and aperture mass profiles of \clustere.  See the Appendix for a description.\label{fig:profilee}}
\end{figure*}

Shear and aperture mass profiles are shown in Figures
\ref{fig:profilea}, \ref{fig:profileb}, \ref{fig:profilec},
\ref{fig:profiled}, and \ref{fig:profilee}.  The shaded region denotes
radii interior to $R_{500,\mathrm{SZ}}$. The left panels show the
binned shear data with best-fit 
% SIS profiles and 
\citetalias{navarro97} profiles
assuming concentrations of \citetalias{duffy08}. 
These models are fit at radii $\theta_1 <
\theta <\theta_2$, where $\theta_1$ corresponds to
$R_{500,\mathrm{SZ}}$.%, and are in good agreement with one another at these radii.  
The right panels show the aperture mass
profiles within circular apertures of radius $\theta_1$.  For all
data, $\theta_2$ has been fixed to $12\arcmin$.  The cyan shaded
region denotes the $\pm 1\sigma_{\mathrm{stat}}$ excursion in the
aperture mass at a given $\theta_1$.  Cluster masses inferred from SZ
data are shown as data points with error bars.  These masses have been
projected assuming spherically symmetric, three dimensional
\citetalias{navarro97} profiles with \citetalias{duffy08}
concentrations, and then filtered using the same kernel as the WL
aperture mass statistic at the given $\theta_1$.

\clearpage

\clearpage

\clearpage

\clearpage

\clearpage

\end{document}